\documentclass[twocolumn]{aastex631}
\newcommand{\kms}{\ifmmode{\,\hbox{km\,s}^{-1}}\else {\rm\,km\,s$^{-1}$}\fi}
\begin{document}
\shorttitle{Globular Cluster Dark Matter}
\shortauthors{Carlberg \& Grillmair}
\title{Testing for Dark Matter in the Outskirts of Globular Clusters}
\author[0000-0002-7667-0081]{Raymond G. Carlberg}
\affiliation{Department of Astronomy \& Astrophysics 
University of Toronto 
Toronto, ON M5S 3H4, Canada} 
\email{raymond.carlberg@utoronto.ca}
\author[0000-0003-4072-169X]{Carl J. Grillmair}
\affiliation{IPAC, California Institute of Technology, Pasadena, CA 91125}
\email{carl@ipac.caltech.edu}

\begin{abstract}
The proper motions of stars in the outskirts of globular clusters are used to estimate cluster velocity dispersion profiles as far as possible within their tidal radii. We use individual color-magnitude diagrams to select high probability cluster stars for 25 metal-poor globular clusters within 20 kpc of the sun, 19 of which have substantial numbers of stars at large radii. Of the 19, 11 clusters have a falling velocity dispersion in the 3-6 half mass radii range, 6 are flat, and 2 plausibly have a rising velocity dispersion. The profiles are all in the range expected from simulated clusters started at high redshift in a zoom-in cosmological simulation. The 11 clusters with falling velocity dispersion profiles are consistent with no dark matter above the Galactic background. The  6 clusters with approximately flat velocity dispersion profiles could have local dark matter, but are ambiguous. The 2 clusters with rising velocity dispersion profiles are consistent with a remnant local dark matter halo, but need membership confirmation and detailed orbital modeling to further test these preliminary results.
\end{abstract}

\section{INTRODUCTION}

Whether or not globular clusters are surrounded by locally bound dark matter sub-halos as they orbit within the galaxy is an important observational question that will test theories of globular cluster origins and to some degree the nature of dark matter \citep{1984ApJ...277..470P}. Star clusters forming in a self-gravitating galactic gas disk rotating within a dark matter halo are not expected to have an excess of dark matter above the galactic background dark matter. The disk population of globular clusters \citep{1978ApJ...225..357S} is relatively metal rich, and younger than the halo clusters \citep{2013ApJ...775..134V}. However, even the metal poor halo clusters may have formed as the high mass, high gas pressure, regime of low metal gas disk star cluster formation at high redshift \citep{2010ARA&A..48..431P,2018ApJ...860..172S}. The halo clusters are very old \citep{2013ApJ...775..134V}, likely forming in pre-galactic sub-halos which later merged into the Milky Way's halo \citep{1978ApJ...225..357S}.

One possibility is that some halo clusters formed at or near the center of small dark matter halos. Tidal fields in an assembling galaxy later remove the outer dark matter, but clusters can retain dark matter within their tidal radii to the present day \citep{Ibata13,2020MNRAS.492.3169B,CK}.  An example is  the cluster NGC~6715 (M54), located at the center of  the Sagittarius dwarf galaxy, an accreted dwarf galaxy being dispersed into the Milky Way \citep{Ibata97,SgrDwarf20}. The velocity dispersion of M54 begins to rise  at about 5 half mass radii  \citep{Bellazzini08,Ibata09} likely due to the dwarf galaxy\rq{}s remnant dark matter halo. M54 is not typical, since it qualifies as a nuclear star cluster, having both an extended old stellar population and younger stars in the inner region \citep{M54MUSE}.  More generally, the lower mass nuclear star clusters are considered to be a globular cluster (or the merger of several)  in the centers of dark halos \cite{NSCReview2020}.  That is, some massive old globular clusters clearly were formed in the central regions of dark halos. On the other hand, the well studied Milky Way halo globular clusters are well fit with purely stellar models \citep{2018MNRAS.478.1520B,VB21} which provide the dynamically expected mass segregation which explains the modest rise in mass-to-light in the outer regions. Nevertheless a significant amount of dark matter is still allowed in the outskirts, though limits on the mass remain poorly constrained   \citep{Ibata13}.

The proper motions of individual globular cluster stars is coming within the range of observational capabilities \citep{2017MNRAS.464.2174B,2018MNRAS.478.1520B,2019MNRAS.482.5138B,2020MNRAS.495.2222S,2021MNRAS.502.4513W}. Current observational results show no evidence for any dark matter above the Galactic background, though these results are generally limited to 5 half-mass radii or less. A primary issue is to be able to discriminate cluster members from background stars.  A second issue is the development of dynamical models that predict stellar velocities to large cluster radii.

Numerical modeling shows that extended stellar density profiles arise as a result of internally driven expansion and the heating from external galactic tides \citep{1987degc.book.....S,2000MNRAS.318..753F,2017MNRAS.468.1453D,2016MNRAS.461..402T,2017MNRAS.471L..31P}. Beyond the tidal radius the outer region loses stars into tidal streams \citep{2003AJ....126.2385O,2006ApJ...643L..17G,2018MNRAS.473.2881K}. Developing a model which describes the entire range and its orbital phase dependence remains an active area of research. This paper analyzes the available {\it Gaia} EDR3 data  \citep{2016A&A...595A...1G, 2021A&A...649A...1G} for metal poor halo globular clusters sufficiently nearby to have good internal velocities from proper motions. The main innovations are color-magnitude weighting of the data and comparison to n-body simulation results that form globular clusters at various locations of high redshift sub-halos, including the sub-halo centers. This paper is a preliminary investigation using currently available data, straightforward measurement methods and a simple comparison of models to the data. 

\section{Globular Cluster Selection}

We select Milky Way clusters that have [Fe/H] below -1 and are within 20 kpc of the sun as promising targets for dark matter in the outskirts. The more distant clusters will have proportionally larger proper motion uncertainties, but we include them to determine where the falloff in internal kinematic measurement precision occurs for the currently available data. Setting a Galactic latitude limit of at least 20 degrees minimizes extinction and extinction variations with position. Color-magnitude filtering requires accurate colors and brightnesses, which can be significantly compromised by high and variable extinction. Higher latitudes also greatly reduce the surface density of contaminating field stars.

The cluster sample is listed in Table~\ref{table_clusters} along with some of their properties as listed in \citet{2019MNRAS.482.5138B} and available at https://\-people.smp.uq.edu.au/\-HolgerBaumgardt/\-globular/. The fifth column is a qualitative assessment of the shape of the outer velocity dispersion-radius relation in the range of 3-6 half-mass radii in Figures~\ref{fig_6752}-\ref{fig_4147} below. The last two columns give the current orbital radius relative to the apocenter and the orbital eccentricity. The clusters in the table are in the same order as the plots below.
\begin{deluxetable}{lrrrrrr}
\tablecaption{Globular Cluster Sample Properties\label{table_clusters}}
\tablewidth{0pt}
\tablehead{
\colhead{Name} & \colhead{$D_\sun$} & \colhead{Mass} & \colhead{$r_{h}$} & \colhead{$\sigma_v$ trend} & \colhead{$r/r_{apo}$} & \colhead{e}\\
\colhead{} & \colhead{Kpc} & \colhead{$M_\sun$} & \colhead{pc} & & &\\
}
\startdata
NGC 6752 & 4.1 & $2.8\times 10^5$ & 5.3 & rising & 0.96&0.25\\
NGC 6218 & 4.6 & $1.1\times 10^5$ & 4.1 & falling &0.98 &0.34\\
NGC 6254 & 5.1 & $2.1 \times 10^5$ & 4.8 & falling & 0.95& 0.40 \\
NGC 6809 & 5.4 & $1.9\times 10^5$ & 7.0 & falling & 0.72&0.55\\
NGC 6171 & 5.6 & $7.5\times 10^4$ & 4.0 & no stars& 0.96 & 0.56\\
NGC 6205 & 7.4 & $5.5\times 10^5$ &5.3 & rising &1.00 & 0.70 \\
NGC 5904 & 7.5 & $3.9\times 10^5$ & 5.7 & falling & 0.26&0.79 \\
NGC 7099 & 8.5 & $1.4\times 10^5$ & 5.0 & falling &0.87& 0.69\\
NGC 6341 & 8.5 & $3.6\times 10^5$ & 4.5 & falling& 0.93& 0.83\\
NGC 288 & 9.0 & $9.3\times 10^4$ & 8.4 & falling& 0.99 & 0.60\\
NGC 362 & 8.8 &$2.8\times 10^5$ & 3.8 & flat/fall& 0.78&0.85\\
NGC 5272 & 10.2 & $4.1\times 10^5$ & 6.3 & flat/fall&0.77&0.47 \\
NGC 4590 & 10.4 & $1.2\times 10^5$ & 7.6 & flat/fall&0.35&0.53\\
NGC 7078 & 10.8 & $6.3\times 10^5$ & 4.3 & flat/fall&1.00&0.49\\
NGC 7089 & 11.7 & $6.2\times 10^5$ & 4.8 & falling&0.58&0.94 \\
NGC 1851 & 12.0 & $3.2\times 10^5$& 2.9 & falling&0.84&0.92 \\
NGC 5897 & 12.6 & $1.6\times 10^5$ & 11.0 & falling&0.80&0.53\\
NGC 1904 & 13.1 & $1.4\times 10^5$ & 3.2 & flat/fall&0.99&0.92\\
NGC 5466 & 16.1 & $6.0\times 10^4$ & 14.0 & no stars&0.26&0.78\\
NGC 6981 & 16.7 & $6.9\times 10^4$ & 6.0 & no stars&0.54&0.90\\
NGC 1261 & 16.4 & $1.8\times 10^5$ & 5.2 & falling&0.88&0.87\\
NGC 5053 & 17.5 & $7.4\times 10^4$ & 17.3 & no stars&1.00&0.26 \\
NGC 5024 & 18.5 & $4.6\times 10^5$ & 10.2 & flat/fall&0.84&0.41\\
IC 4499 & 18.9 & $1.5\times 10^5$ & 15.0 & no stars&0.55&0.63 \\
NGC 4147 & 18.5 & $3.9\times 10^4$ & 4.0 & no stars&0.83&0.86\\
\enddata
\label{tab_sample}
\end{deluxetable}

\begin{figure}
\begin{center}
{\includegraphics[angle=0,scale=0.35,trim=30 10 50 100,clip=true]{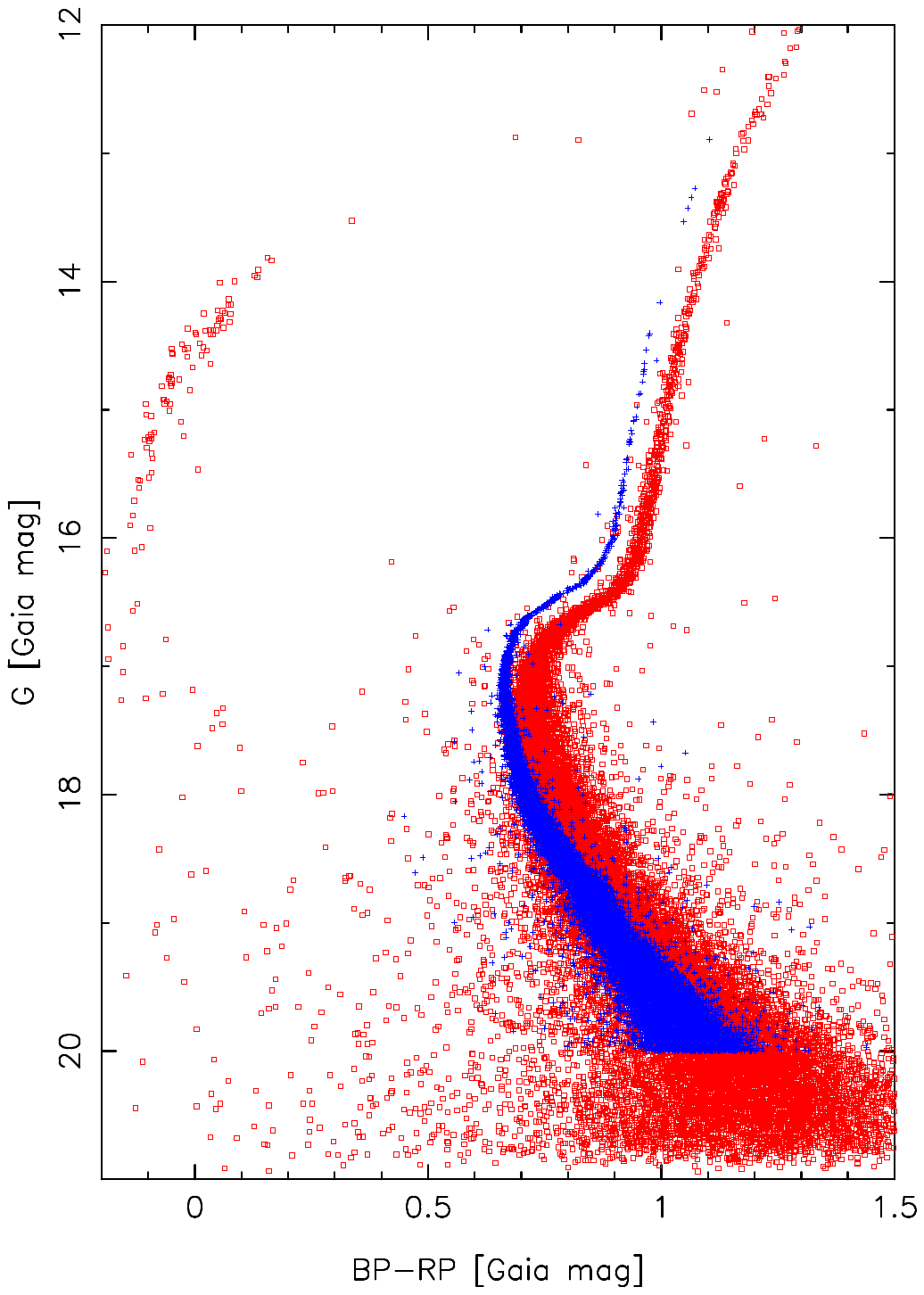}}
\caption{The color-magnitude diagram of the extinction corrected stars in the NGC~6752 cluster with less than $1\sigma$ color offset used here are shown as blue points. The kinematic mixture modeling \citep{VB21} 90\% probability cluster members, with no extinction correction, are shown as red points. Our color-magnitude selection does not include horizontal branch stars.
}
\end{center}
\label{fig_cm}
\end{figure}

\section{Cluster Star Photometric Selection}

Stars are selected using a matched filter technique, which depends on how close a star is to a color-magnitude locus at the distance and metallicity of the cluster \citep{2002AJ....124..349R,2009ApJ...693.1118G}. Specifically, we created a color-magnitude locus for each cluster using the observed {\it Gaia G}, $G_{BP} - G_{RP}$ measurements, extinction corrected using the reddening maps of \citet{1998ApJ...500..525S}, themselves corrected using the prescription of \citet{2011ApJ...737..103S}. We employed theoretical isochrones with appropriate metallicities from http://stev.oapd.inaf.it/cgi-bin/cmd \citep{2004A&A...422..205G}, though we adjusted these isochrones to better match the observed red giant branches, where the photometric uncertainties are particularly small. The color offset from the color-magnitude relation normalized to the quoted photometric uncertainties of every star less than $2\sigma$ from the locus is included as a measurement. Only stars brighter than $G=20$ mag are included in our sample. 

\begin{figure}
\begin{center}
{\includegraphics[angle=0,scale=0.35,trim=30 10 100 70,clip=true]{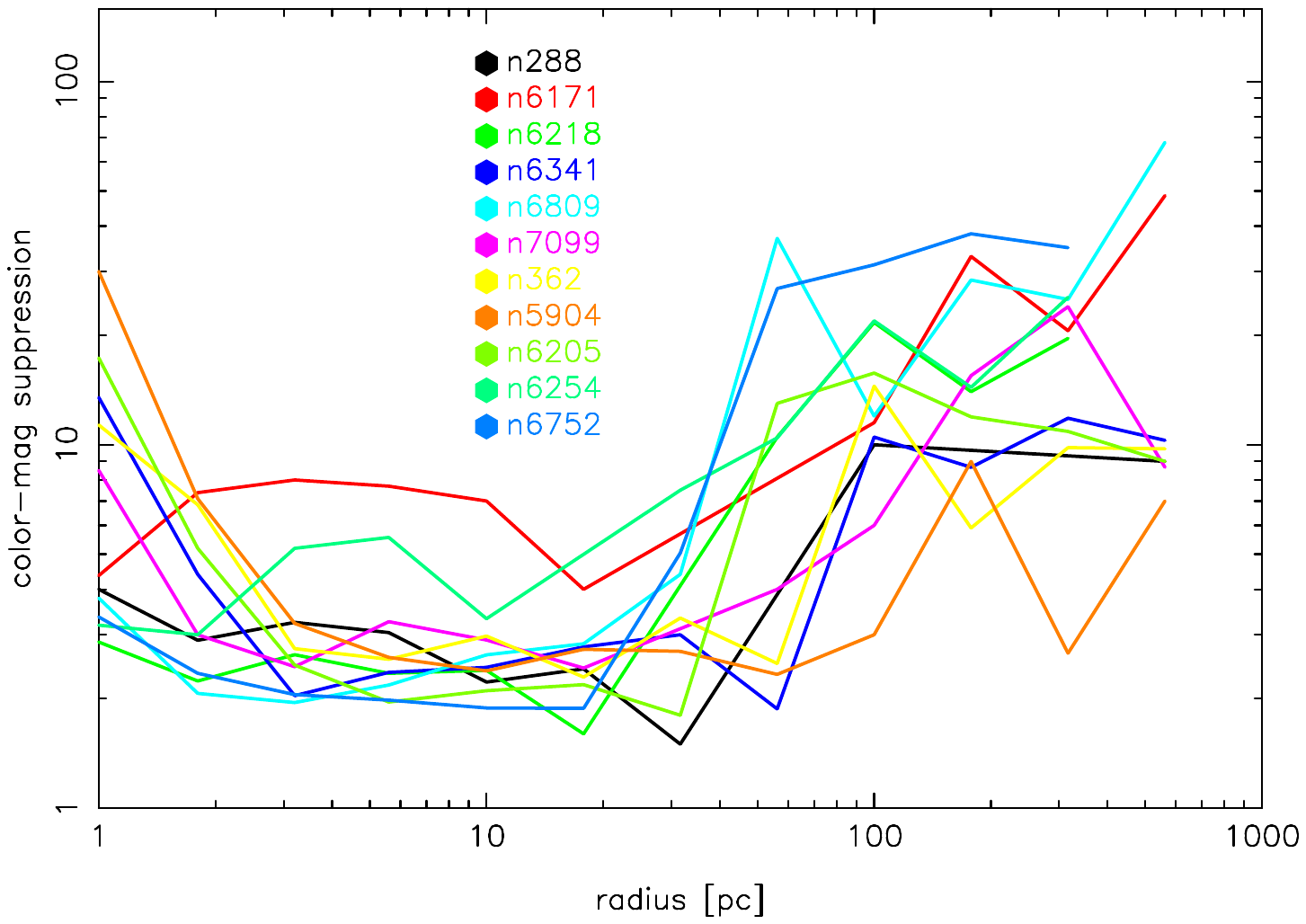}}
\caption{Ratio of  the numbers of all stars below the velocity cut in the radial range of interest to those selected to be close to the cluster's color magnitude relation for the 11 clusters within 10 kpc of the sun.
}
\end{center}
\label{fig_bg}
\end{figure}

Figure~\ref{fig_cm} compares our sample of stars for NGC~6752 to the 90\% probability cluster stars as found in a kinematically-based Gaussian mixture model \citep{VB21} for stars between projected radii of 5 and 100 pc of the cluster center. Stellar crowding in the inner 5 pc increases the photometric errors and reduces the limiting magnitude \citep{2017MNRAS.467..412P,2021A&A...649A...3R}, but these innermost stars are largely irrelevant to the goals of this paper. The horizontal branch stars can not be sufficiently narrowly defined to be usefully included in the photometric model of the cluster, but their numbers are sufficiently small that little is lost. The effectiveness of our photometric selection technique in reducing the number of unrelated field stars is shown in Figure~2 
where we plot the ratio of the total number of stars in a radial bin the number of stars photometrically classified as cluster members at $1\sigma$.  

Every star in the sample has an estimated error in each of the two components of proper motion, which we combine in quadrature for a total proper motion error. Only stars with proper motion errors less 0.15 mas/year are included in the analysis. The proper motion variance is also used with weighted measures of the velocity dispersion.  The color offset is used to construct a Gaussian weight $\exp{[-(c/\sigma_c)^2/2]}$, where $\sigma_c$ is set at 1.5 which gives a weight at the color edge of 0.41. The same analysis with $\sigma_c=1$ gives very similar results. For weighted velocity dispersion measurements, the color weight and the proper motion weight are multiplied together to give a total weight for each star. The radial component of the proper motion velocity is corrected with the viewing angle velocity, $\Delta v_r = V_r \cdot r/D_\sun$, where $V_r$ is the radial velocity of cluster measured from the sun at distance $D_\sun$ and $r$ is the projected radial distance of the star from the cluster center. The correction has no effect on the velocity dispersion.

\begin{figure*}
{\includegraphics[angle=0,scale=0.36,trim=30 10 100 80, clip=true]{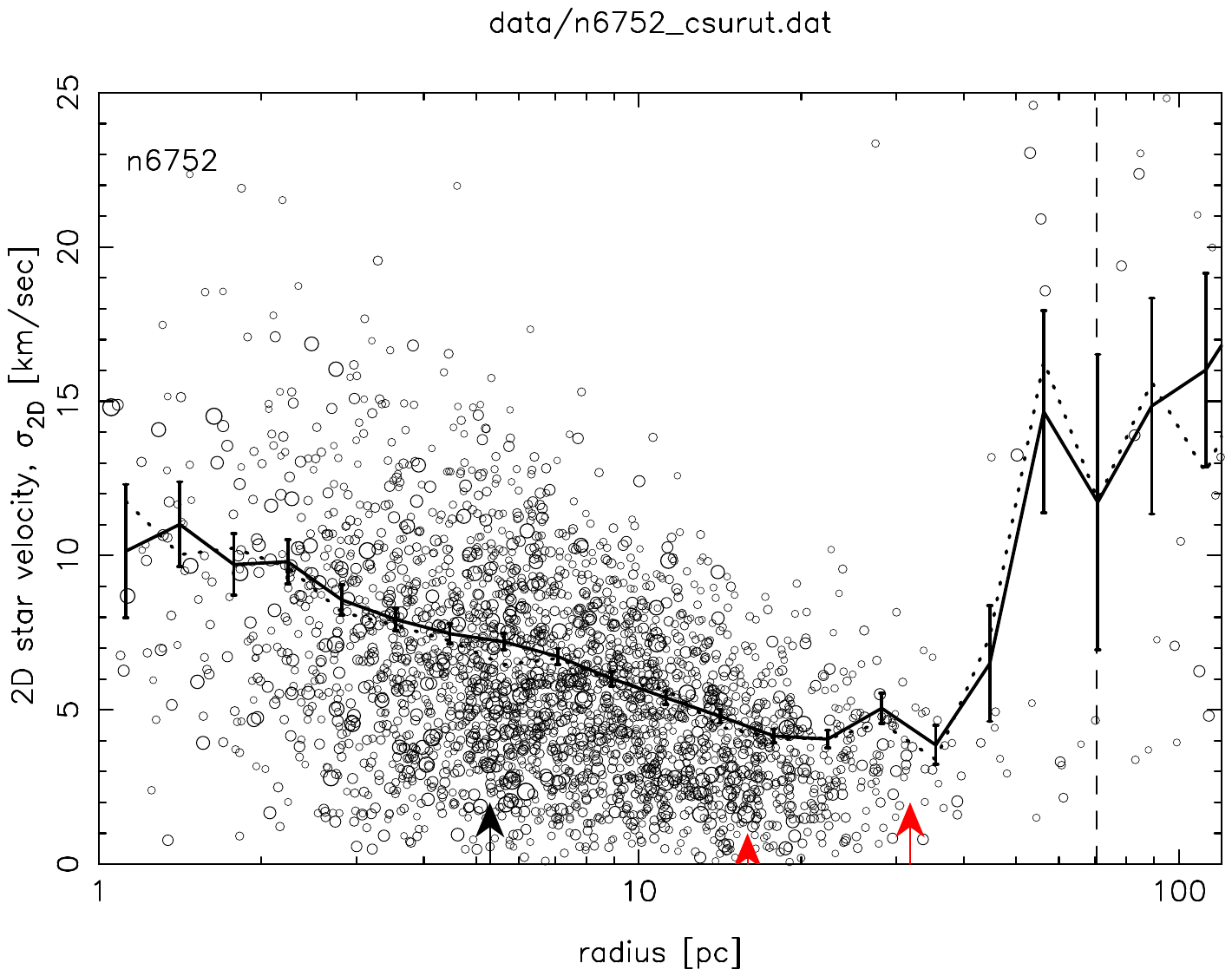}}
\put(5,0){\includegraphics[angle=0,scale=0.36,trim=30 10 100 80, clip=true]{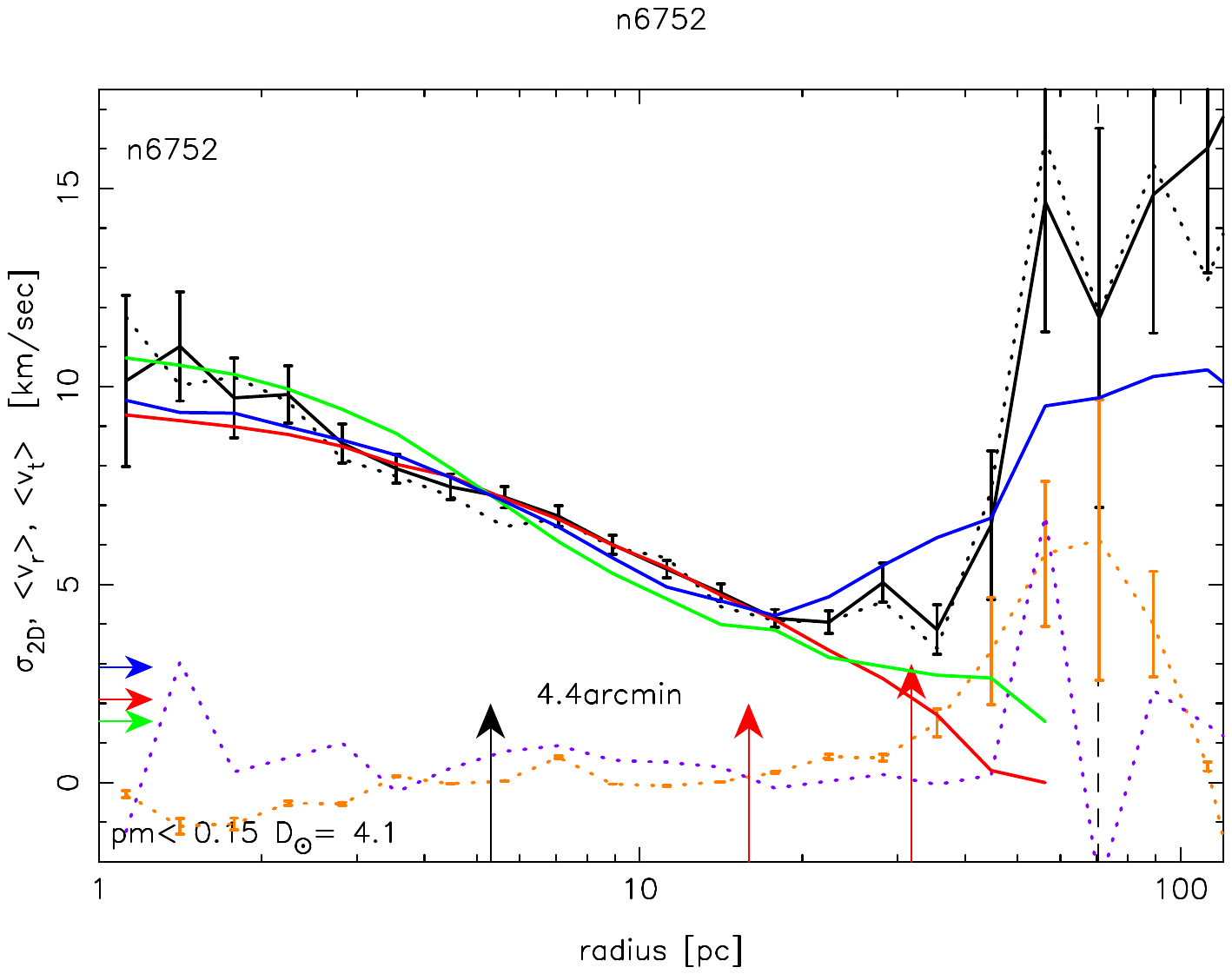}}
\put(120,120){\includegraphics[angle=0,scale=0.12,trim=30 10 100 80, clip=true]{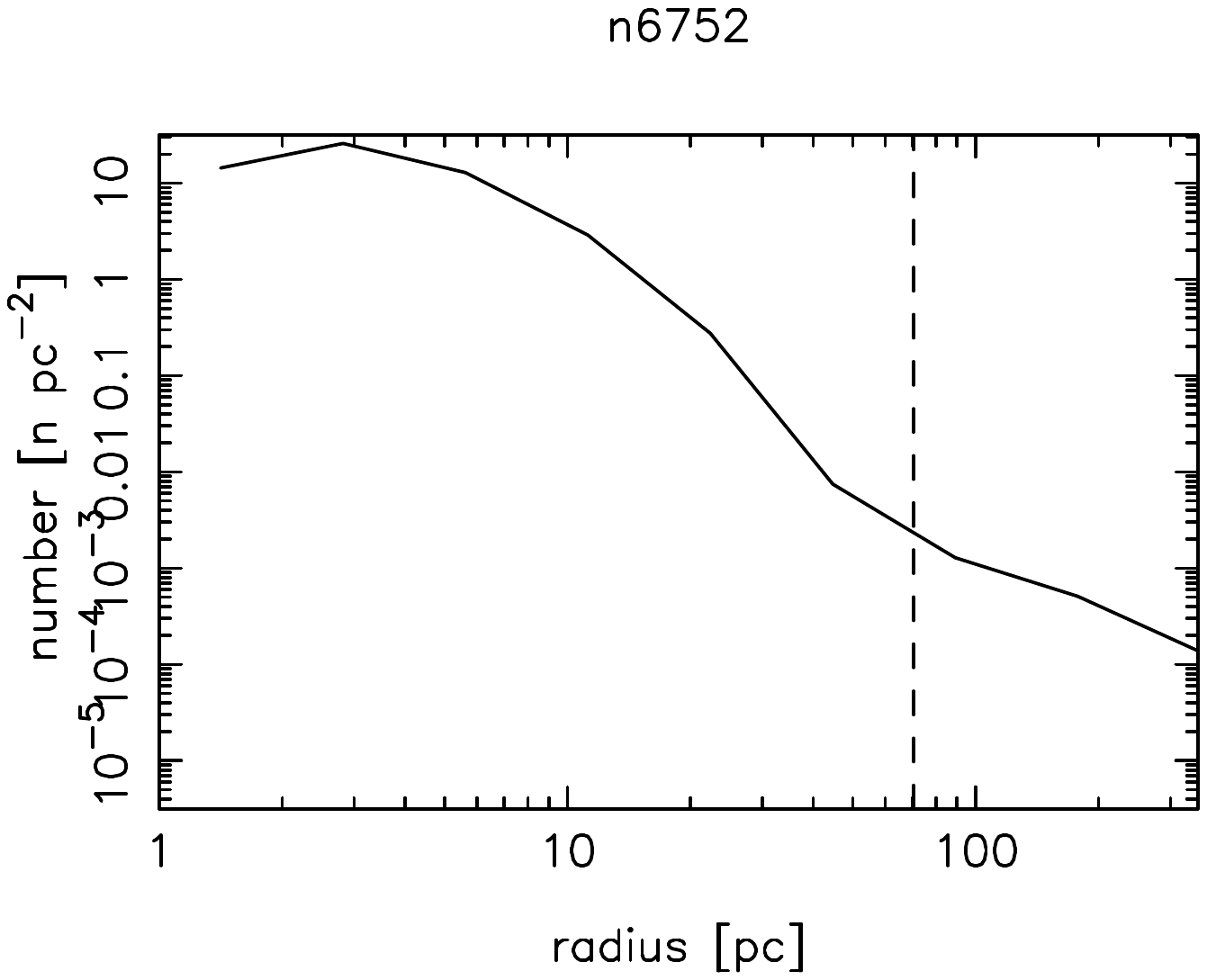}}
\caption{Measured velocities for NGC 6752. The left panel shows the projected velocities of selected stars within the velocity cut vs radius, while the right panel compares the radially binned results with dynamical models. Note that the radial range is larger in the left panel. The weighted and un-weighted RMS velocities are shown as the  black solid and dotted lines, respectively. The mean radial and tangential velocities are the dotted orange and purple lines, respectively. The plotted model profiles are:  a W=7 King model (red), a simulated cluster in the galactic dark matter background (green), and a cluster at its sub-halo center (blue).  The upward arrows on the horizontal axis indicate the 3D half-mass radius, and 3 and 6 times its value. The vertical dashed line is the tidal radius. The rightward facing arrows on the vertical axis indicate the velocity corresponding to the average proper motion error (red), the weighted average error (green) and the maximum allowed proper motion error (blue). The sizes of the circles in the left panel are proportional to the proper-motion weight. The inset shows the surface density of selected stars which flattens to the local background at large radius.
}
\label{fig_6752}
\end{figure*}

\begin{figure*}
{\includegraphics[angle=0,scale=0.36,trim=30 10 100 80,clip=true]{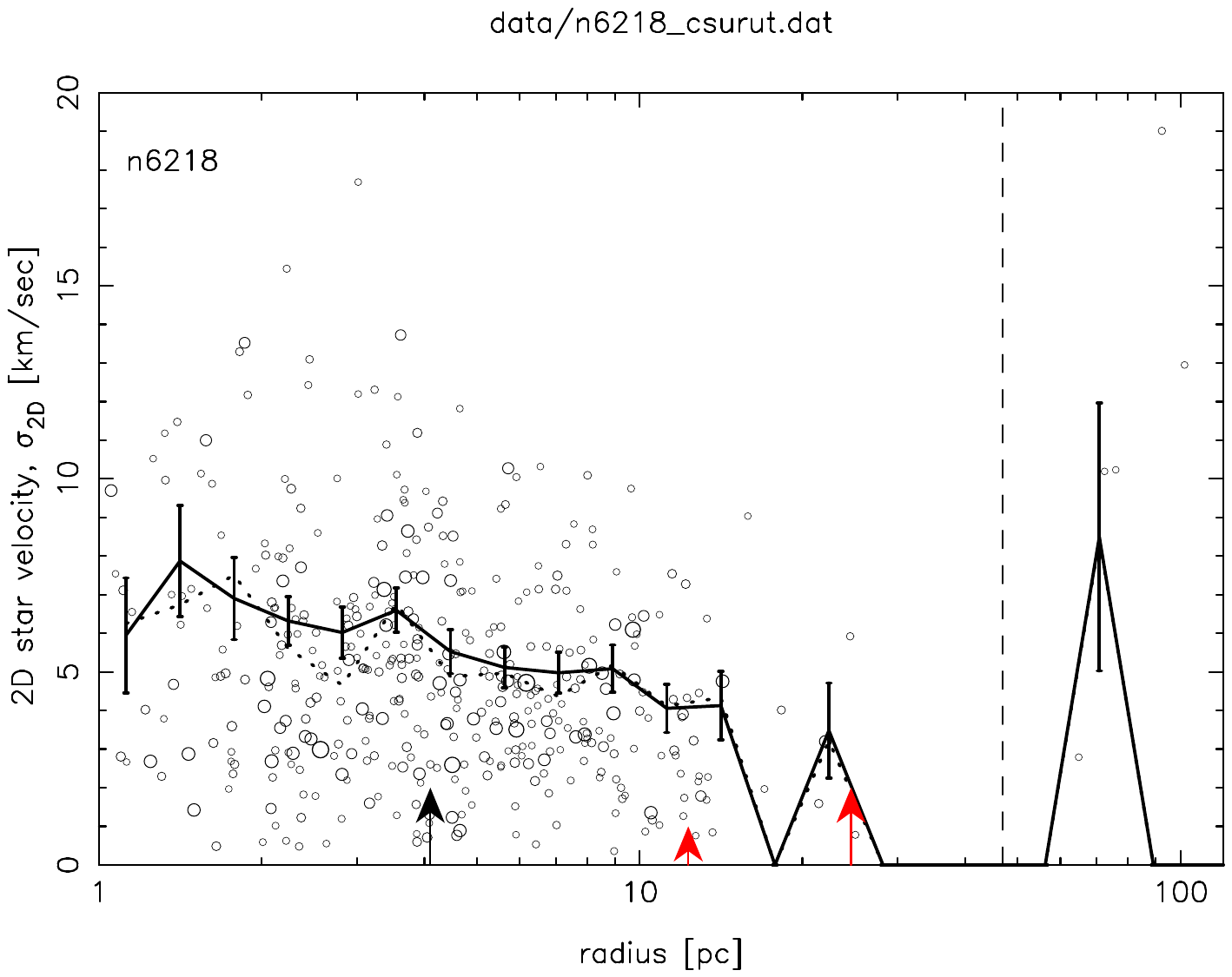}}
\put(5,0){\includegraphics[angle=0,scale=0.36,trim=30 10 100 80,clip=true]{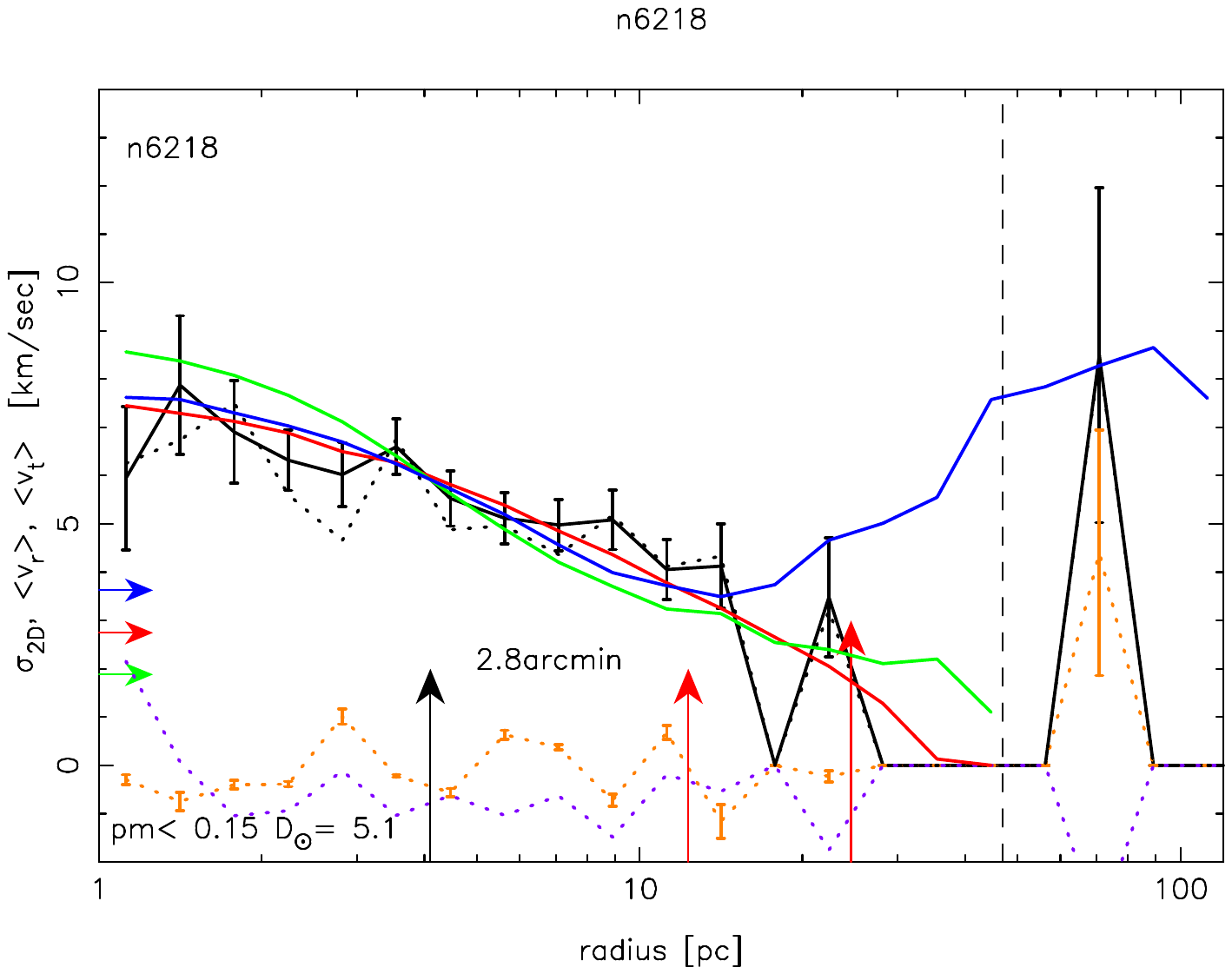}}
\put(120,120){\includegraphics[angle=0,scale=0.12,trim=30 10 100 80, clip=true]{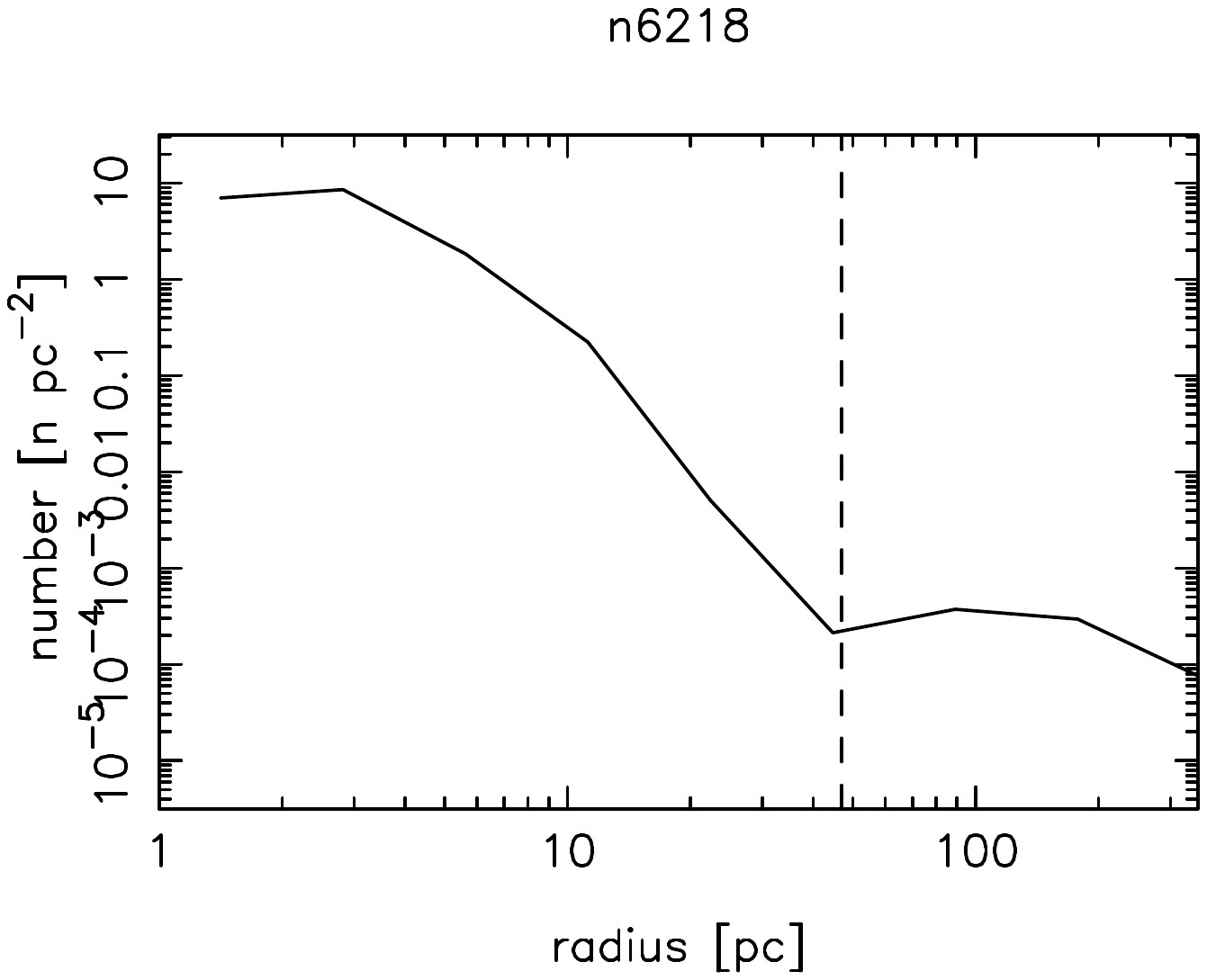}}
\caption{Velocities for the NGC 6218 cluster. Symbols are as in Figure~\ref{fig_6752}.
}
\label{fig_6218}
\end{figure*}
\begin{figure*}
{\includegraphics[angle=0,scale=0.36,trim=30 10 100 80, clip=true]{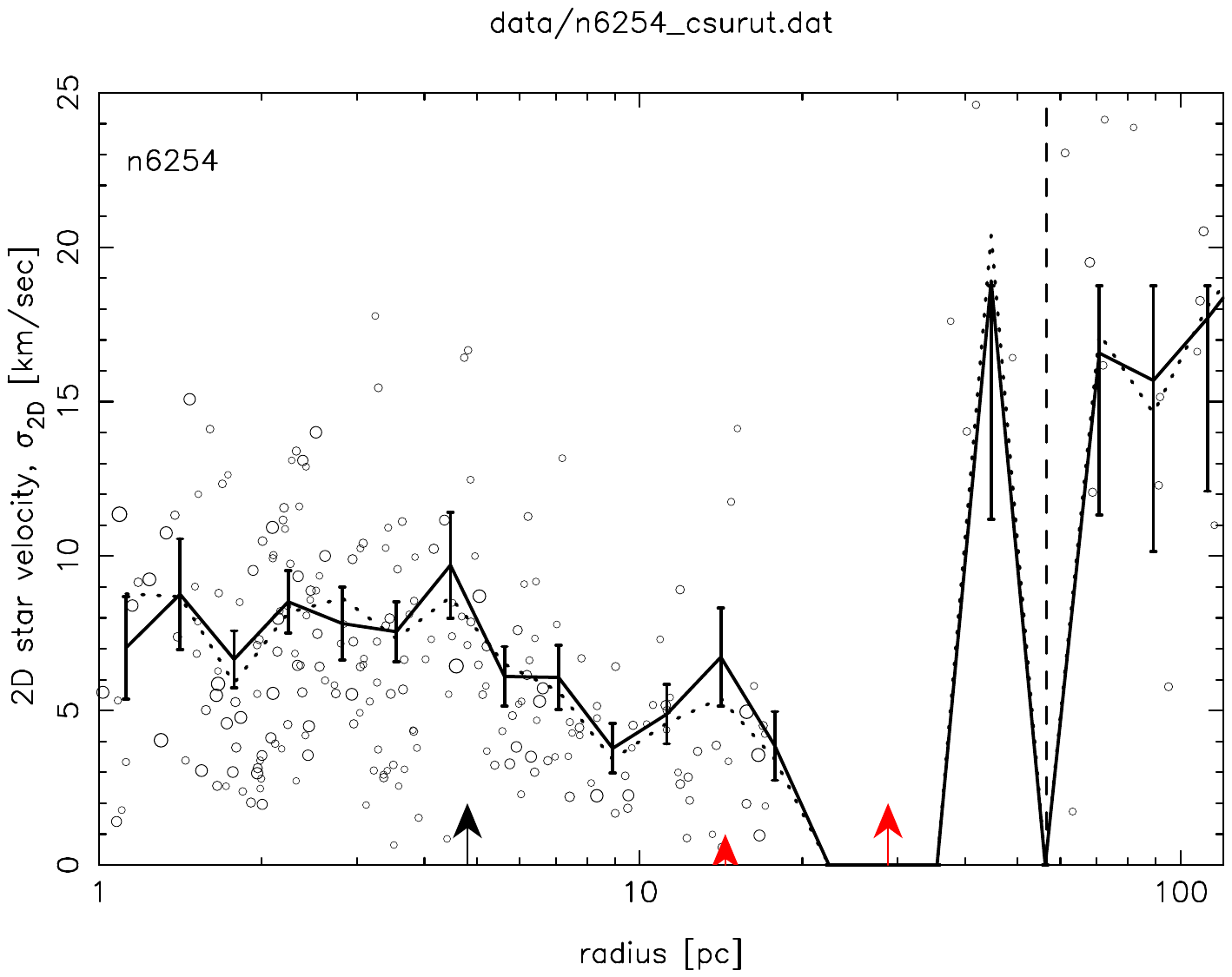}}
\put(5,0){\includegraphics[angle=0,scale=0.36,trim=30 10 100 80, clip=true]{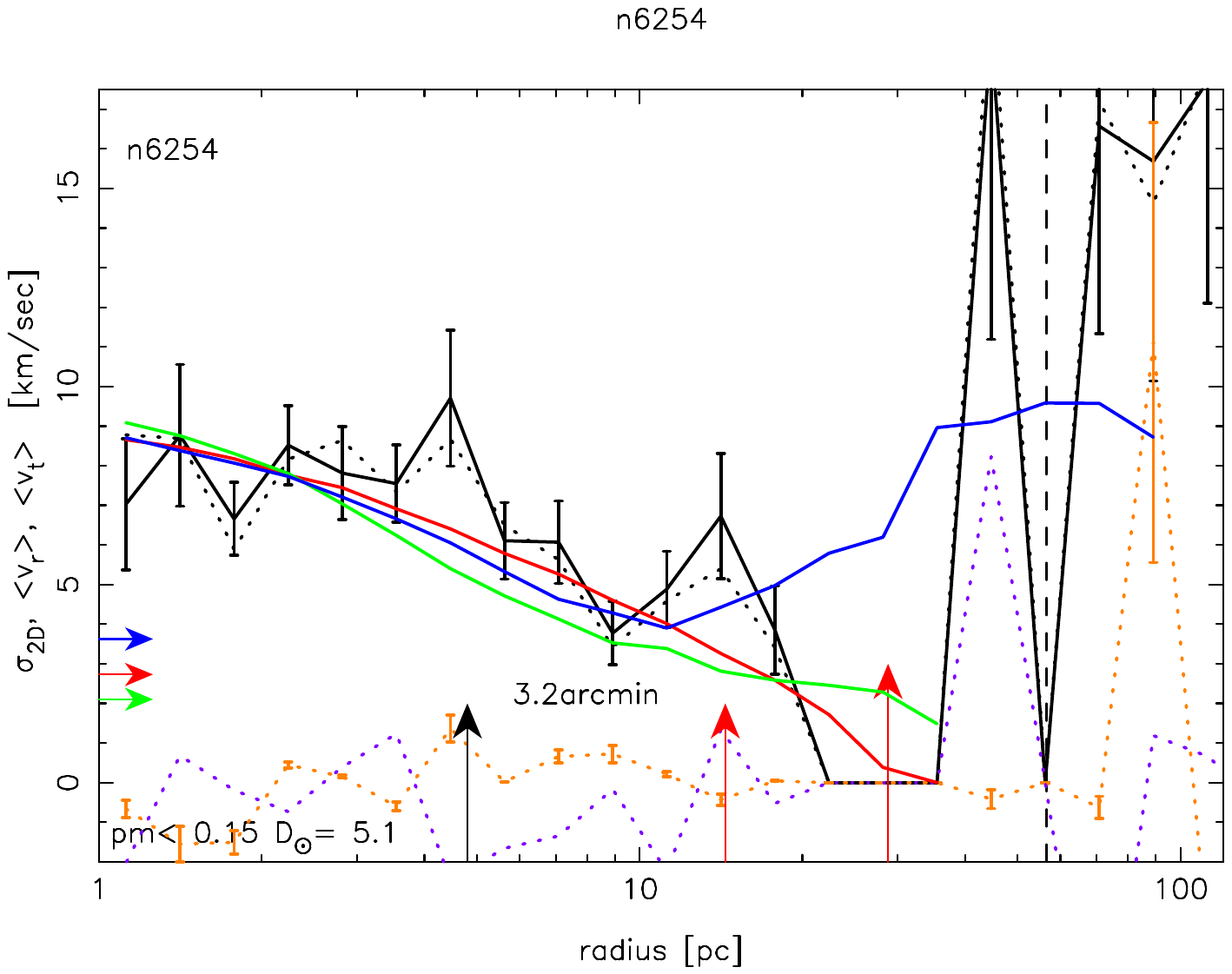}}
\put(120,120){\includegraphics[angle=0,scale=0.12,trim=30 10 100 80, clip=true]{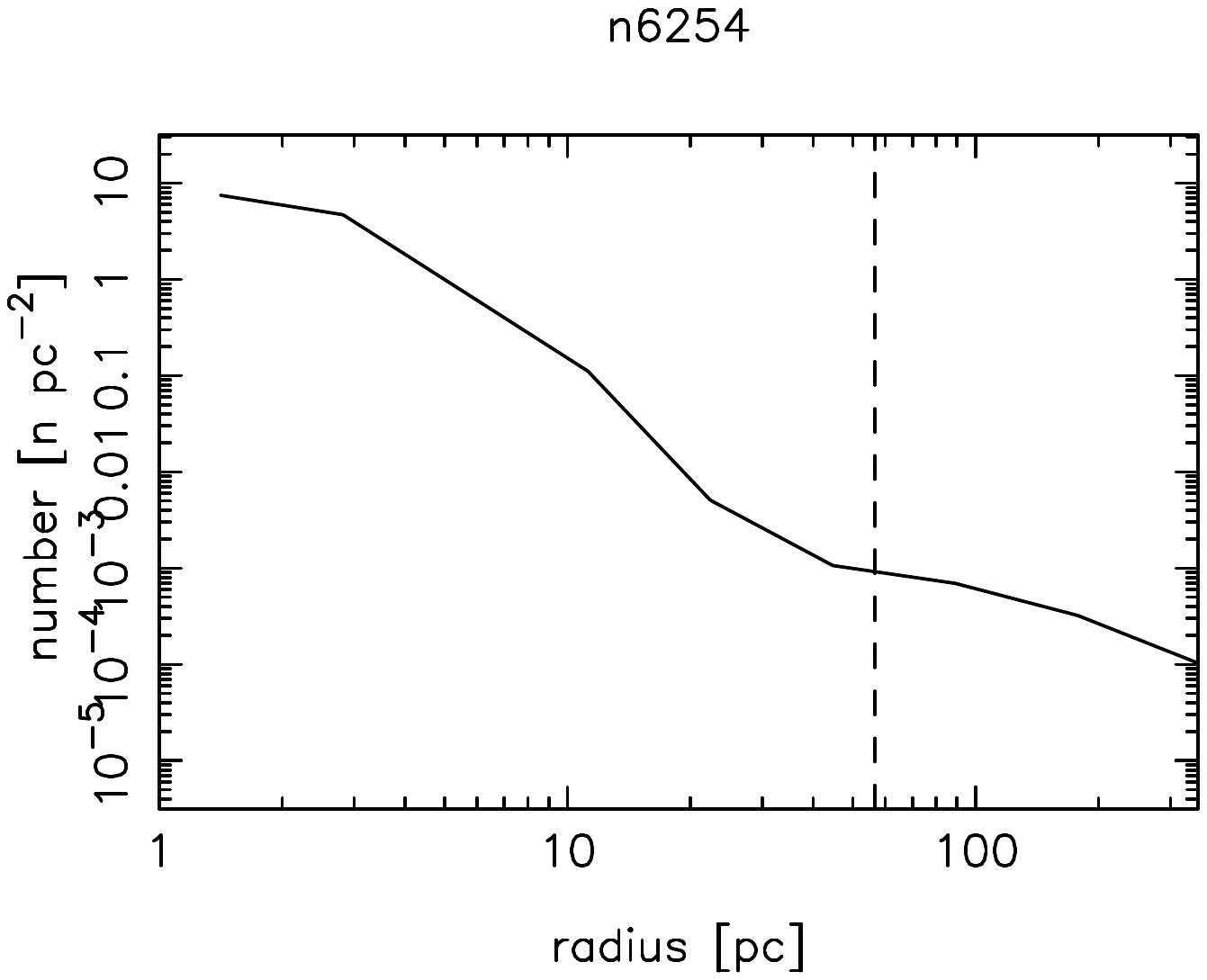}}

{\includegraphics[angle=0,scale=0.36,trim=30 10 100 80, clip=true]{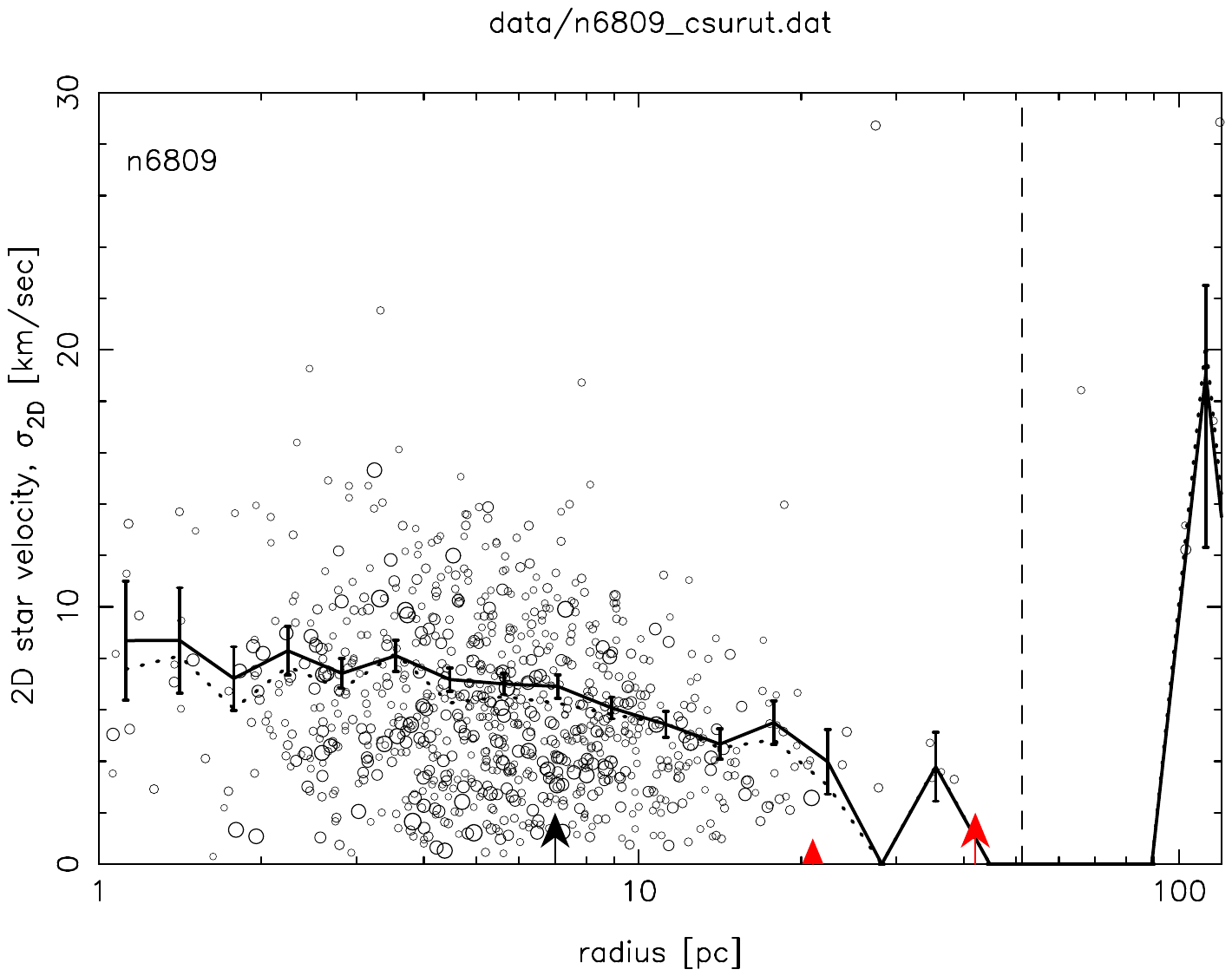}}
\put(5,0){\includegraphics[angle=0,scale=0.36,trim=30 10 100 80, clip=true]{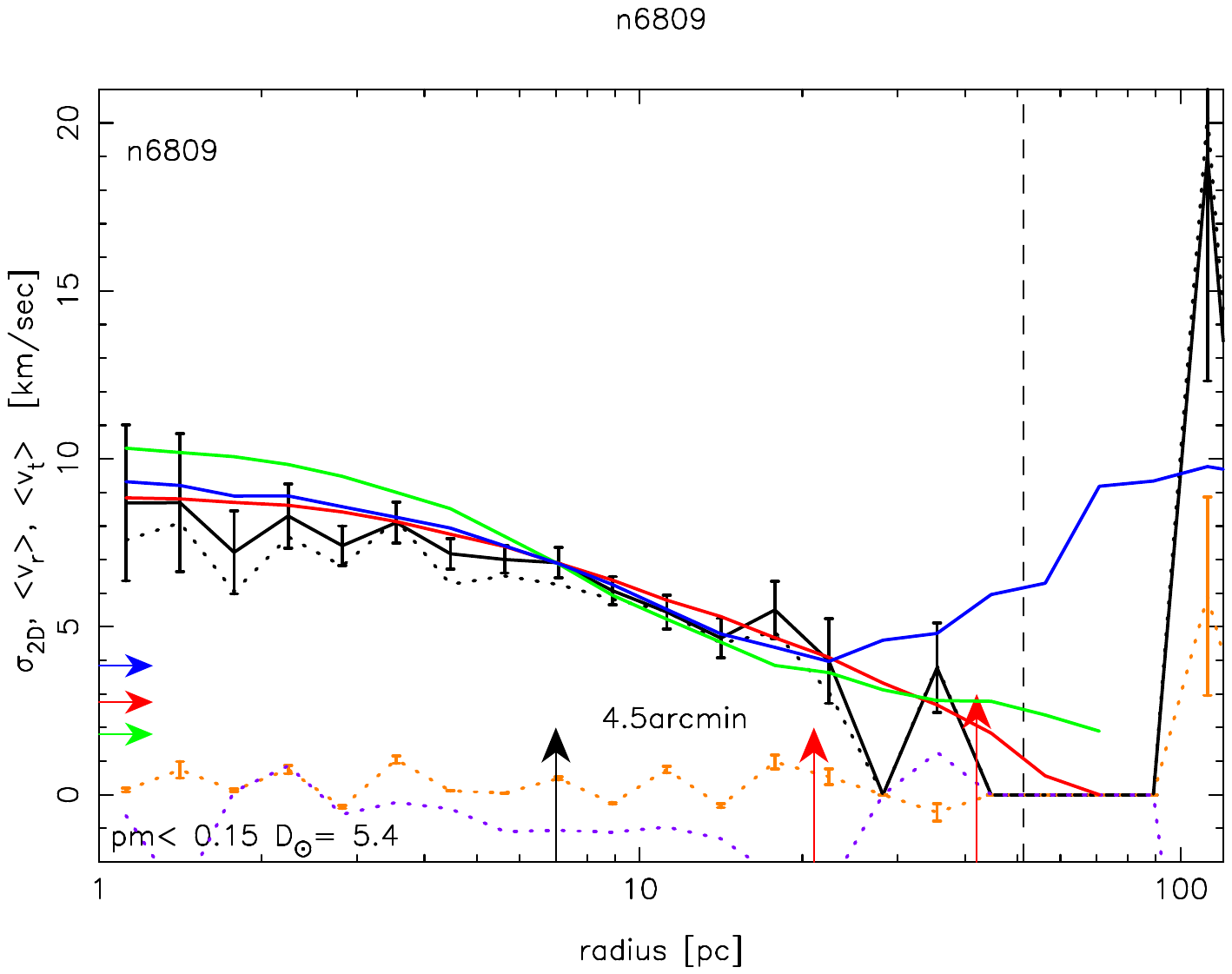}}
\put(120,120){\includegraphics[angle=0,scale=0.12,trim=30 10 100 80, clip=true]{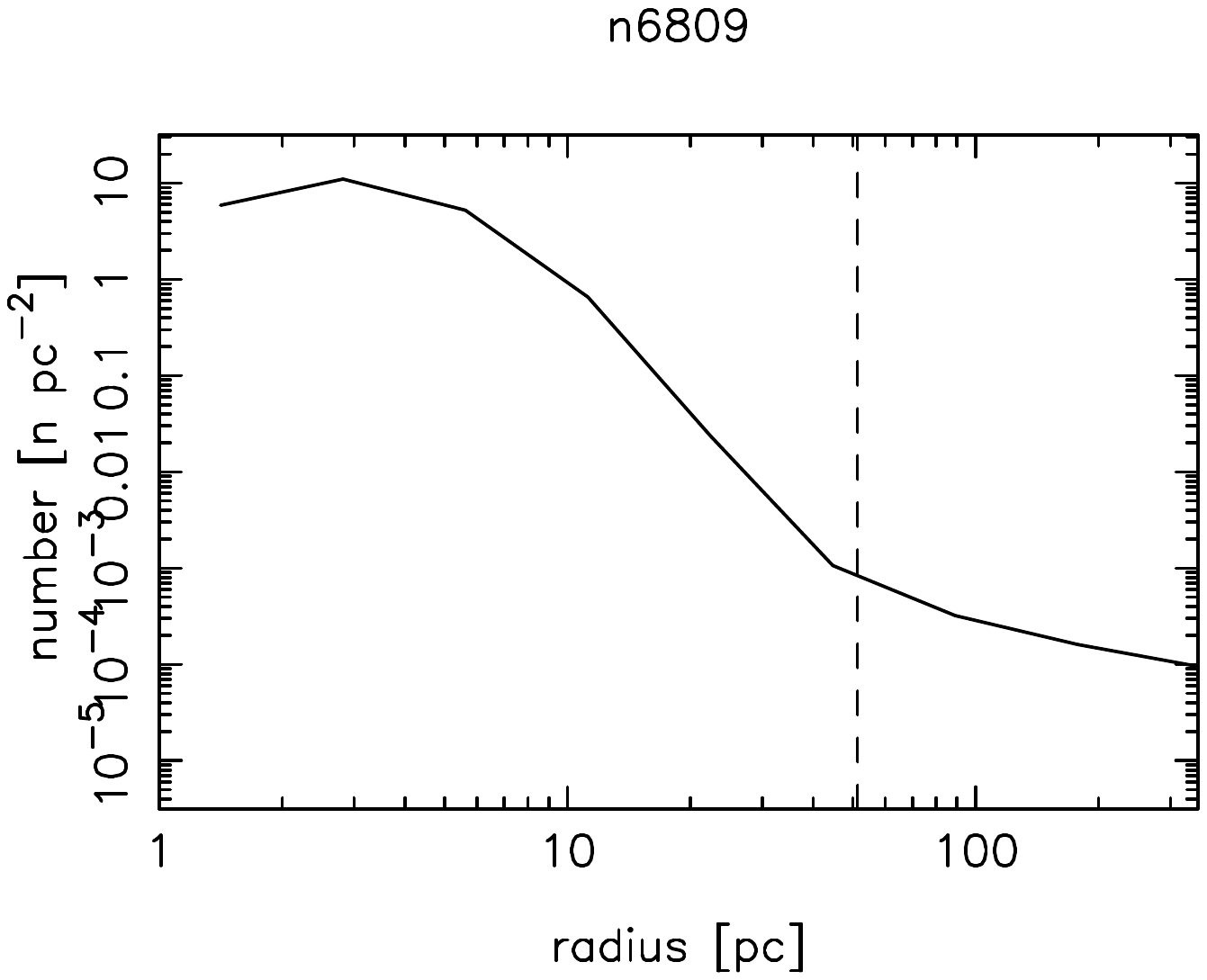}}

{\includegraphics[angle=0,scale=0.36,trim=30 10 100 80, clip=true]{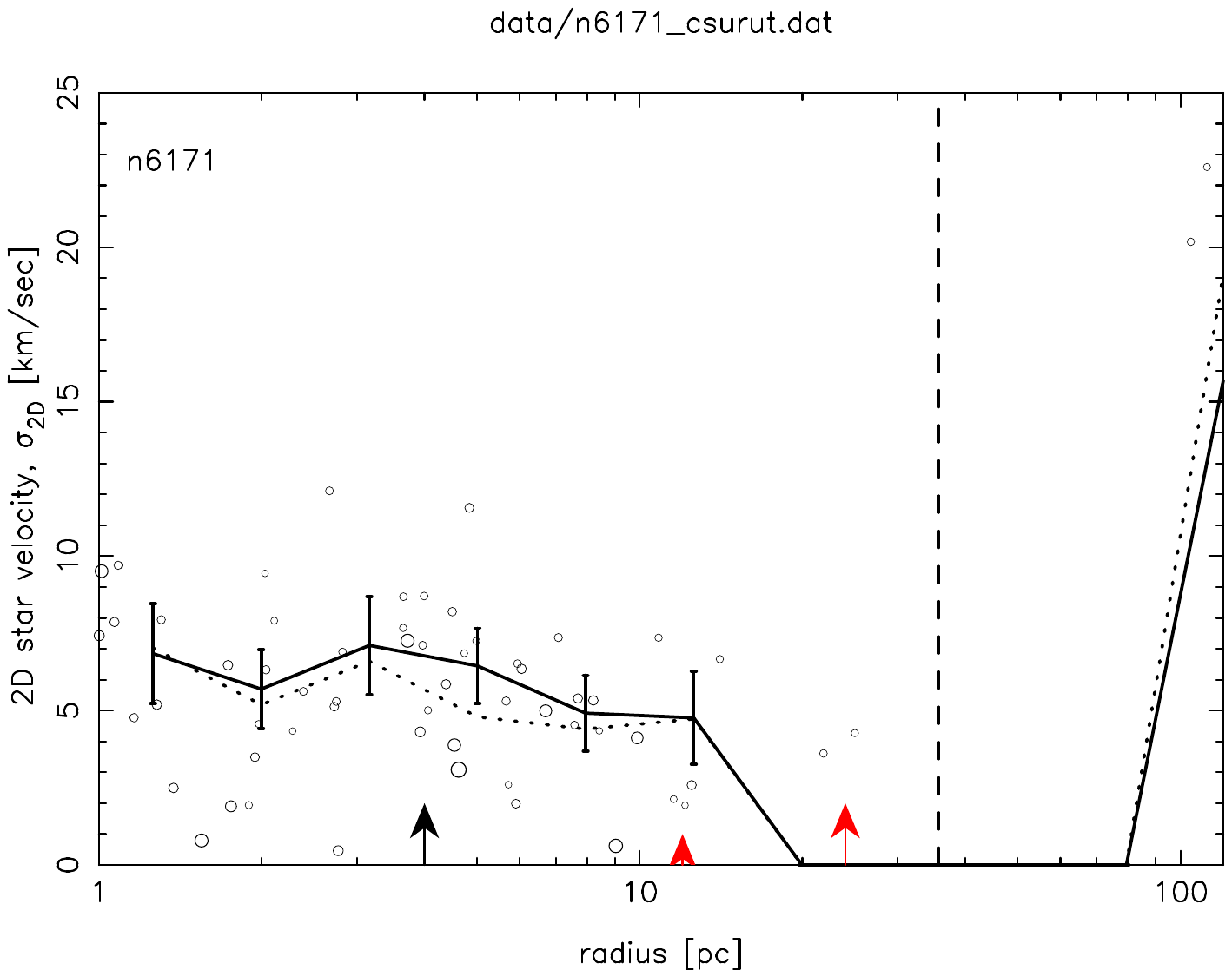}}
\put(5,0){\includegraphics[angle=0,scale=0.36,trim=30 10 100 80, clip=true]{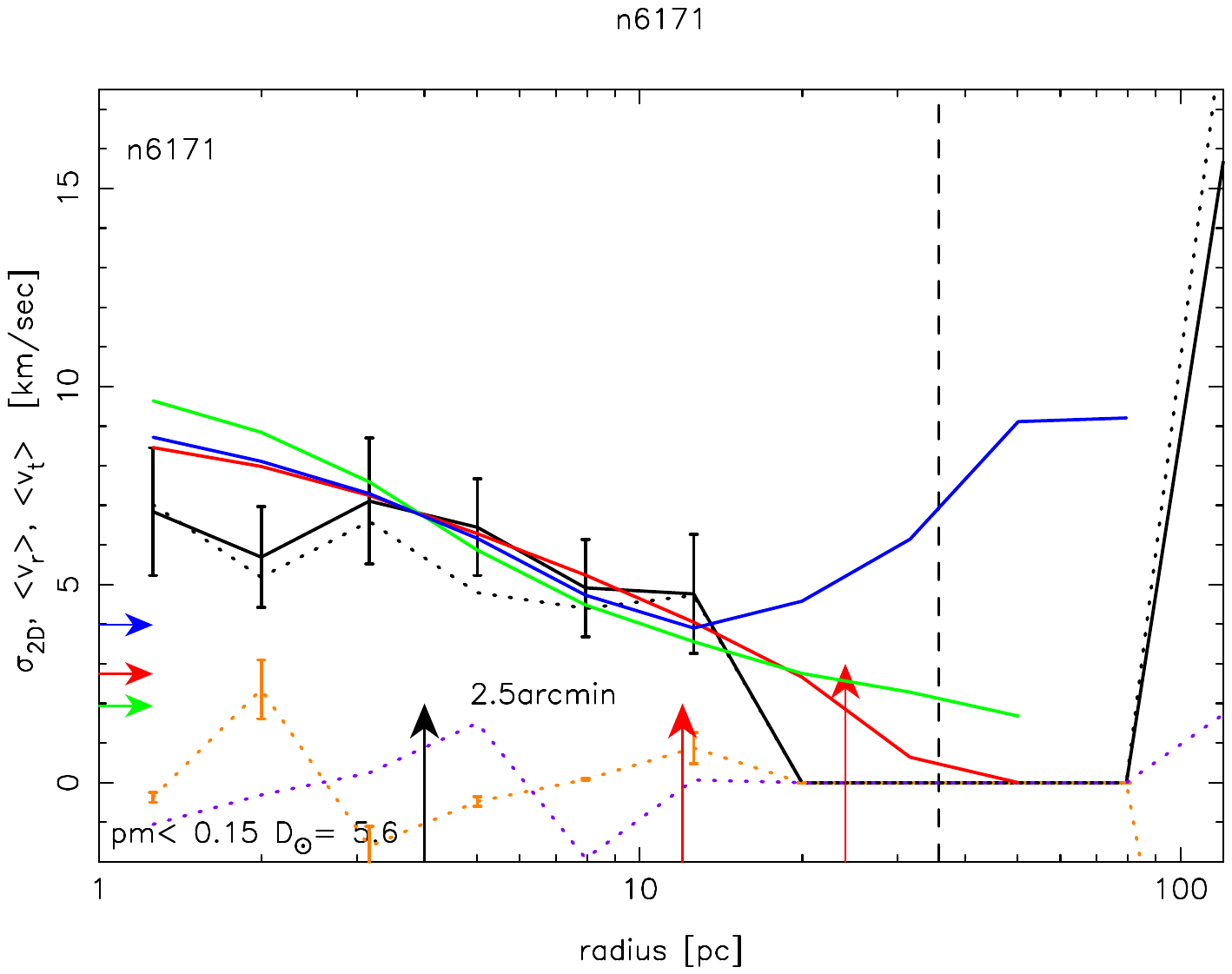}}
\put(120,120){\includegraphics[angle=0,scale=0.12,trim=30 10 100 80, clip=true]{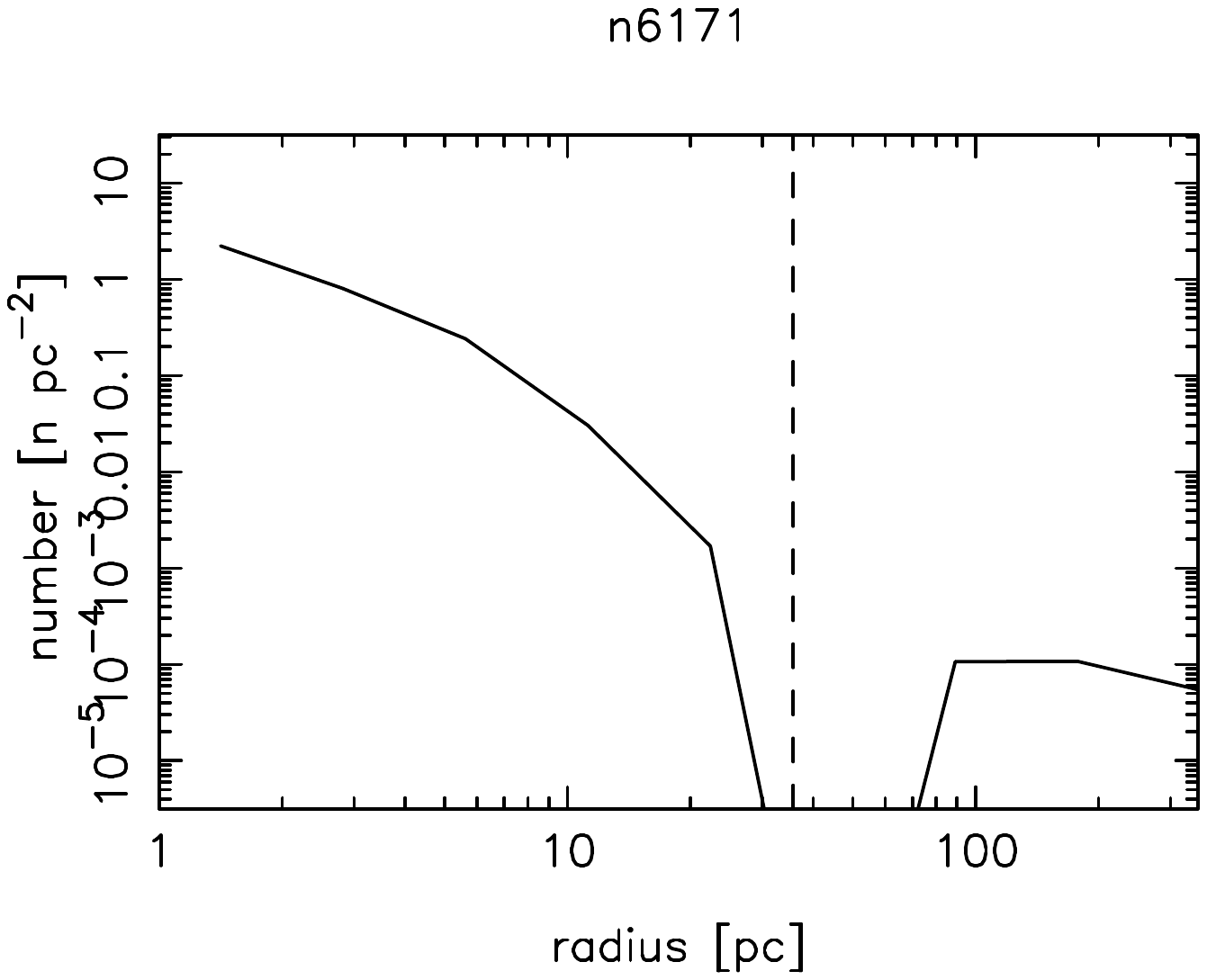}}
\caption{Velocities for the NGC 6254, 6809 and 6171 clusters, top to bottom. Symbols and lines as in Figure~\ref{fig_6752}.
}
\label{fig_6171}
\end{figure*}

\begin{figure*}
{\includegraphics[angle=0,scale=0.36,trim=30 10 100 80, clip=true]{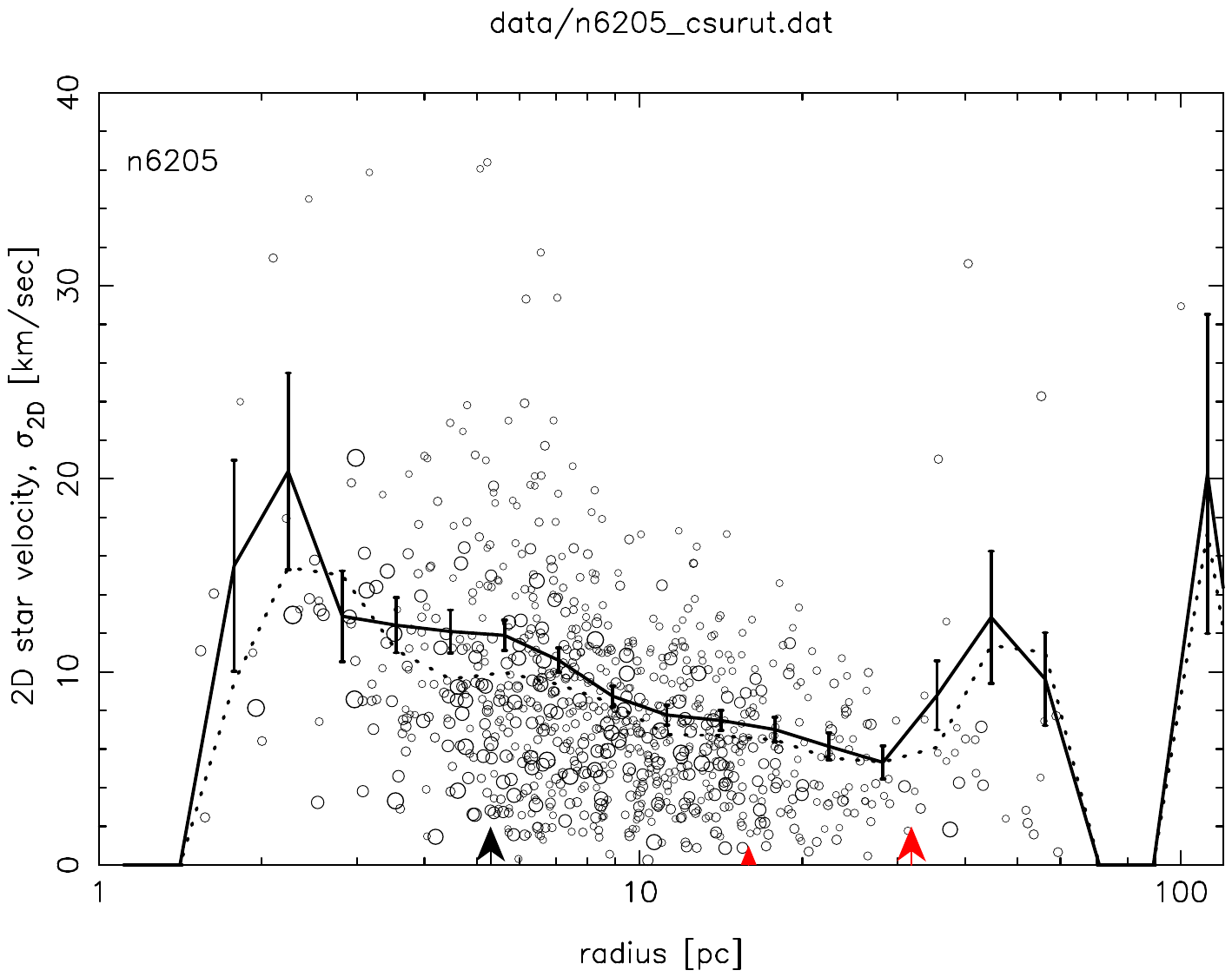}}
\put(5,0){\includegraphics[angle=0,scale=0.36,trim=30 10 100 80, clip=true]{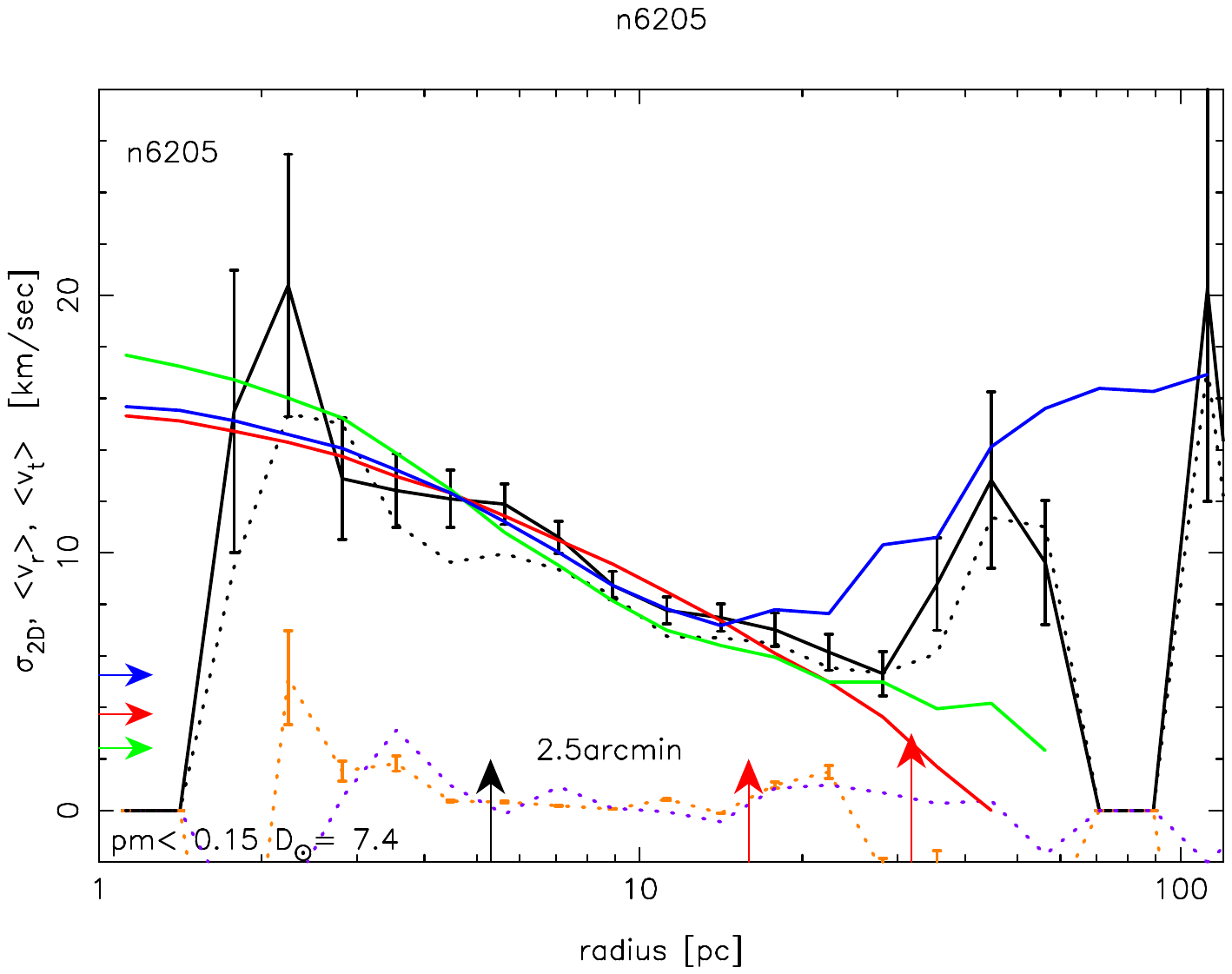}}
\put(120,120){\includegraphics[angle=0,scale=0.12,trim=30 10 100 80, clip=true]{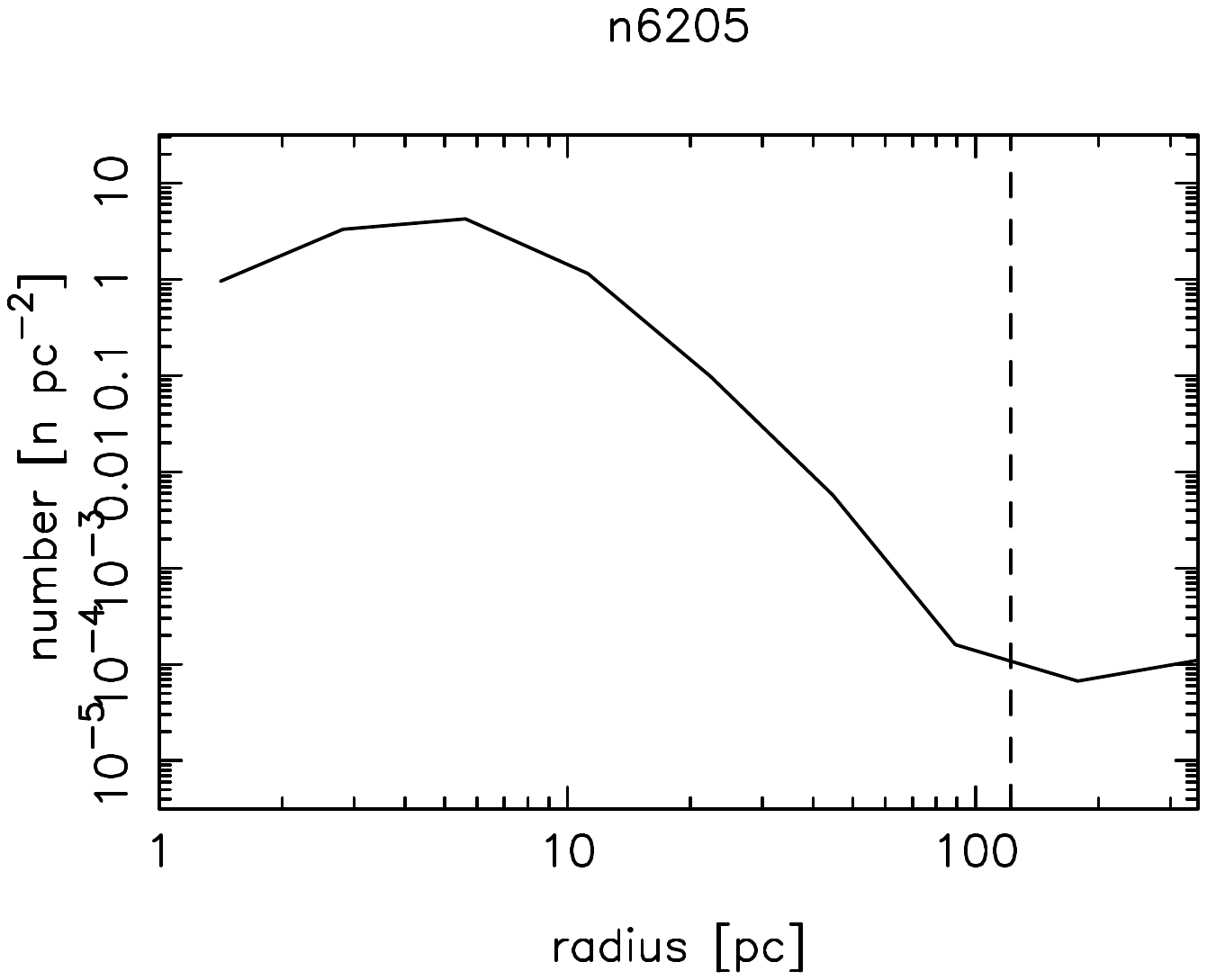}}

{\includegraphics[angle=0,scale=0.36,trim=30 10 100 80, clip=true]{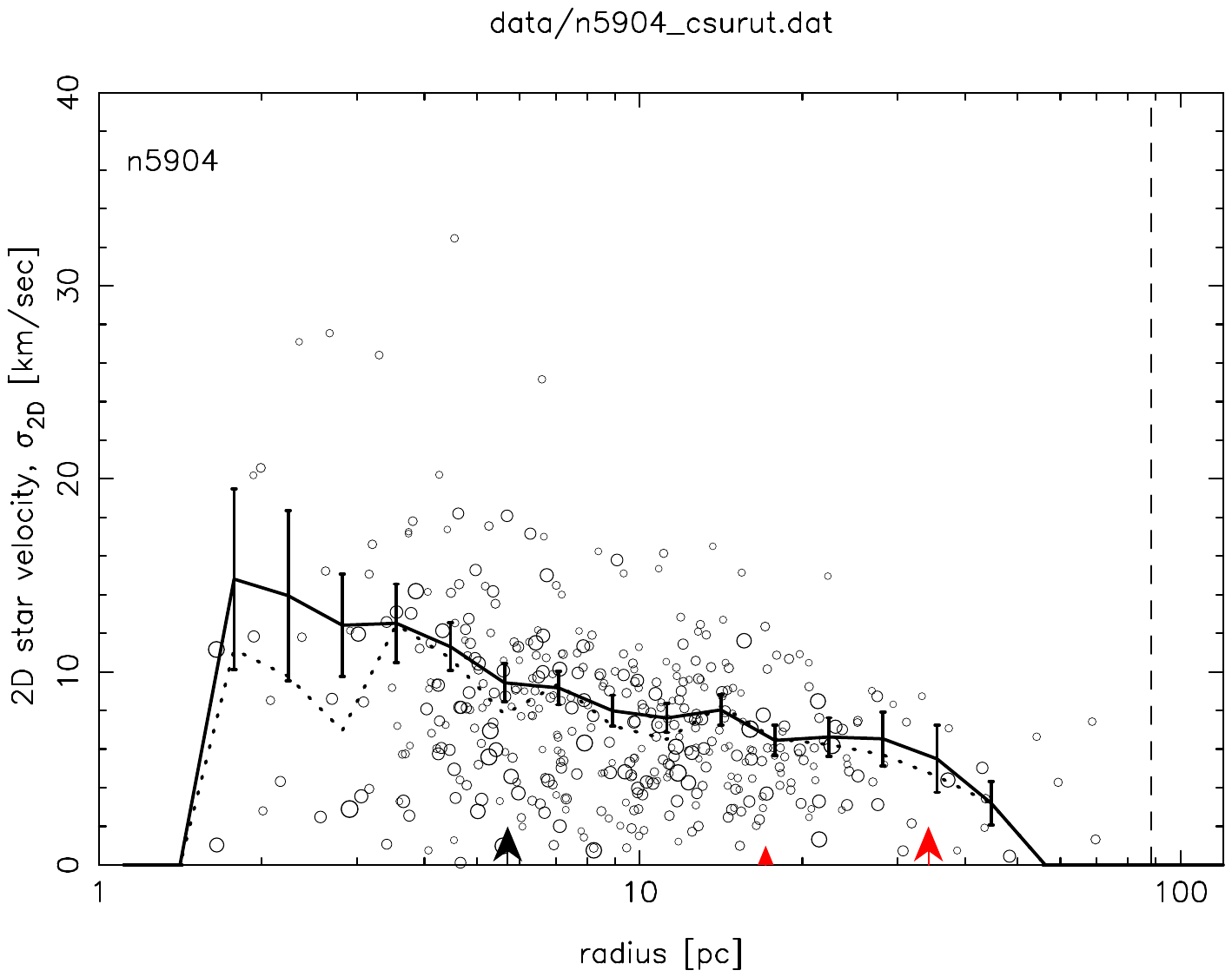}}
\put(5,0){\includegraphics[angle=0,scale=0.36,trim=30 10 100 80, clip=true]{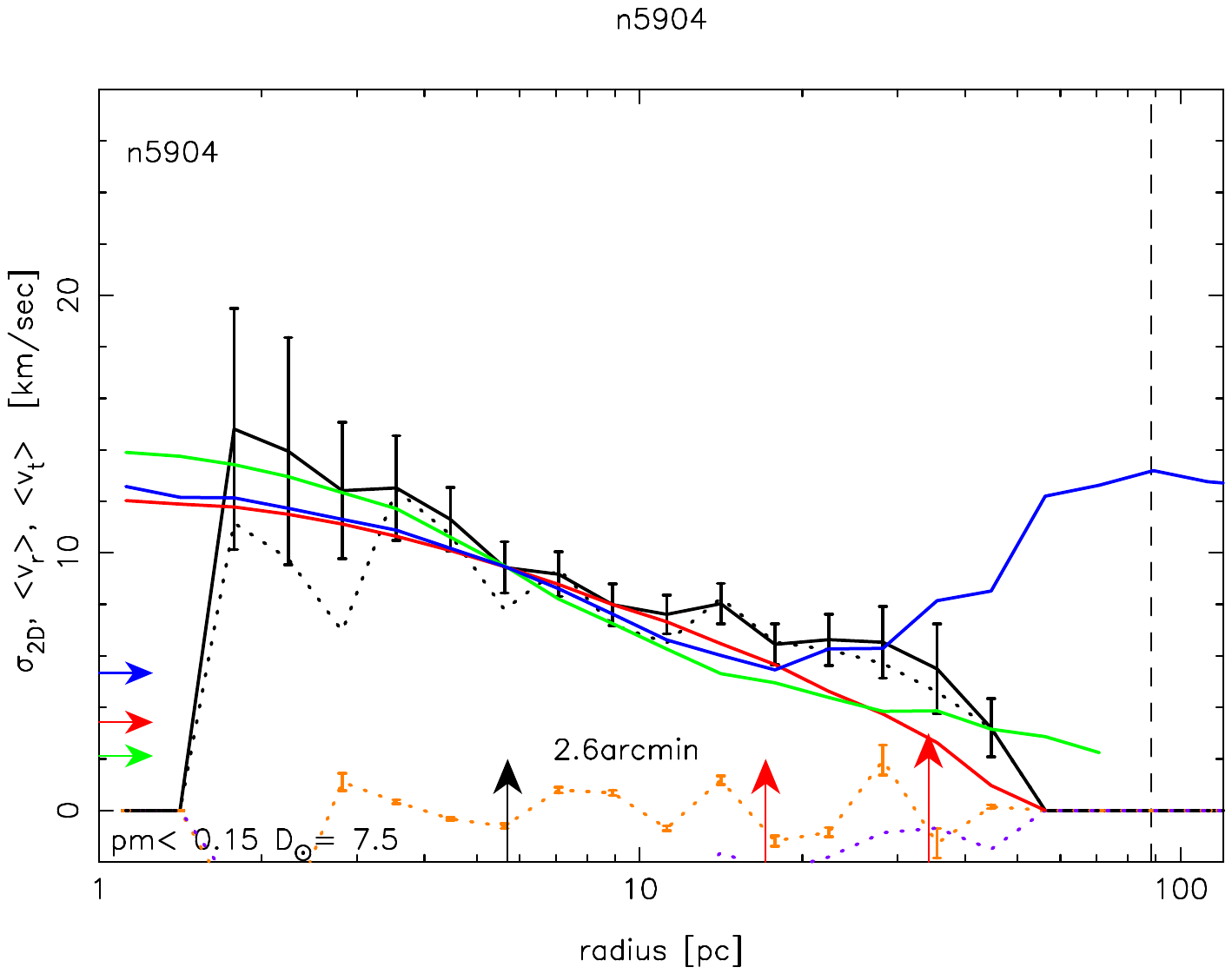}}
\put(120,120){\includegraphics[angle=0,scale=0.12,trim=30 10 100 80, clip=true]{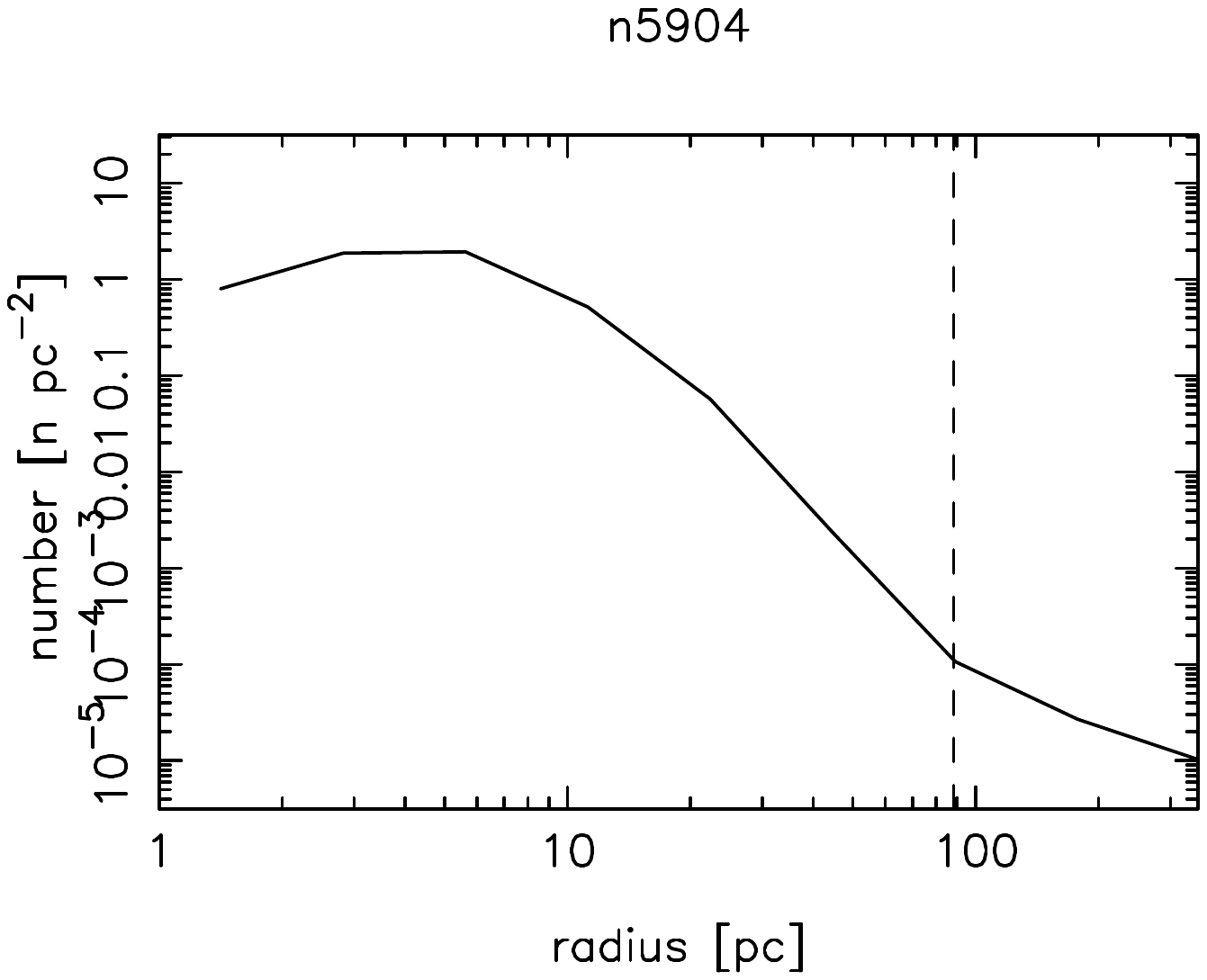}}

{\includegraphics[angle=0,scale=0.36,trim=30 10 100 80, clip=true]{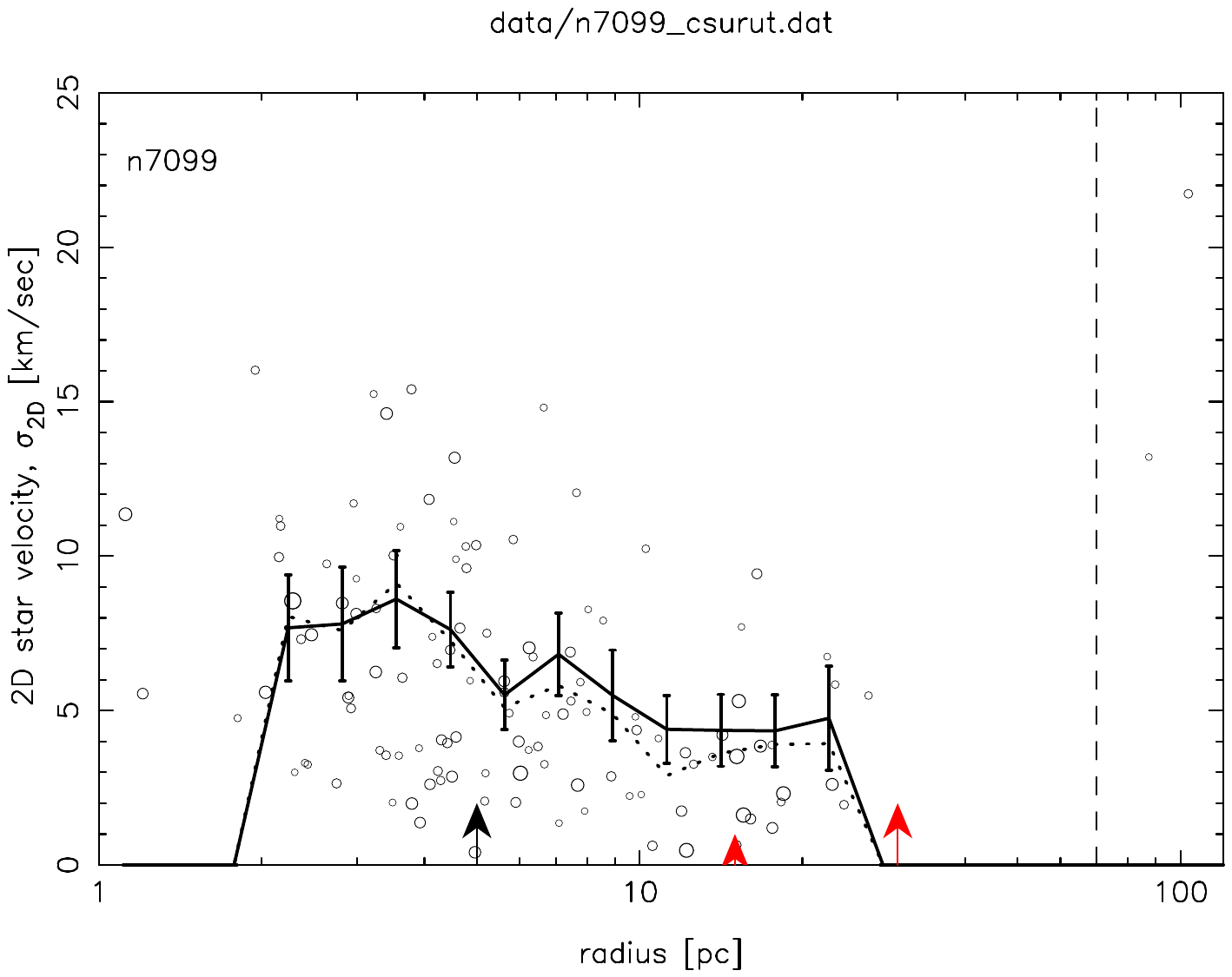}}
\put(5,0){\includegraphics[angle=0,scale=0.36,trim=30 10 100 80, clip=true]{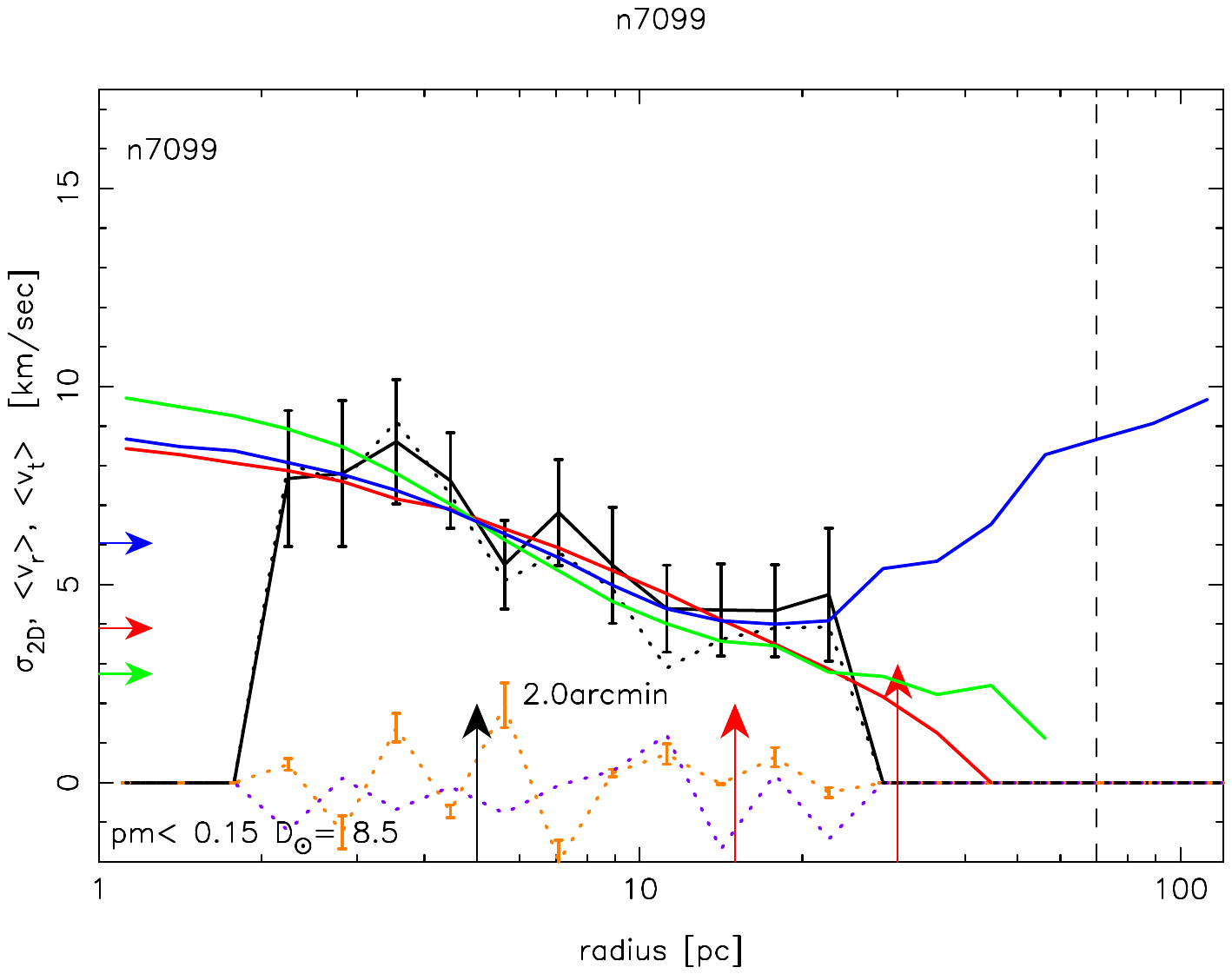}}
\put(120,120){\includegraphics[angle=0,scale=0.12,trim=30 10 100 80, clip=true]{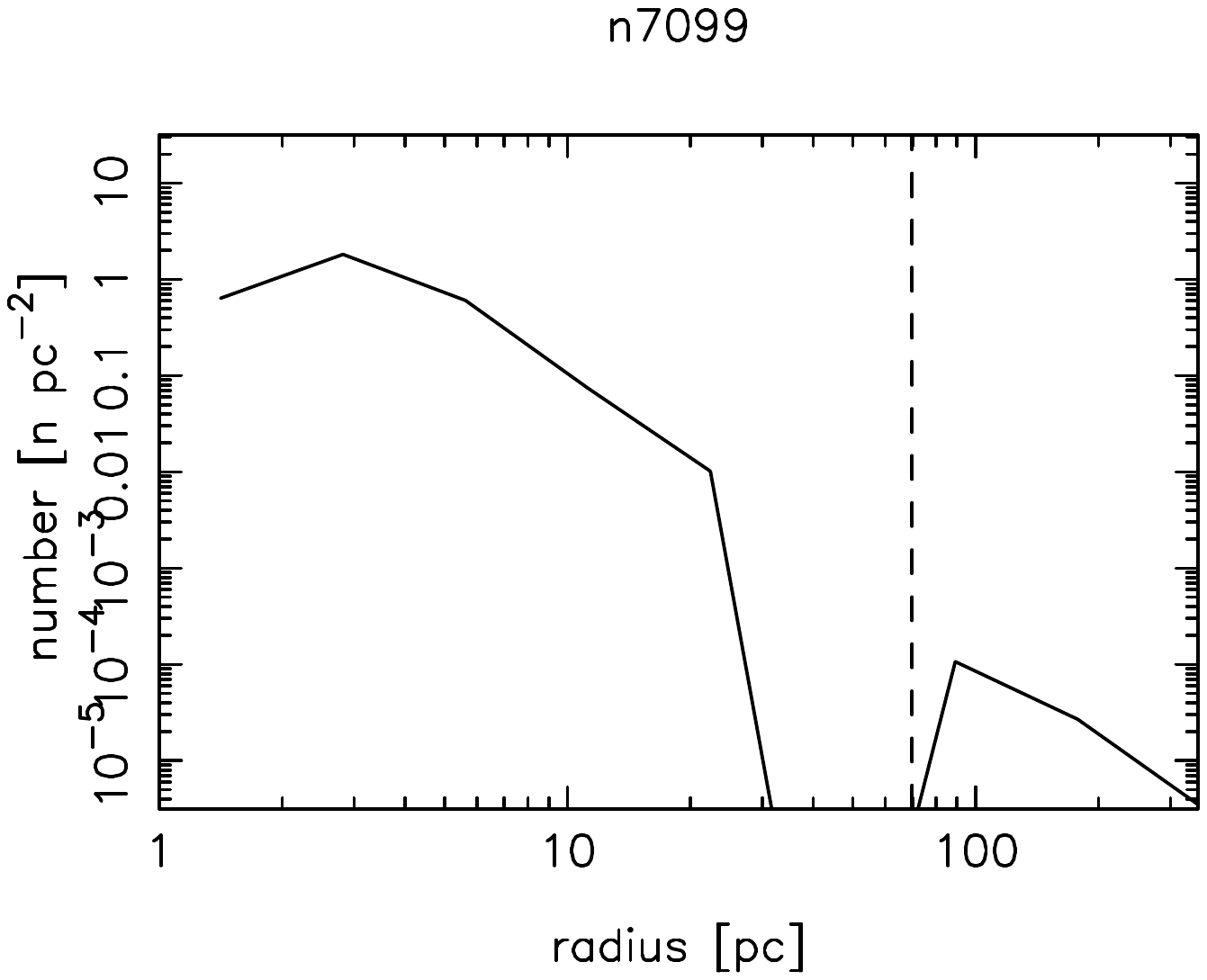}}
\caption{Velocities for the NGC 6205, 5904 and 7099 clusters, top to bottom. Symbols as in Figure~\ref{fig_6752}.
}
\label{fig_7099}
\end{figure*}

\begin{figure*}
{\includegraphics[angle=0,scale=0.36,trim=30 10 100 80, clip=true]{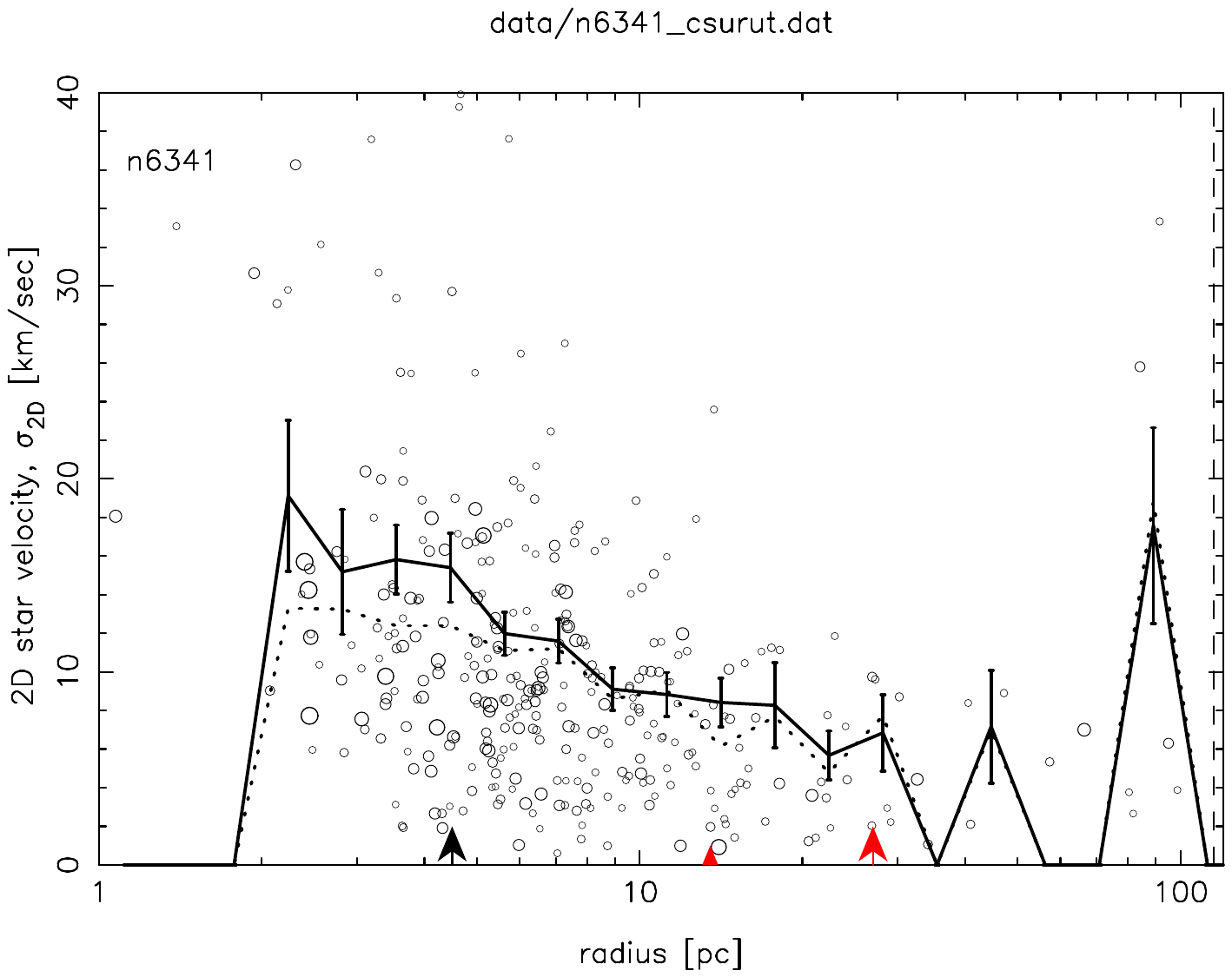}}
\put(5,0){\includegraphics[angle=0,scale=0.36,trim=30 10 100 80, clip=true]{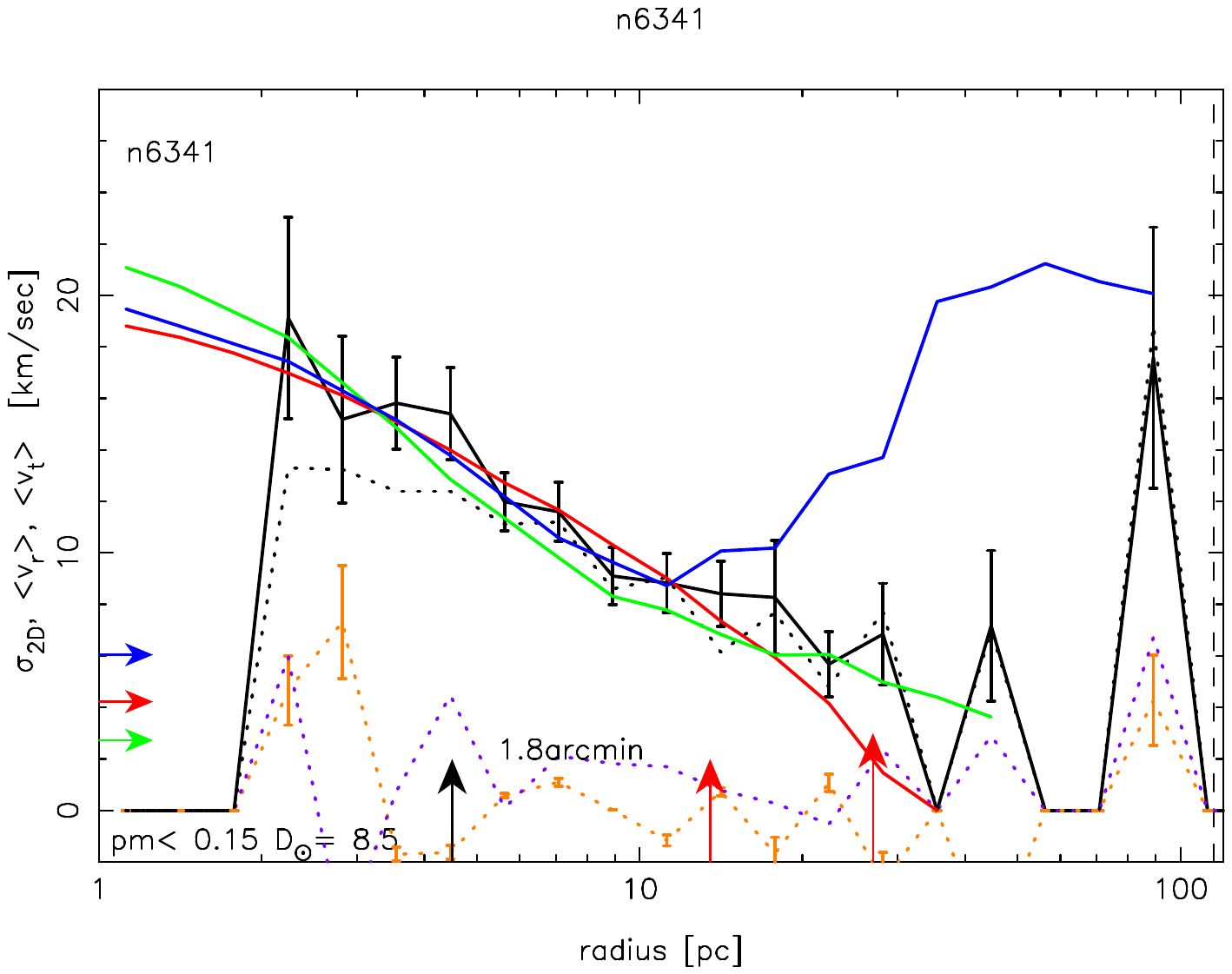}}
\put(120,120){\includegraphics[angle=0,scale=0.12,trim=30 10 100 80, clip=true]{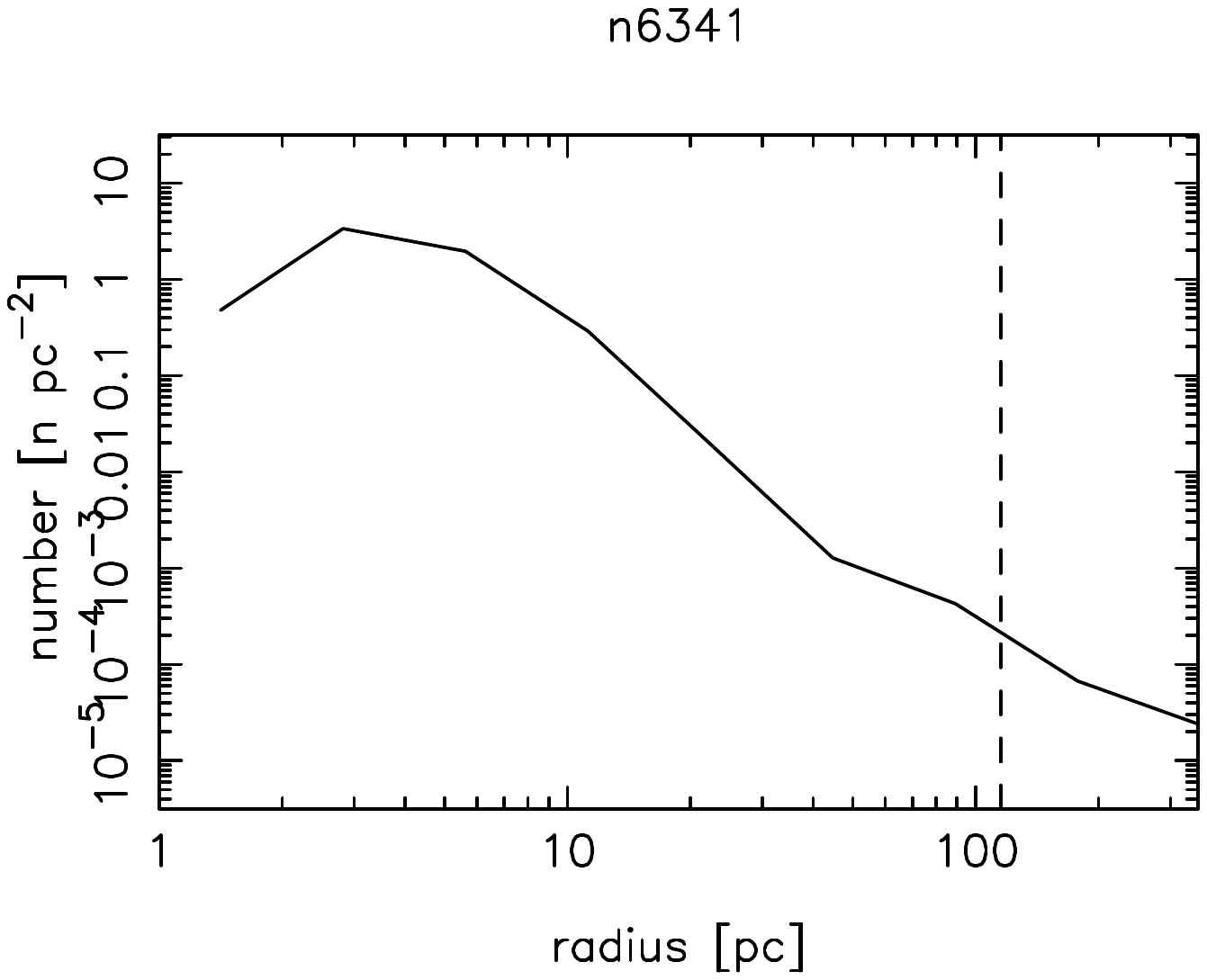}}

{\includegraphics[angle=0,scale=0.36,trim=30 10 100 80, clip=true]{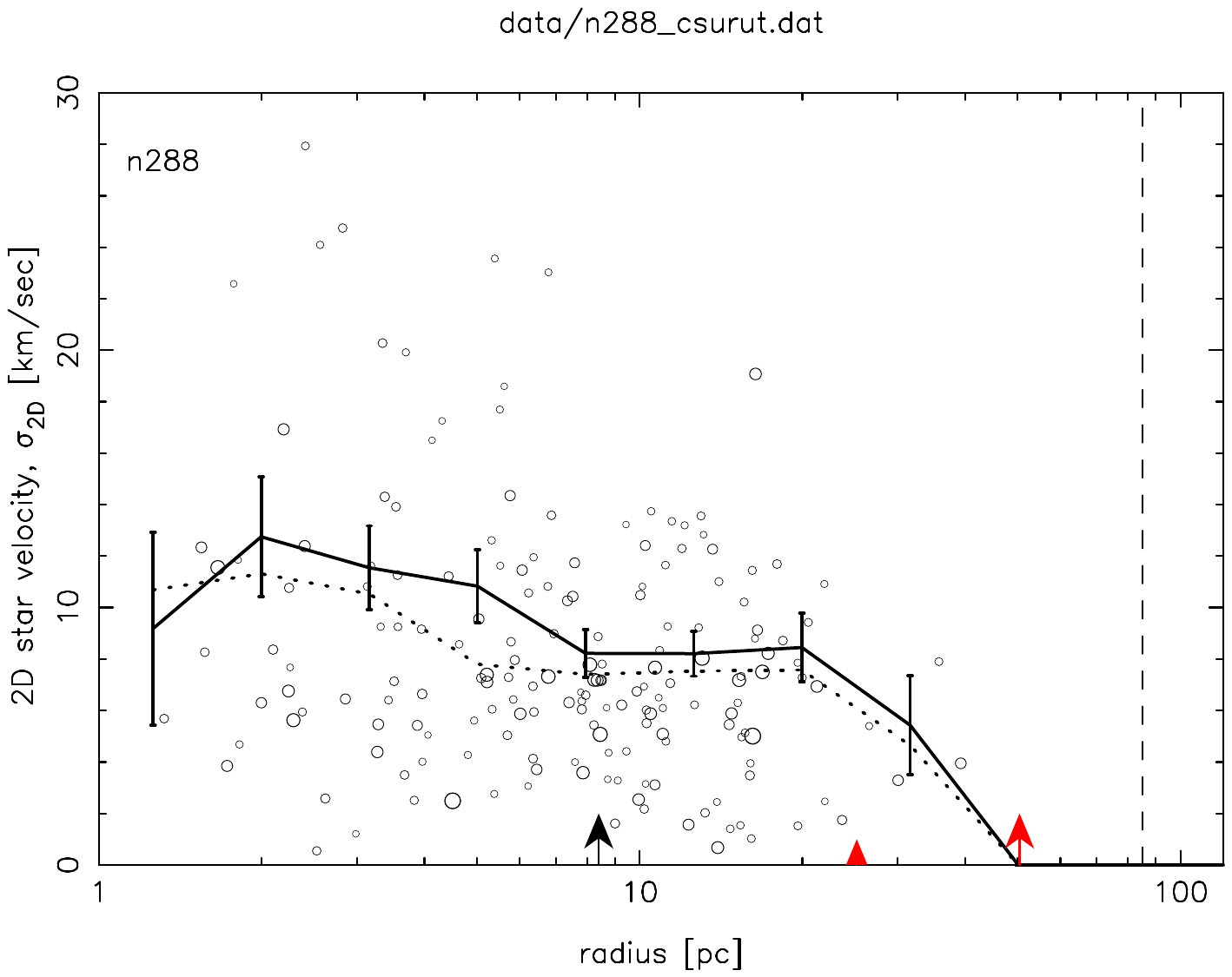}}
\put(5,0){\includegraphics[angle=0,scale=0.36,trim=30 10 100 80, clip=true]{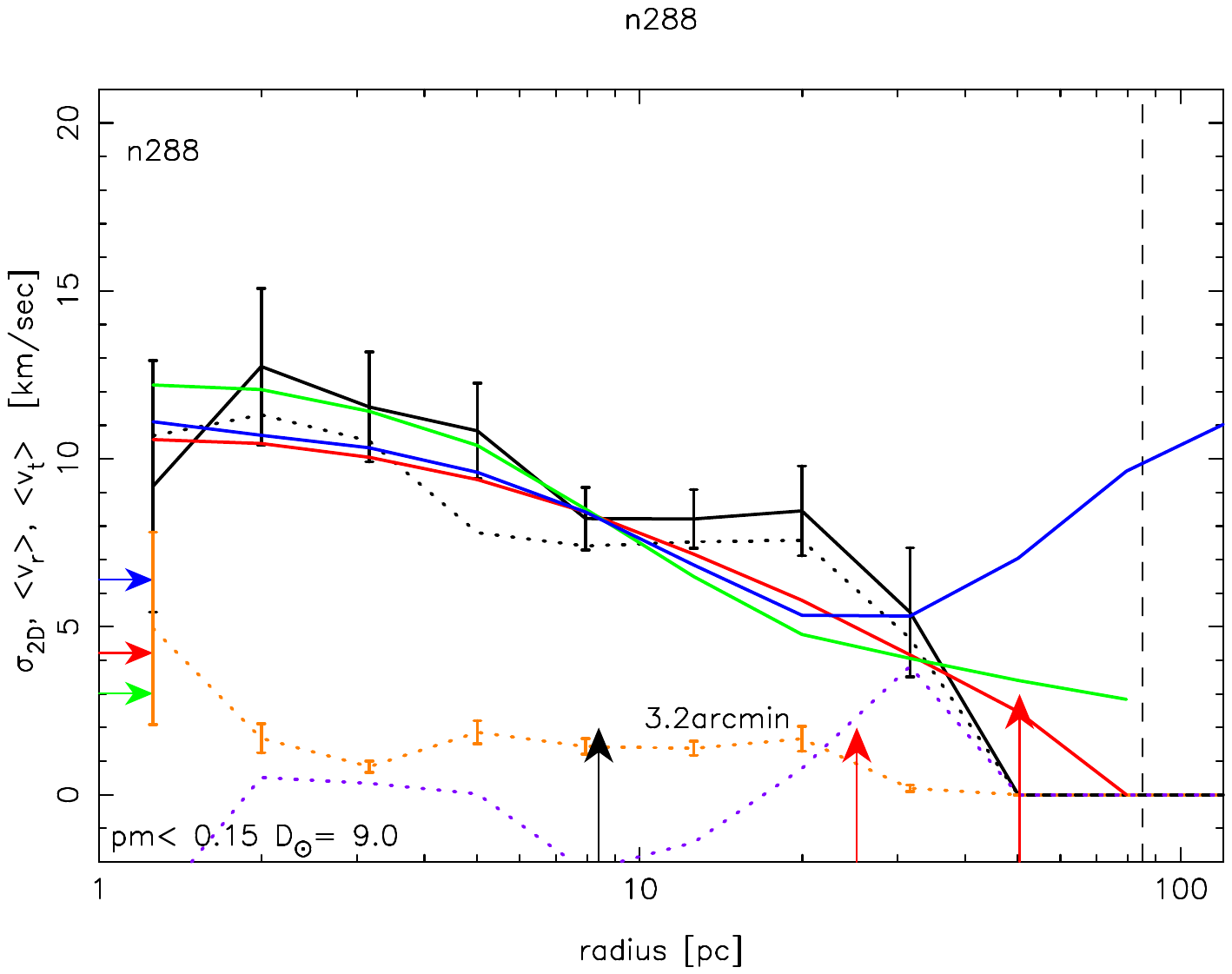}}
\put(120,120){\includegraphics[angle=0,scale=0.12,trim=30 10 100 80, clip=true]{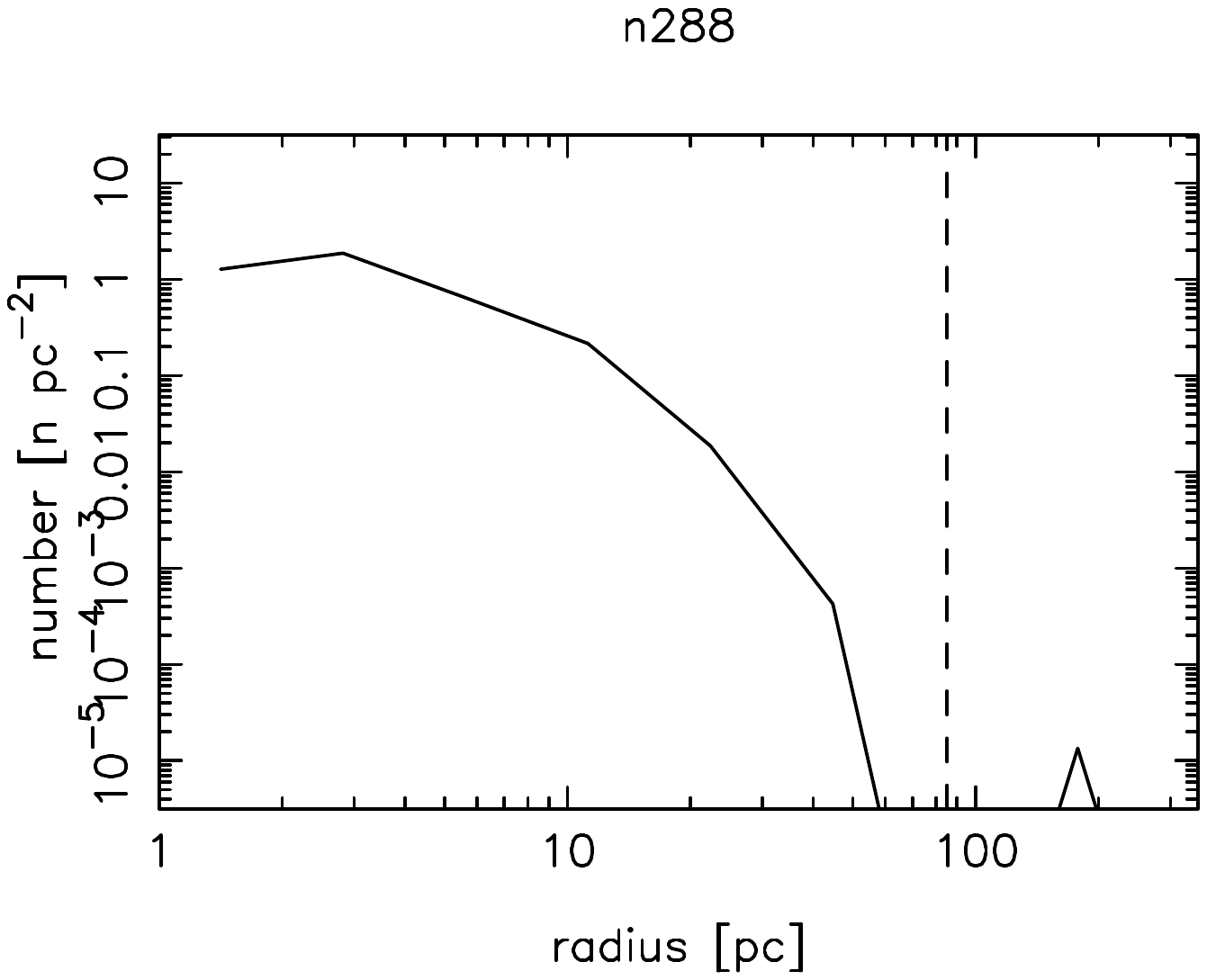}}

{\includegraphics[angle=0,scale=0.36,trim=30 10 100 80, clip=true]{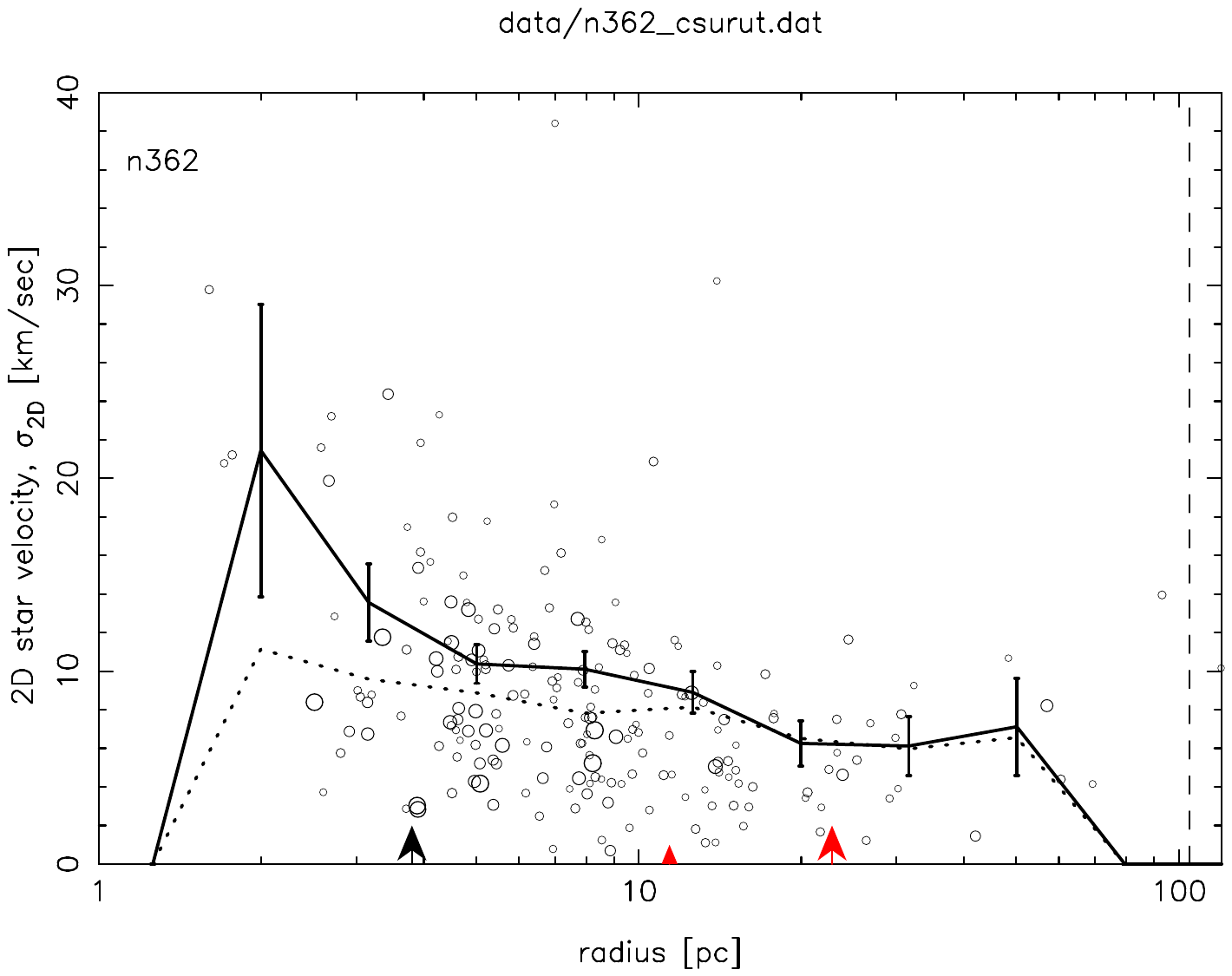}}
\put(5,0){\includegraphics[angle=0,scale=0.36,trim=30 10 100 80, clip=true]{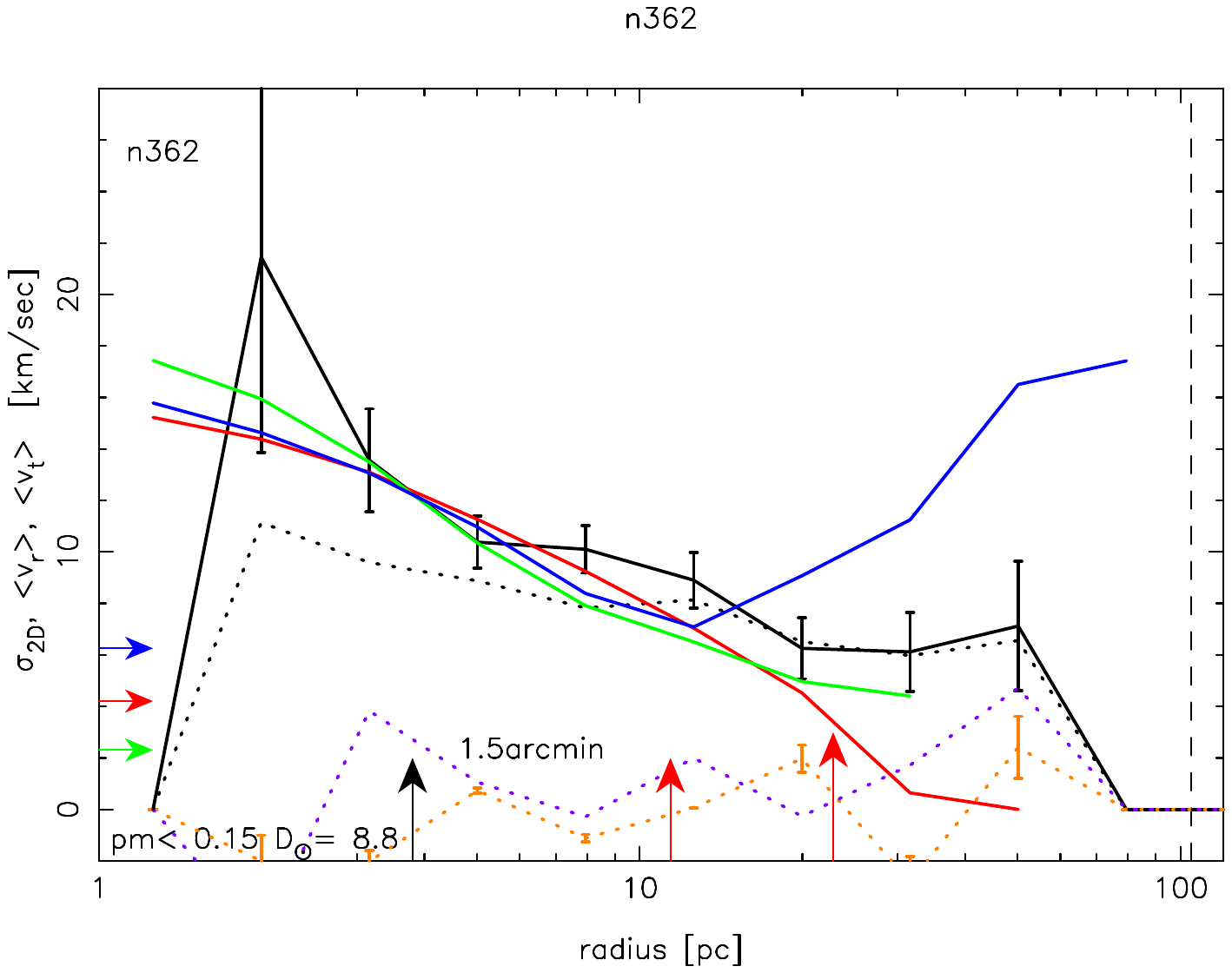}}
\put(120,120){\includegraphics[angle=0,scale=0.12,trim=30 10 100 80, clip=true]{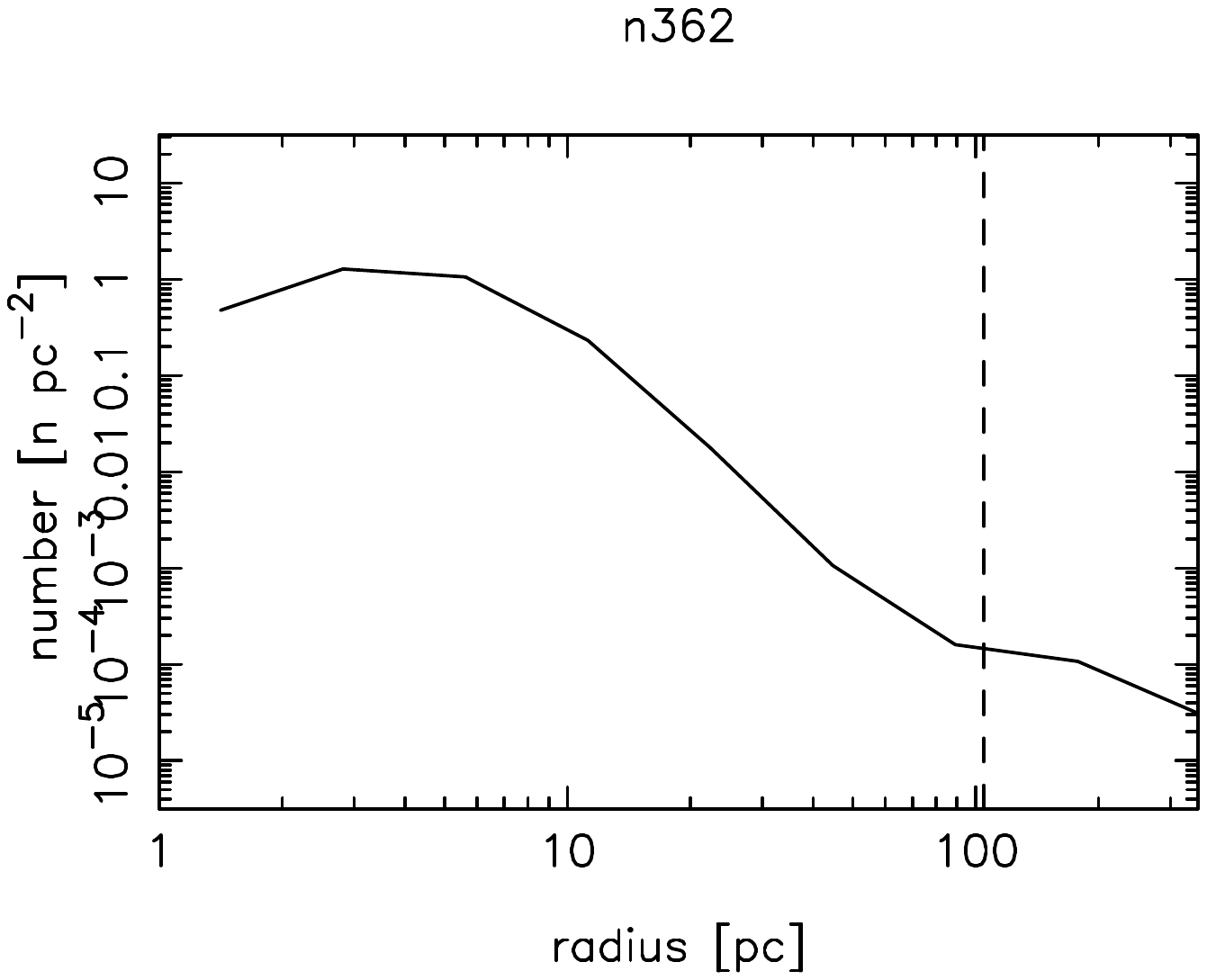}}
\caption{Velocities for the clusters NGC 6341, 288 and 362, top to bottom. Symbols as in Figure~\ref{fig_6752}.
}
\label{fig_362}
\end{figure*}

\begin{figure*}
{\includegraphics[angle=0,scale=0.36,trim=30 10 100 80, clip=true]{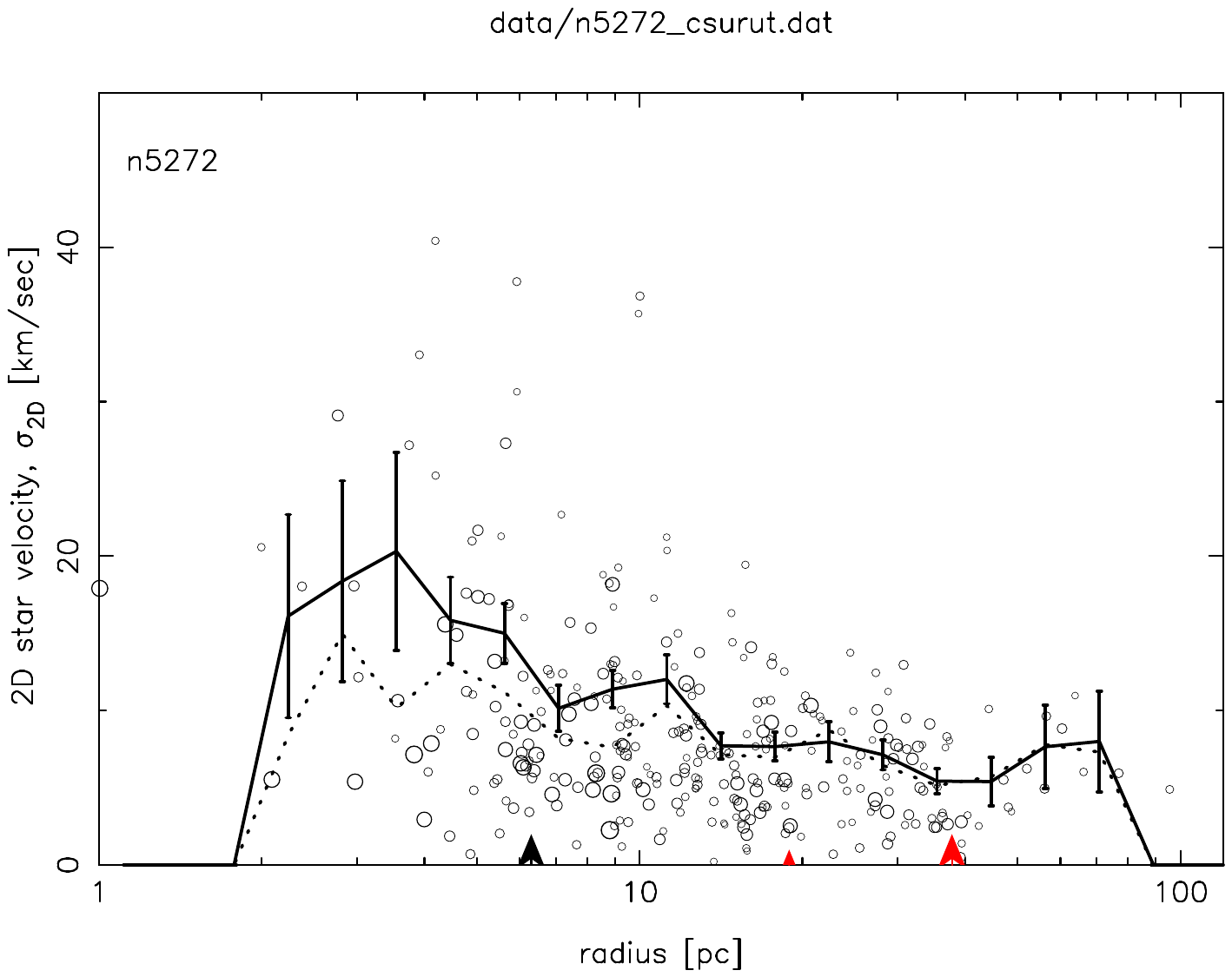}}
\put(5,0){\includegraphics[angle=0,scale=0.36,trim=30 10 100 80, clip=true]{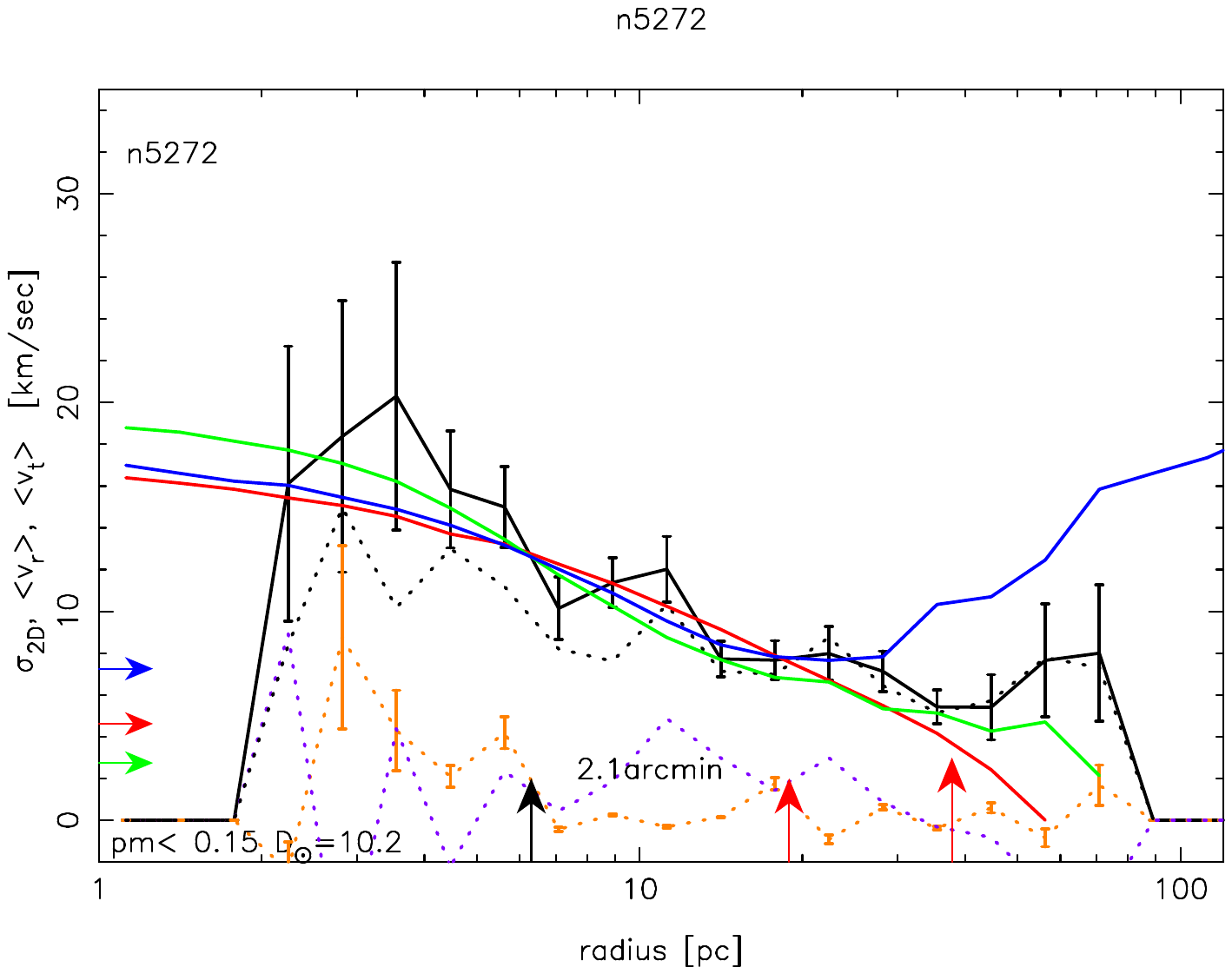}}
\put(120,120){\includegraphics[angle=0,scale=0.12,trim=30 10 100 80, clip=true]{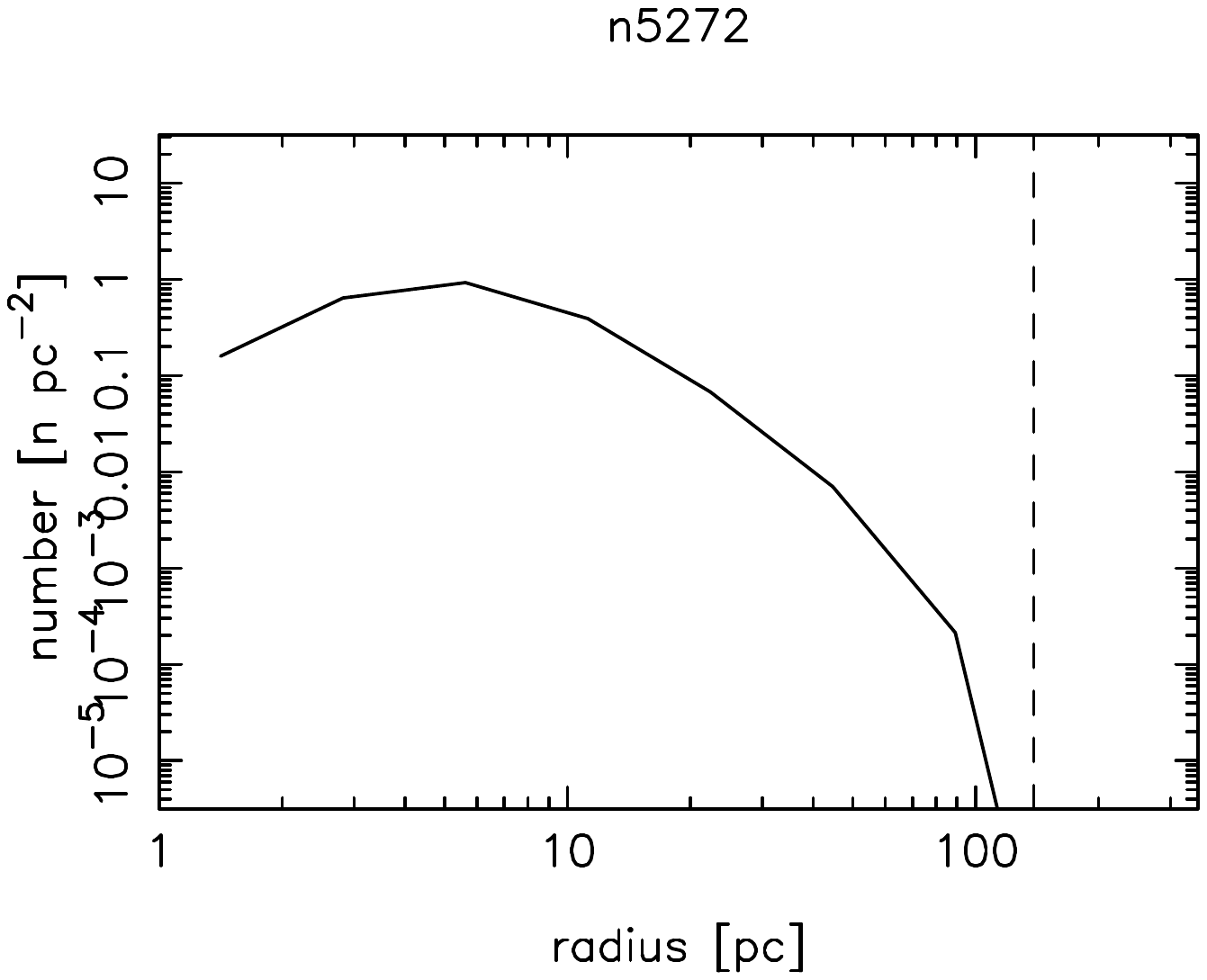}}

{\includegraphics[angle=0,scale=0.36,trim=30 10 100 80, clip=true]{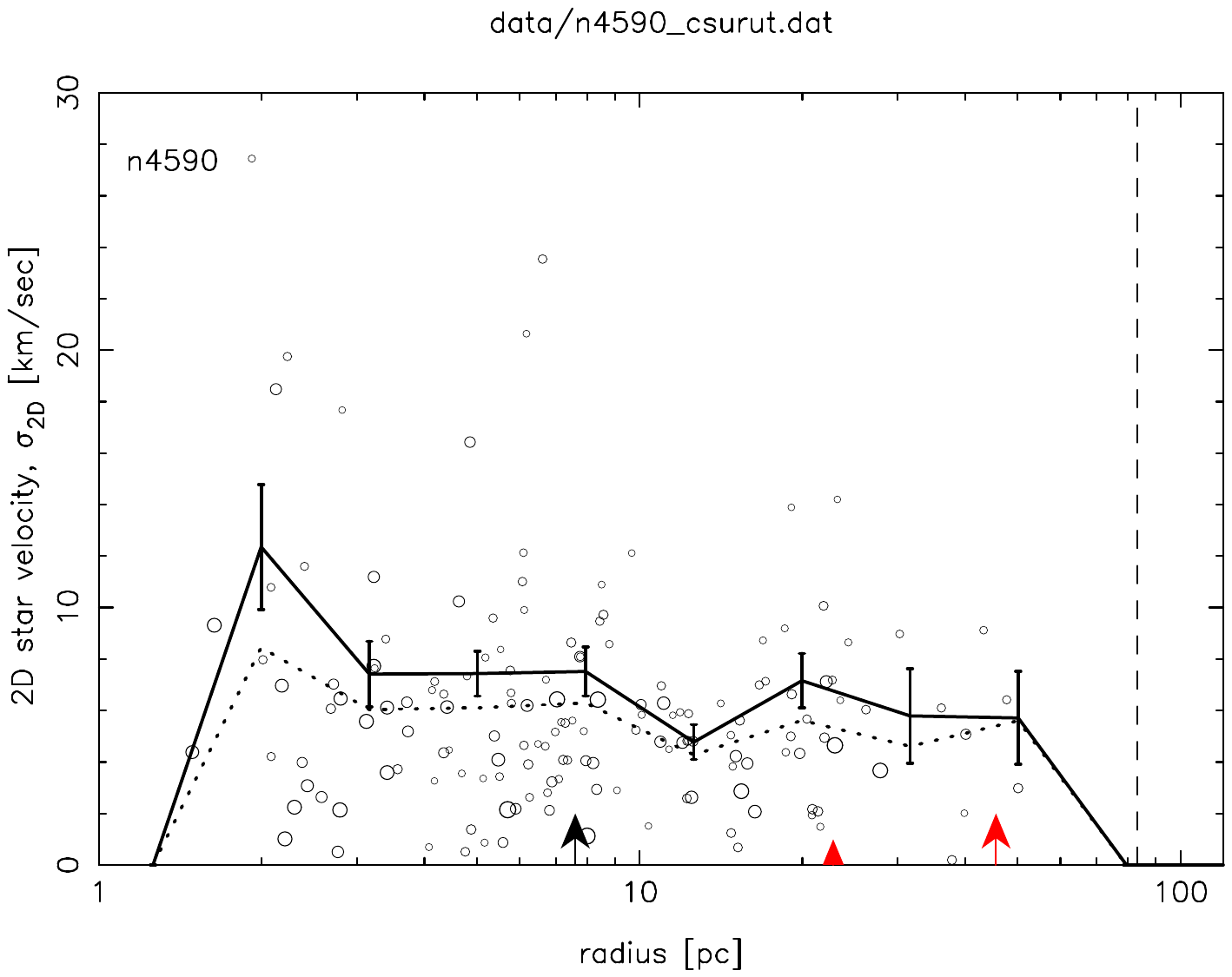}}
\put(5,0){\includegraphics[angle=0,scale=0.36,trim=30 10 100 80, clip=true]{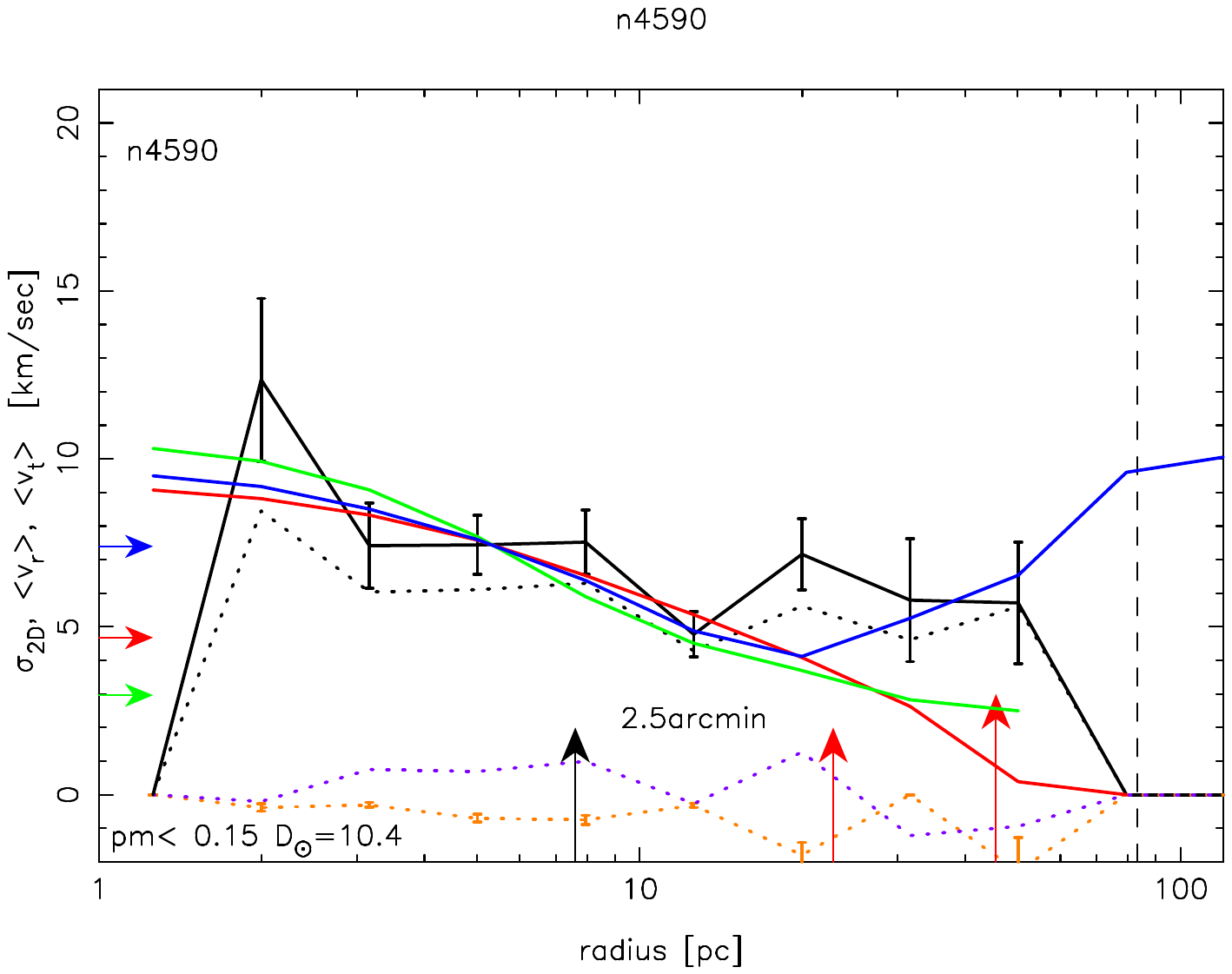}}
\put(120,120){\includegraphics[angle=0,scale=0.12,trim=30 10 100 80, clip=true]{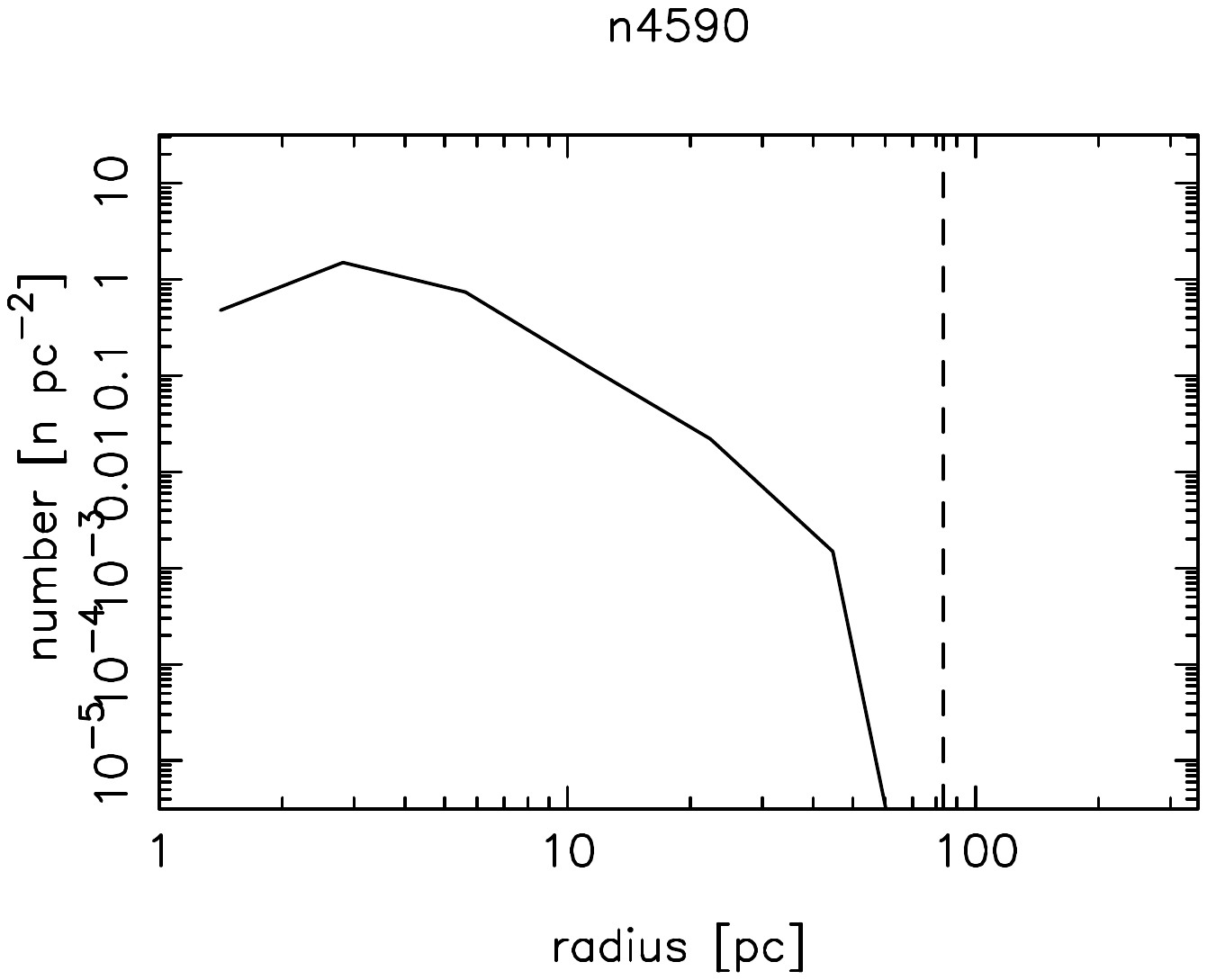}}

{\includegraphics[angle=0,scale=0.36,trim=30 10 100 80, clip=true]{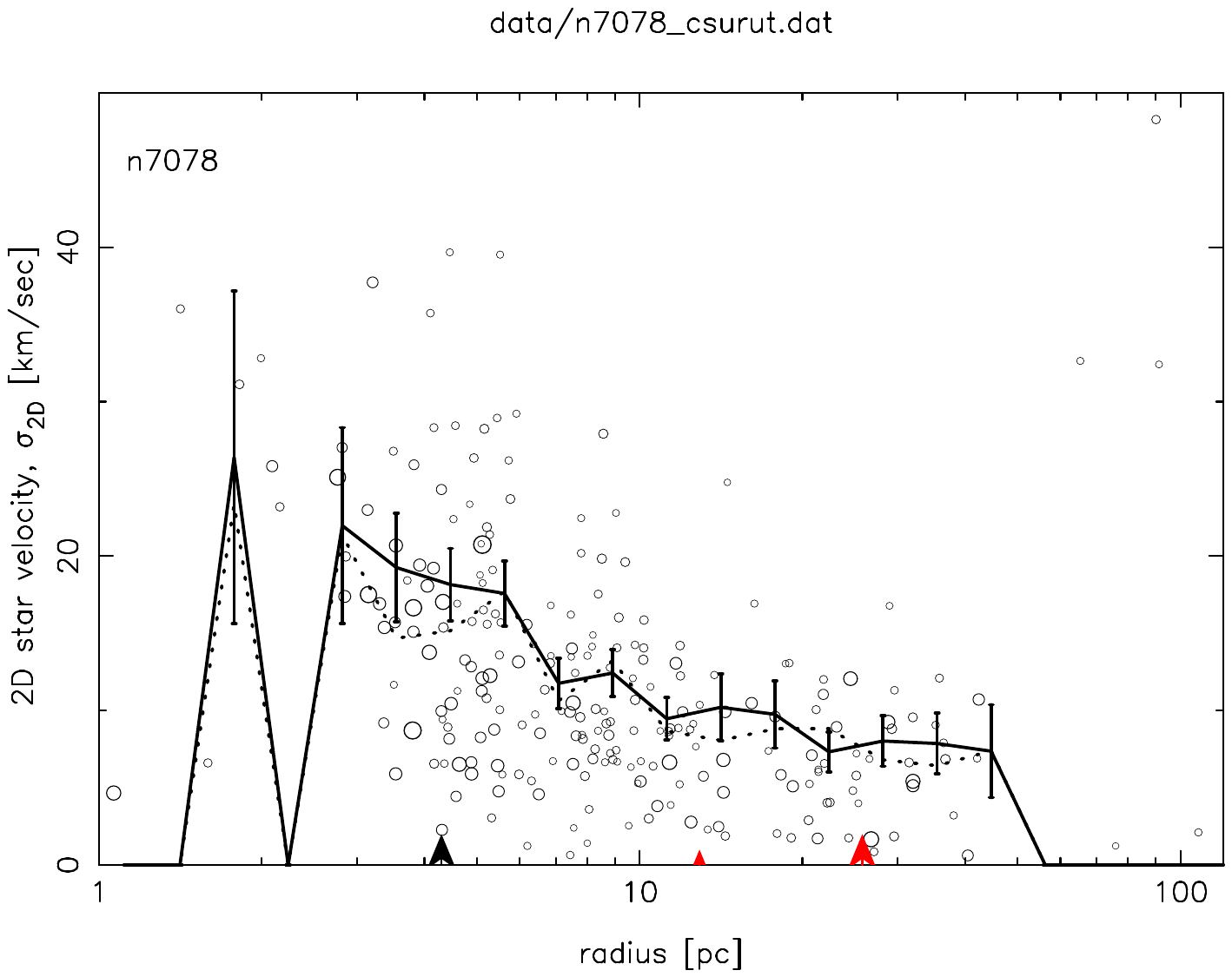}}
\put(5,0){\includegraphics[angle=0,scale=0.36,trim=30 10 100 80, clip=true]{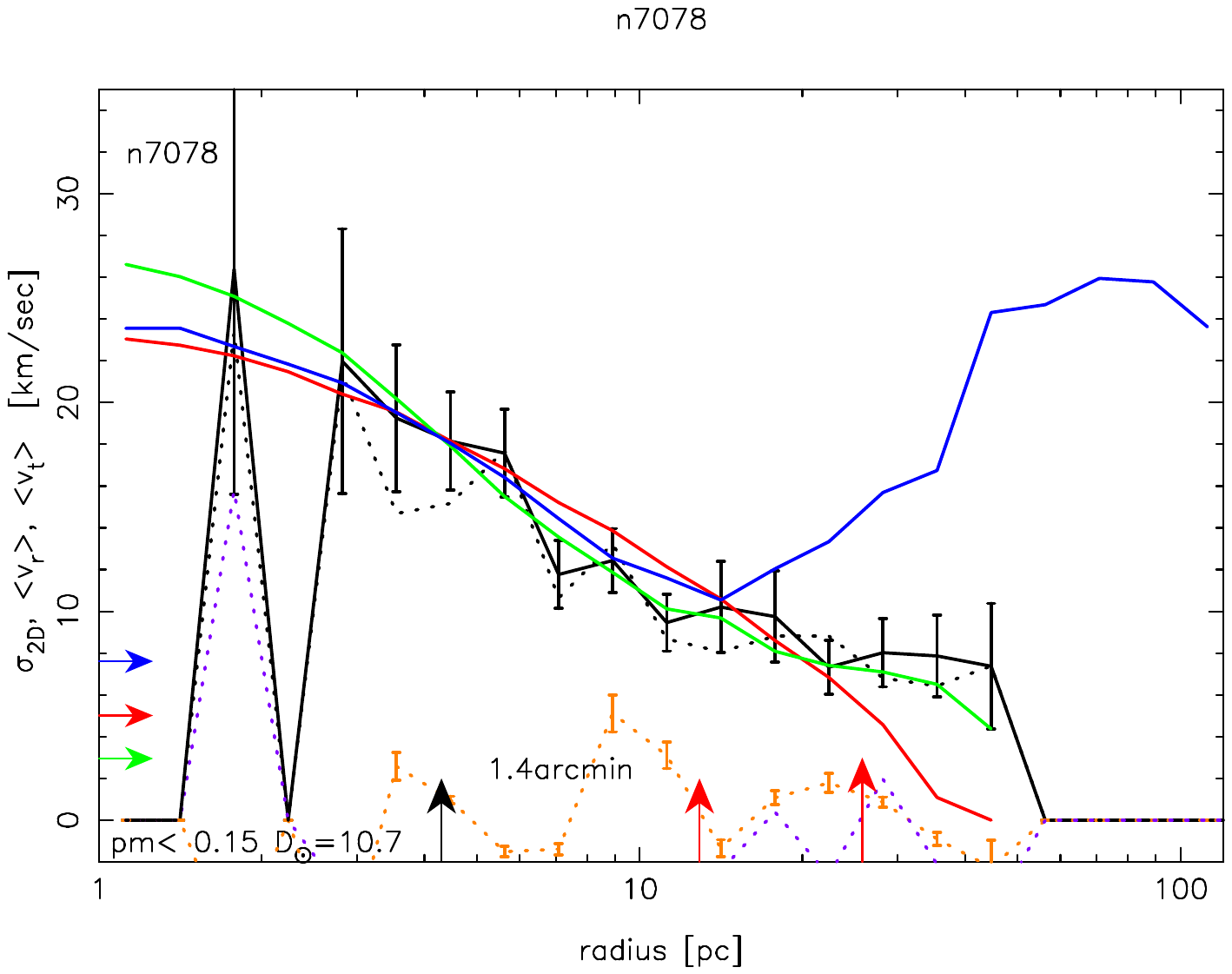}}
\put(120,120){\includegraphics[angle=0,scale=0.12,trim=30 10 100 80, clip=true]{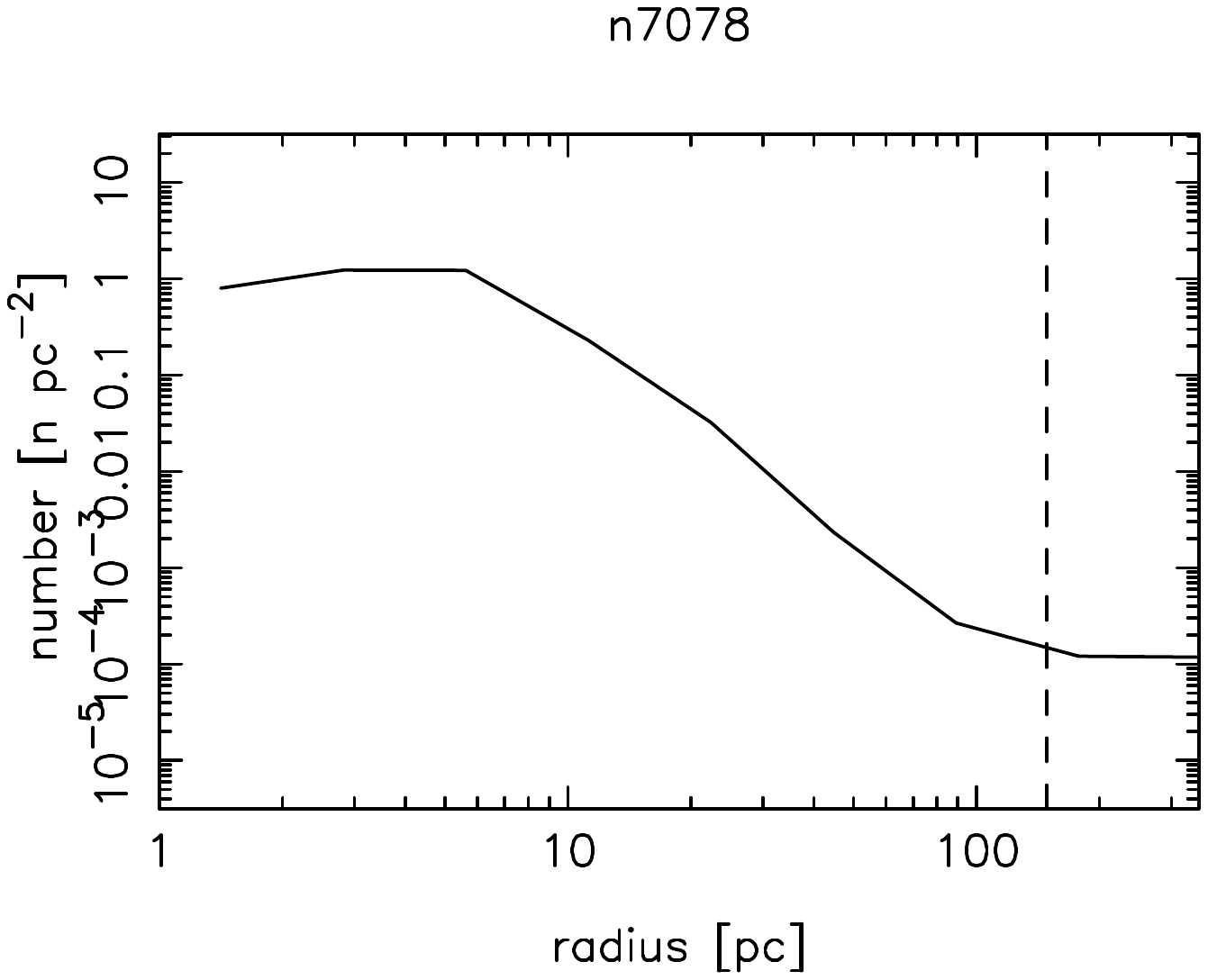}}
\caption{Velocities for the clusters NGC 5272, 4590, 7078, top to bottom. Symbols as in Figure~\ref{fig_6752}.
}
\label{fig_7078}
\end{figure*}
\begin{figure*}
{\includegraphics[angle=0,scale=0.36,trim=30 10 100 80, clip=true]{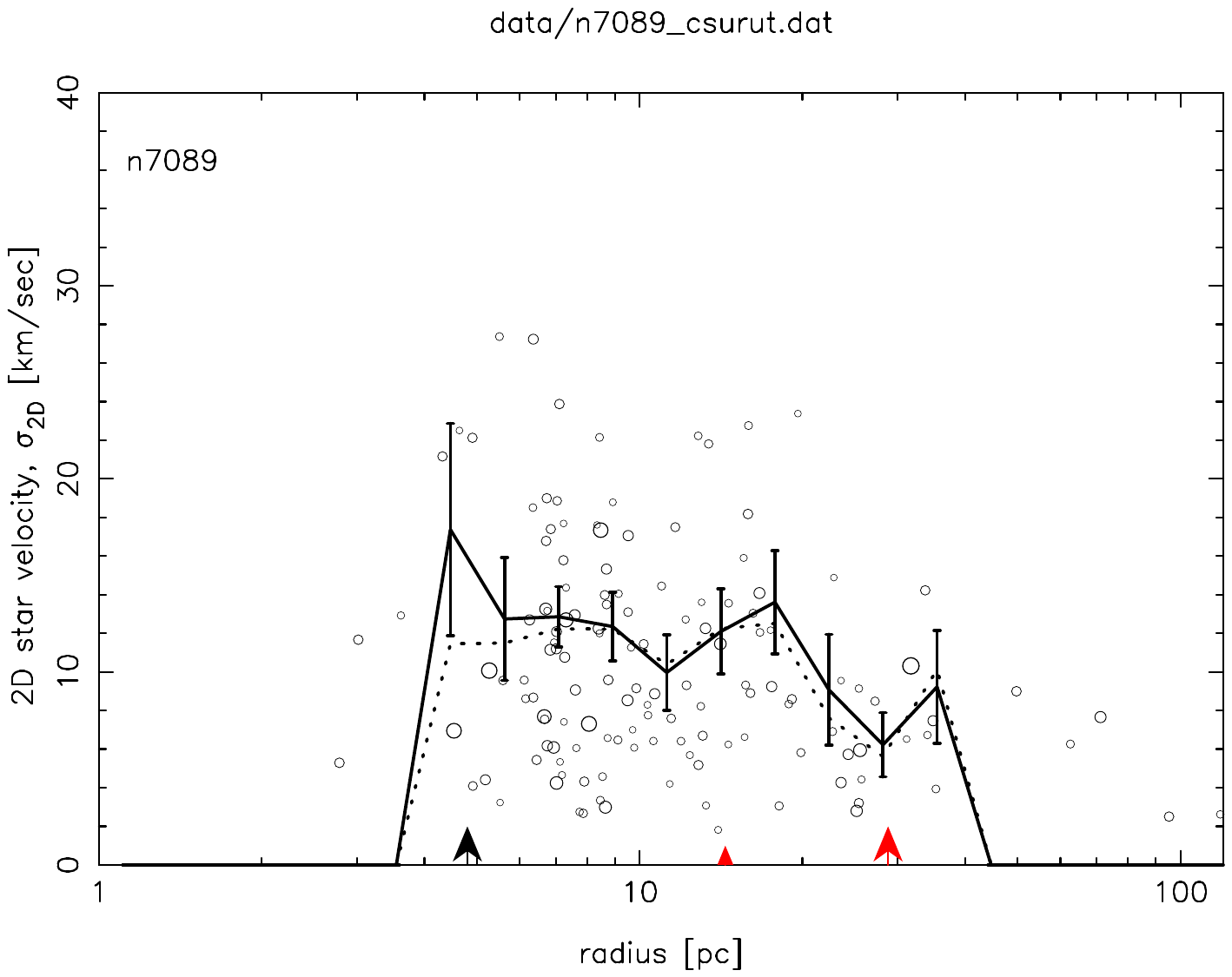}}
\put(5,0){\includegraphics[angle=0,scale=0.36,trim=30 10 100 80, clip=true]{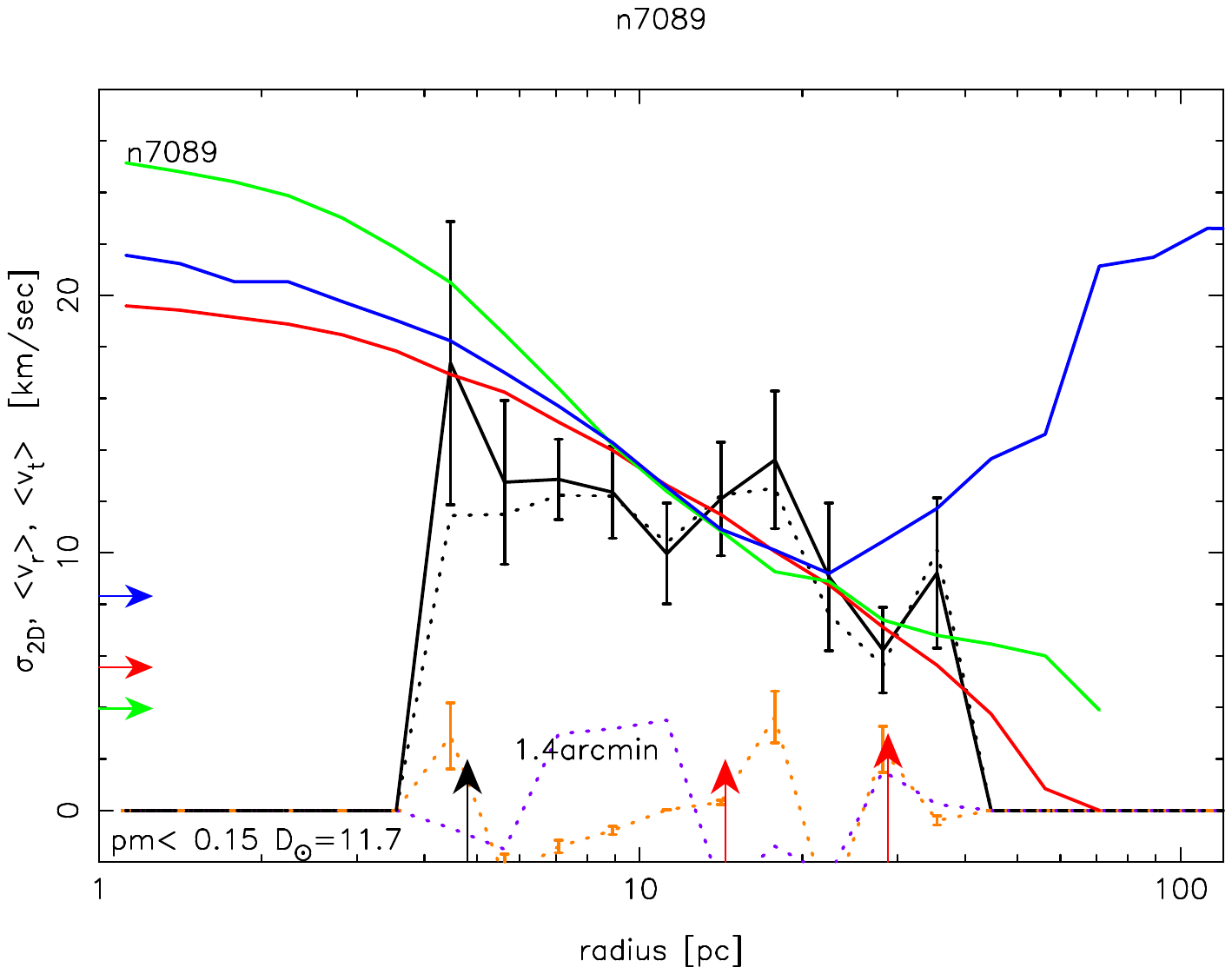}}
\put(120,120){\includegraphics[angle=0,scale=0.12,trim=30 10 100 80, clip=true]{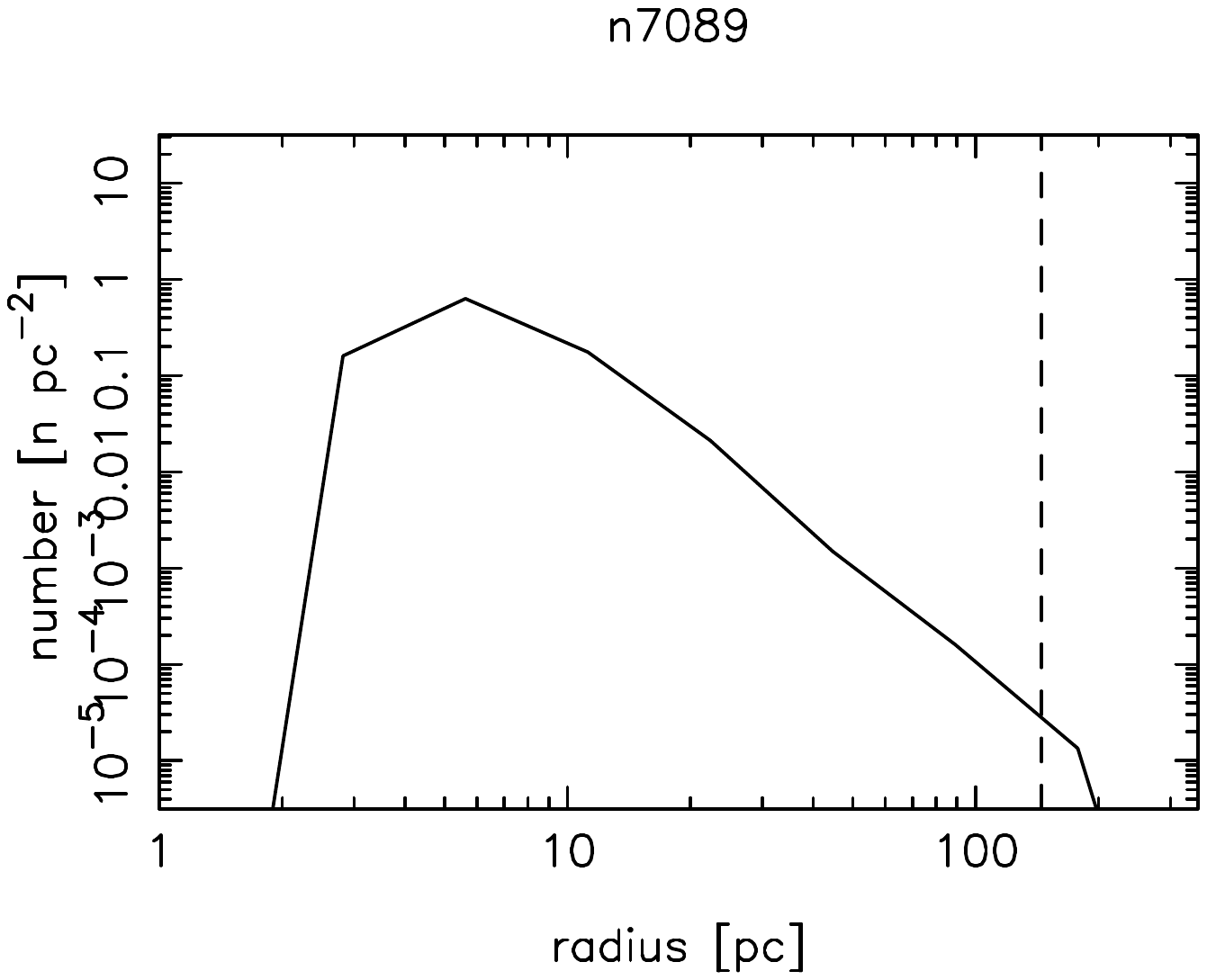}}

{\includegraphics[angle=0,scale=0.36,trim=30 10 100 80, clip=true]{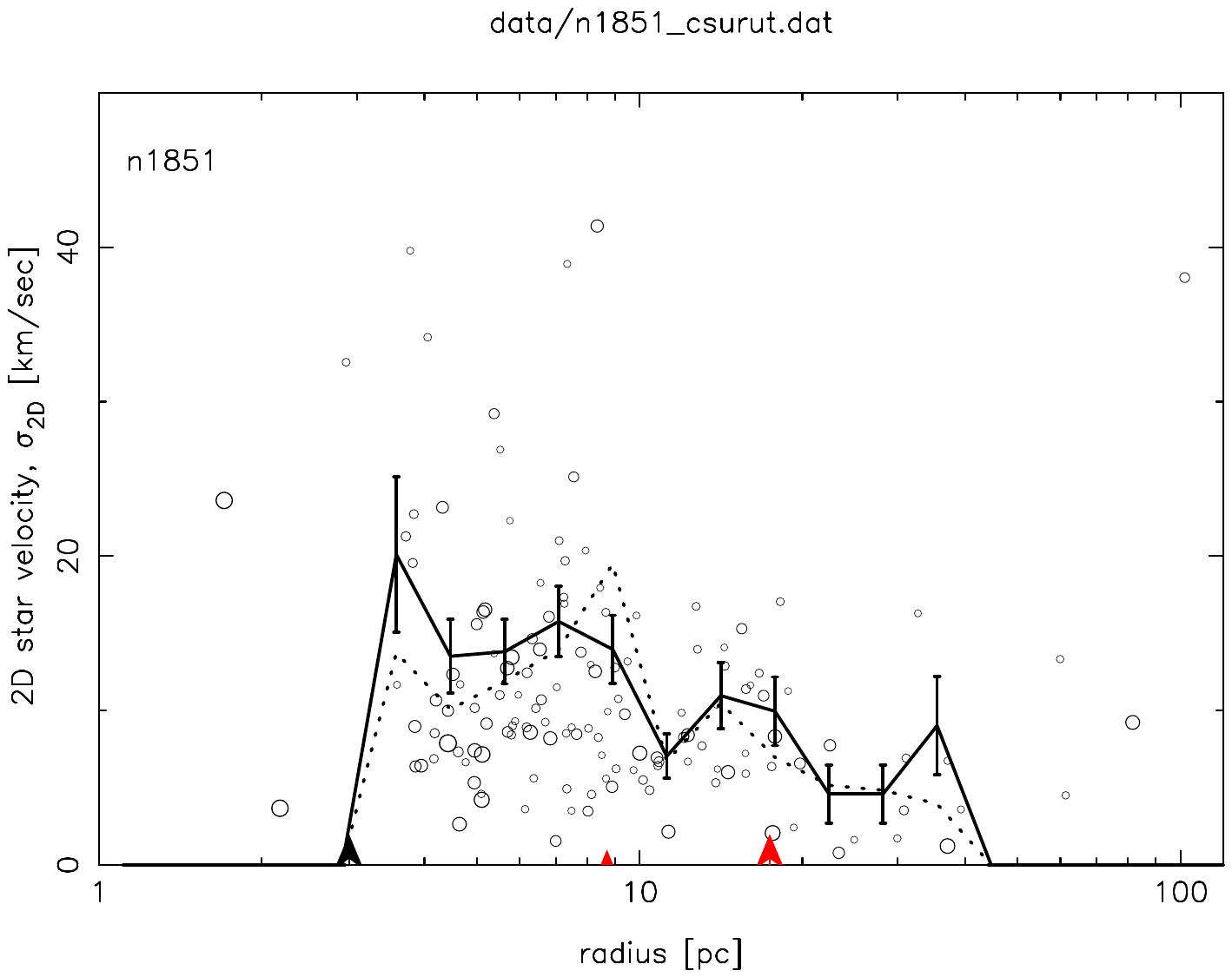}}
\put(5,0){\includegraphics[angle=0,scale=0.36,trim=30 10 100 80, clip=true]{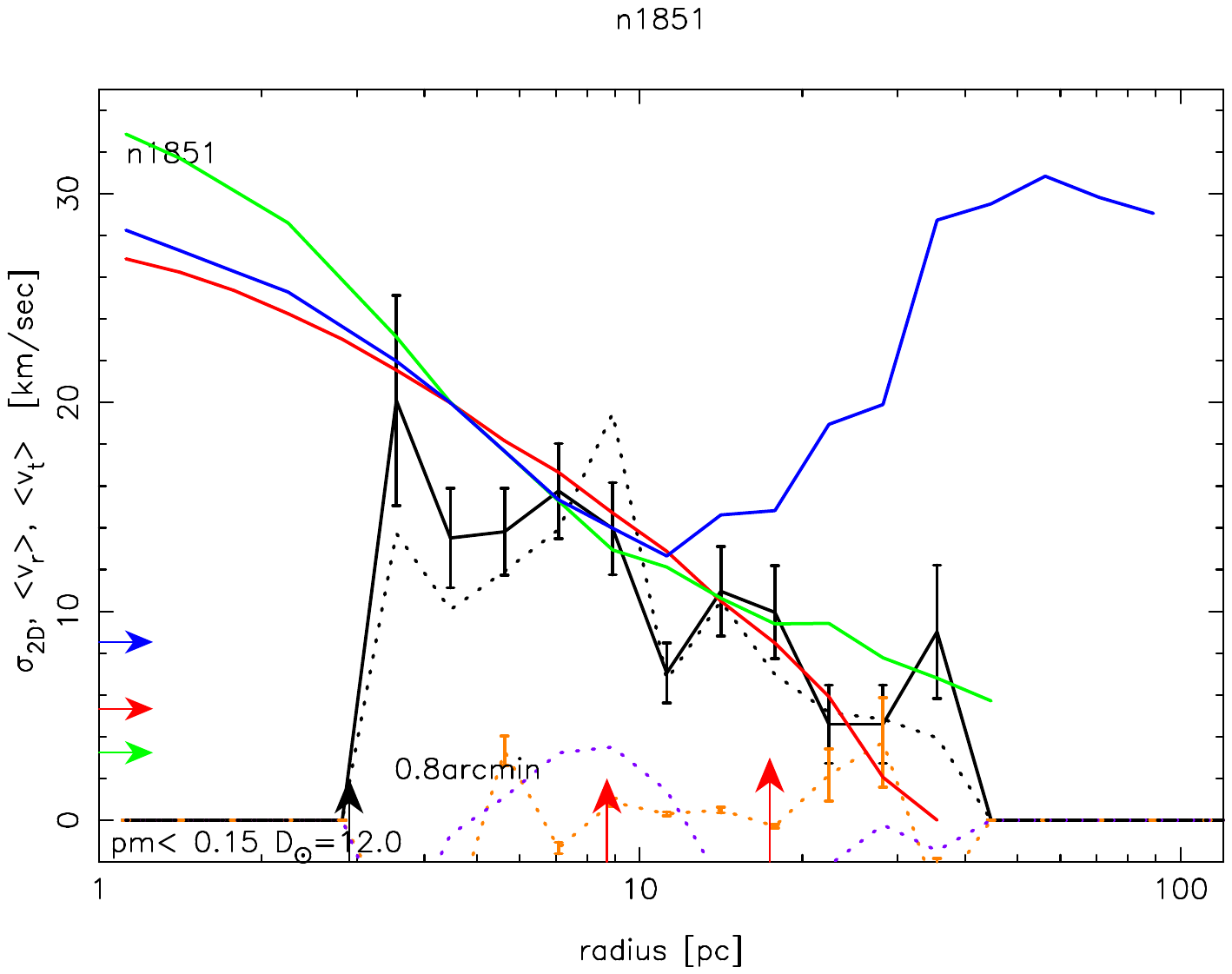}}
\put(120,120){\includegraphics[angle=0,scale=0.12,trim=30 10 100 80, clip=true]{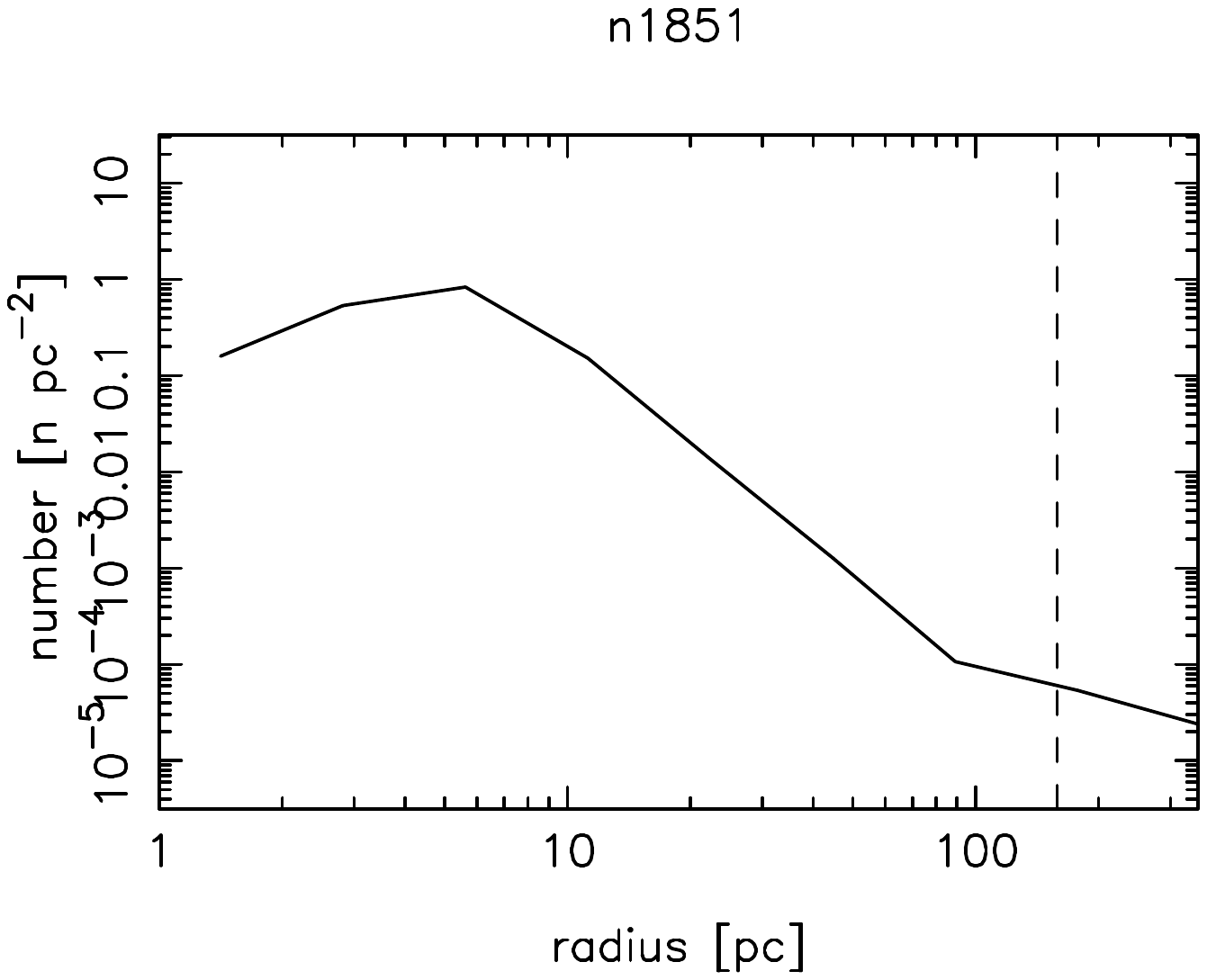}}

{\includegraphics[angle=0,scale=0.36,trim=30 10 100 80, clip=true]{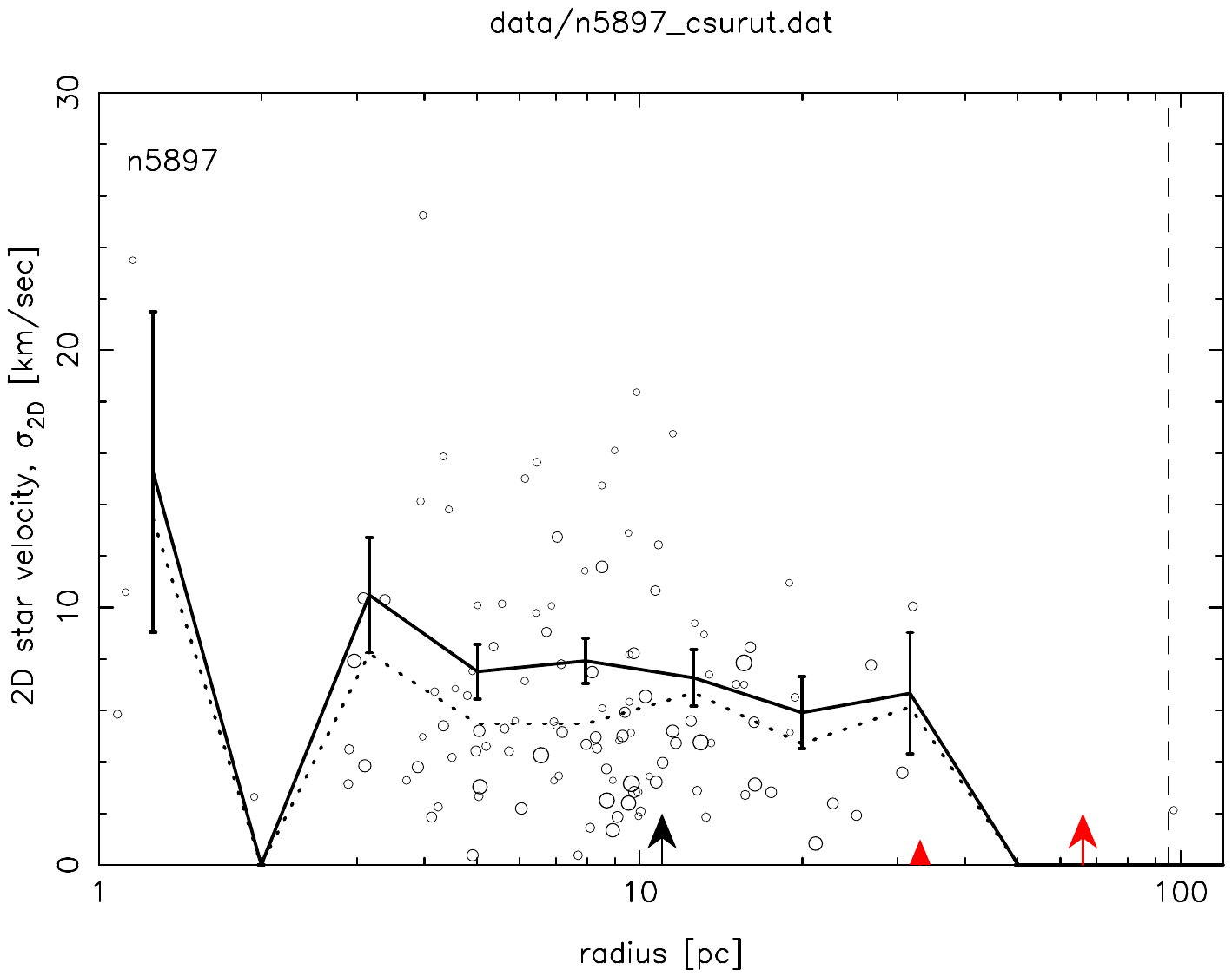}}
\put(5,0){\includegraphics[angle=0,scale=0.36,trim=30 10 100 80, clip=true]{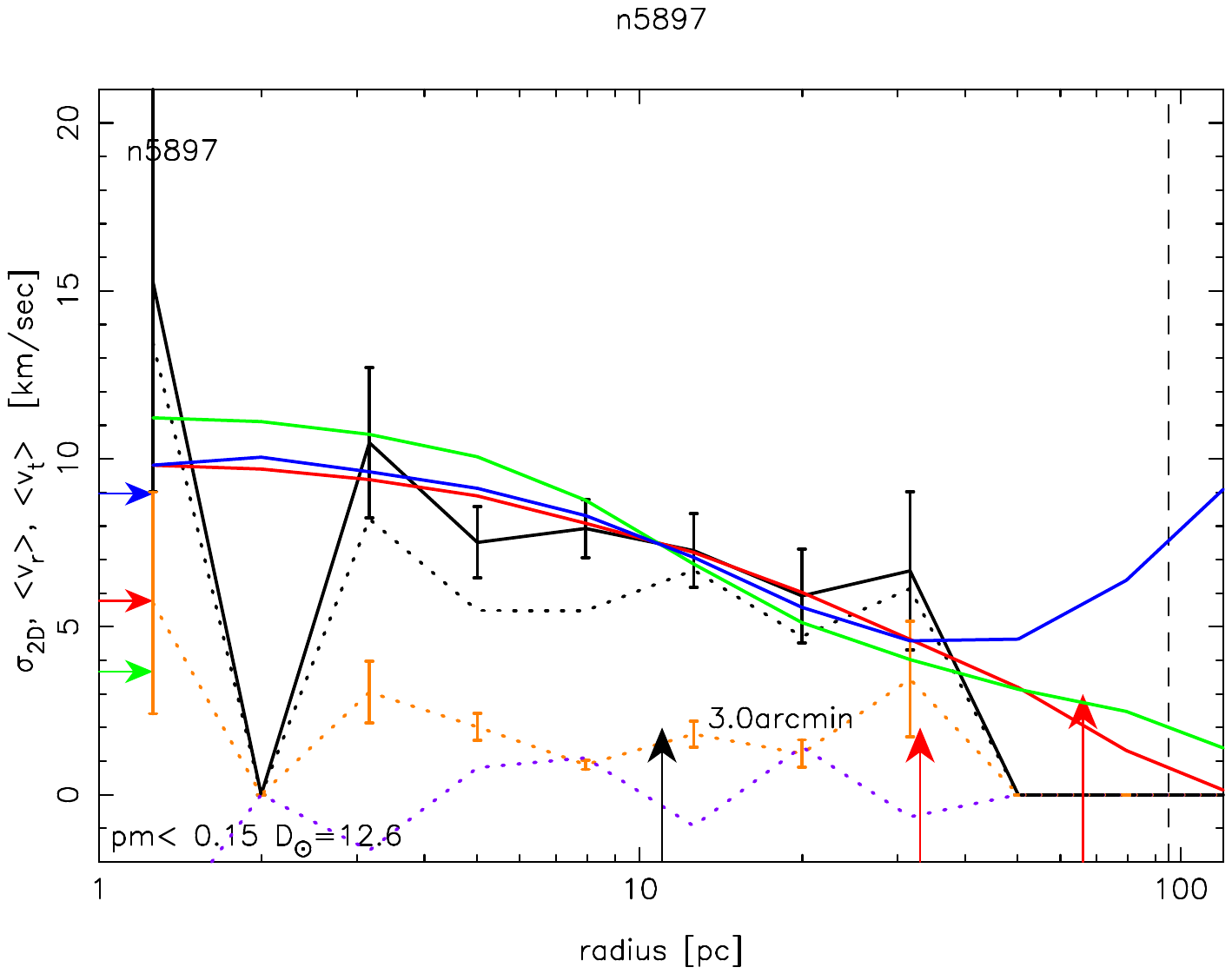}}
\put(120,120){\includegraphics[angle=0,scale=0.12,trim=30 10 100 80, clip=true]{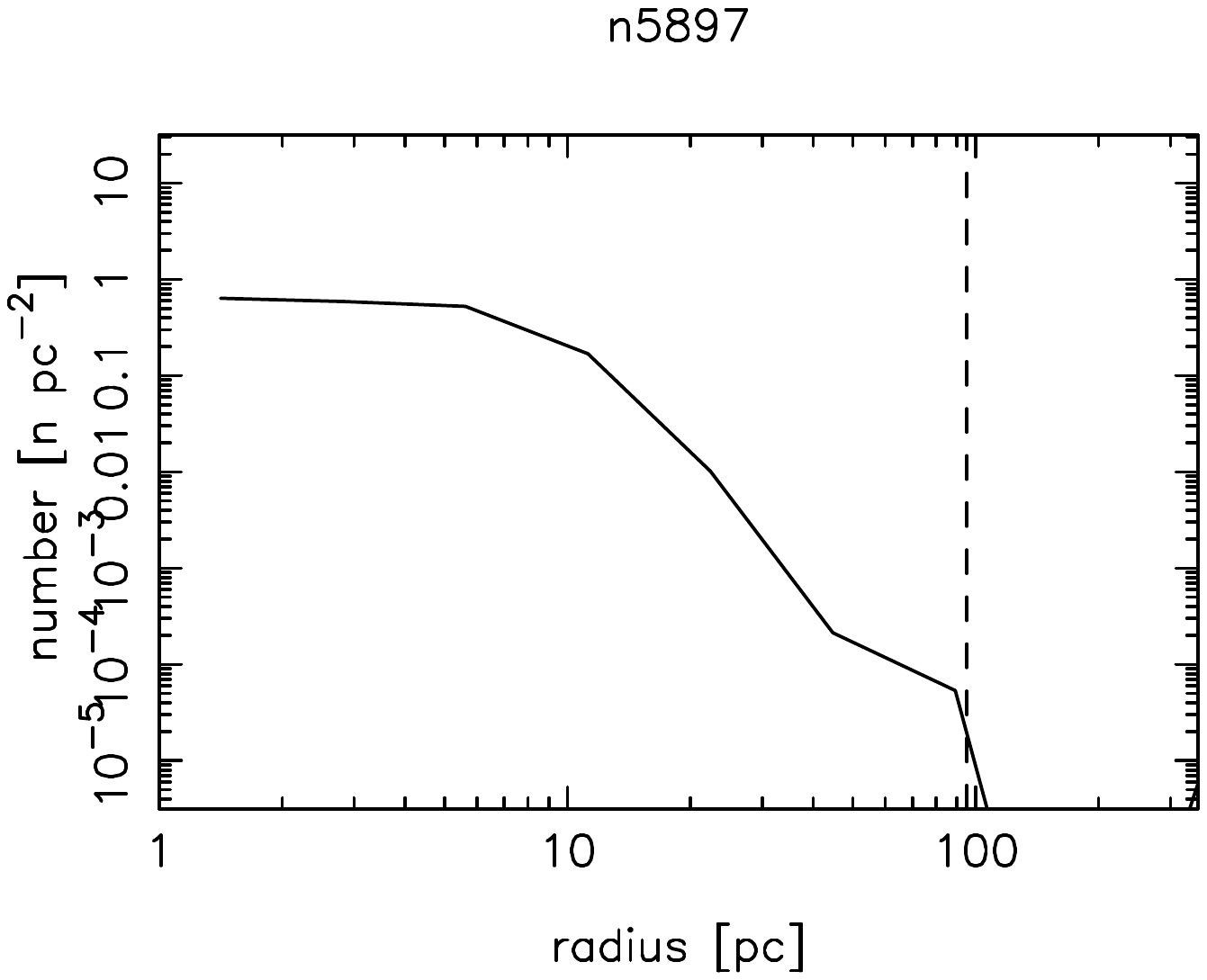}}
\caption{Velocities for the clusters NGC 7089, 1851 and 5897, top to bottom. Symbols as in Figure~\ref{fig_6752}.
}
\label{fig_5897}
\end{figure*}

\begin{figure*}
{\includegraphics[angle=0,scale=0.36,trim=30 10 100 80, clip=true]{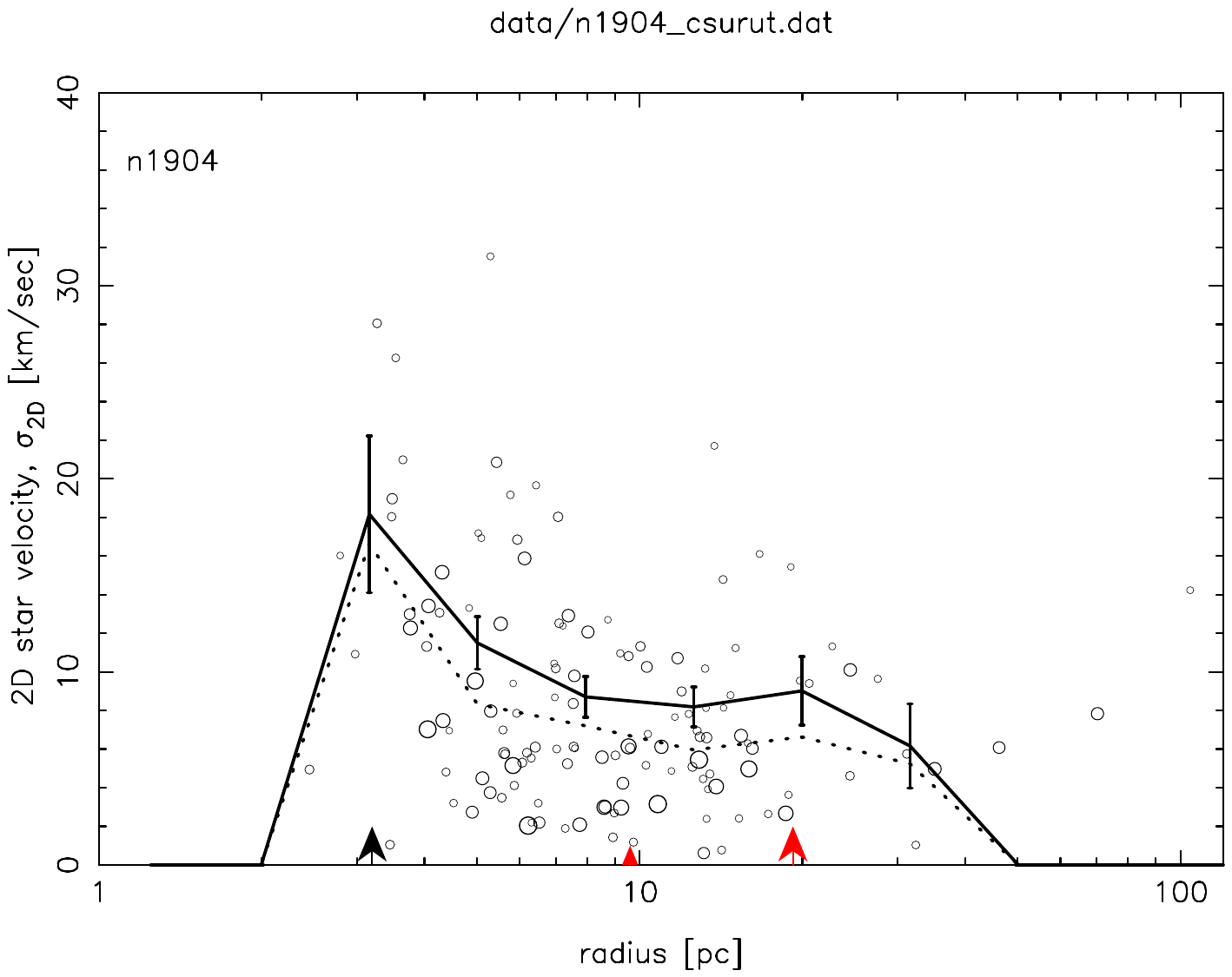}}
\put(5,0){\includegraphics[angle=0,scale=0.36,trim=30 10 100 80, clip=true]{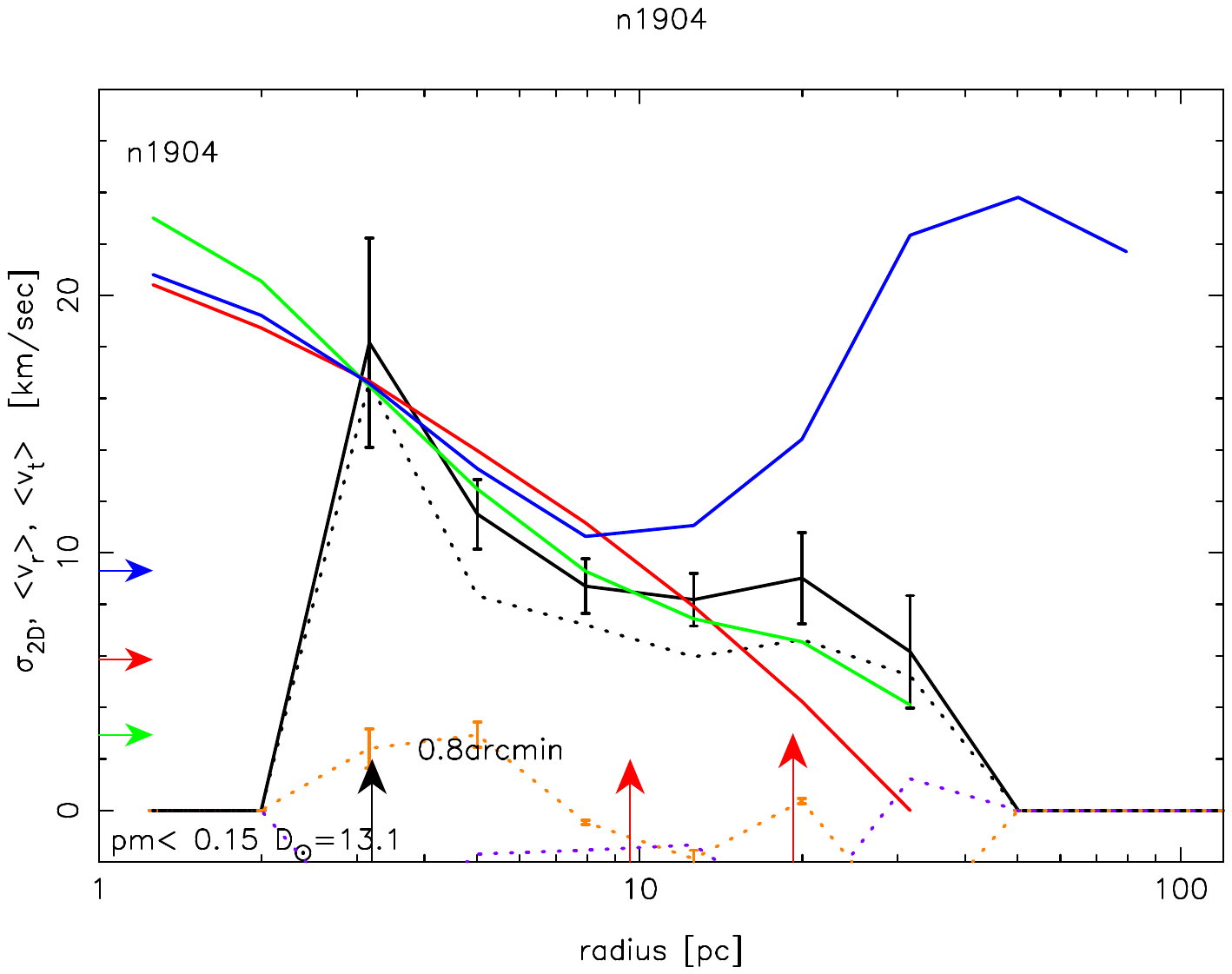}}
\put(120,120){\includegraphics[angle=0,scale=0.12,trim=30 10 100 80, clip=true]{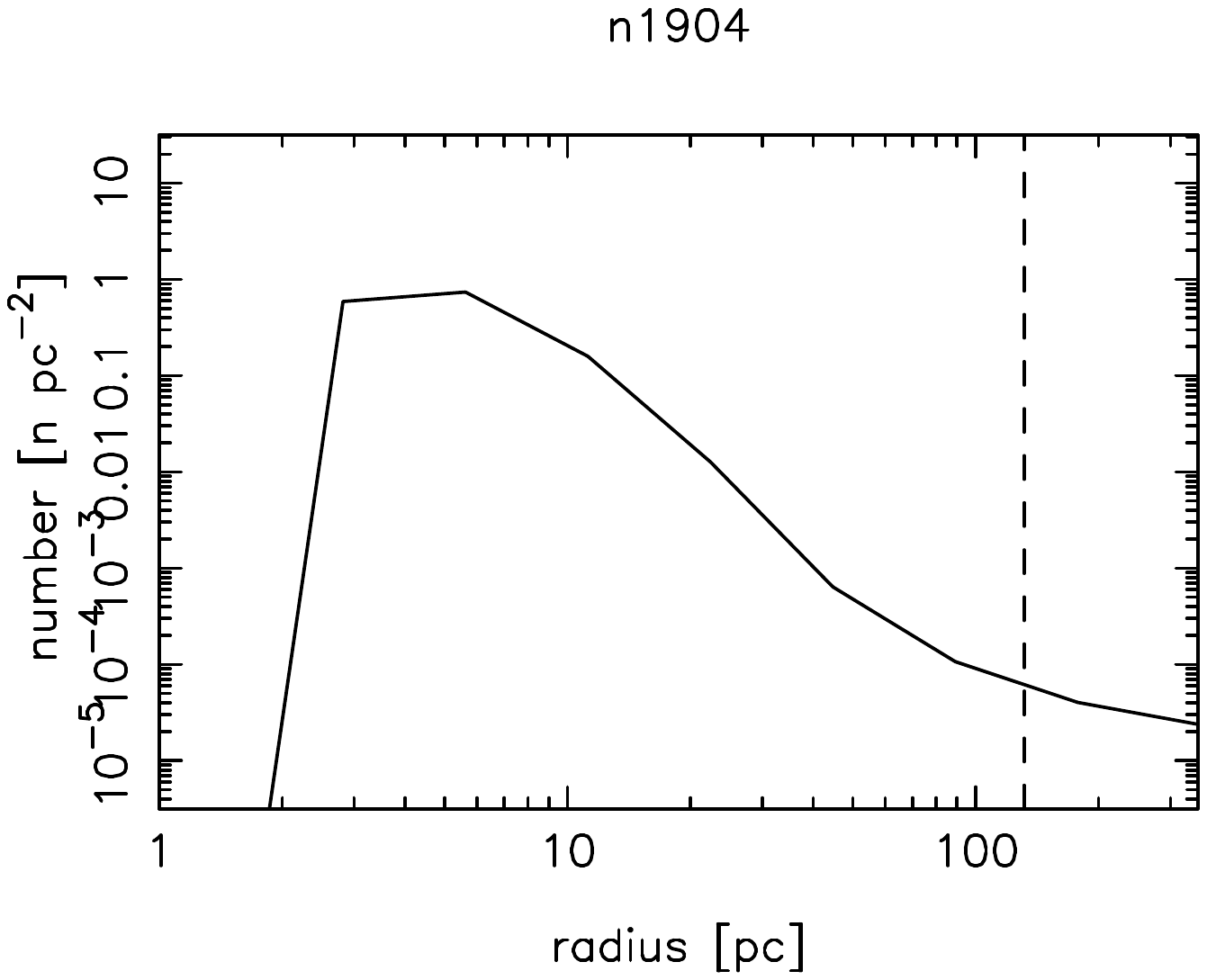}}

{\includegraphics[angle=0,scale=0.36,trim=30 10 100 80, clip=true]{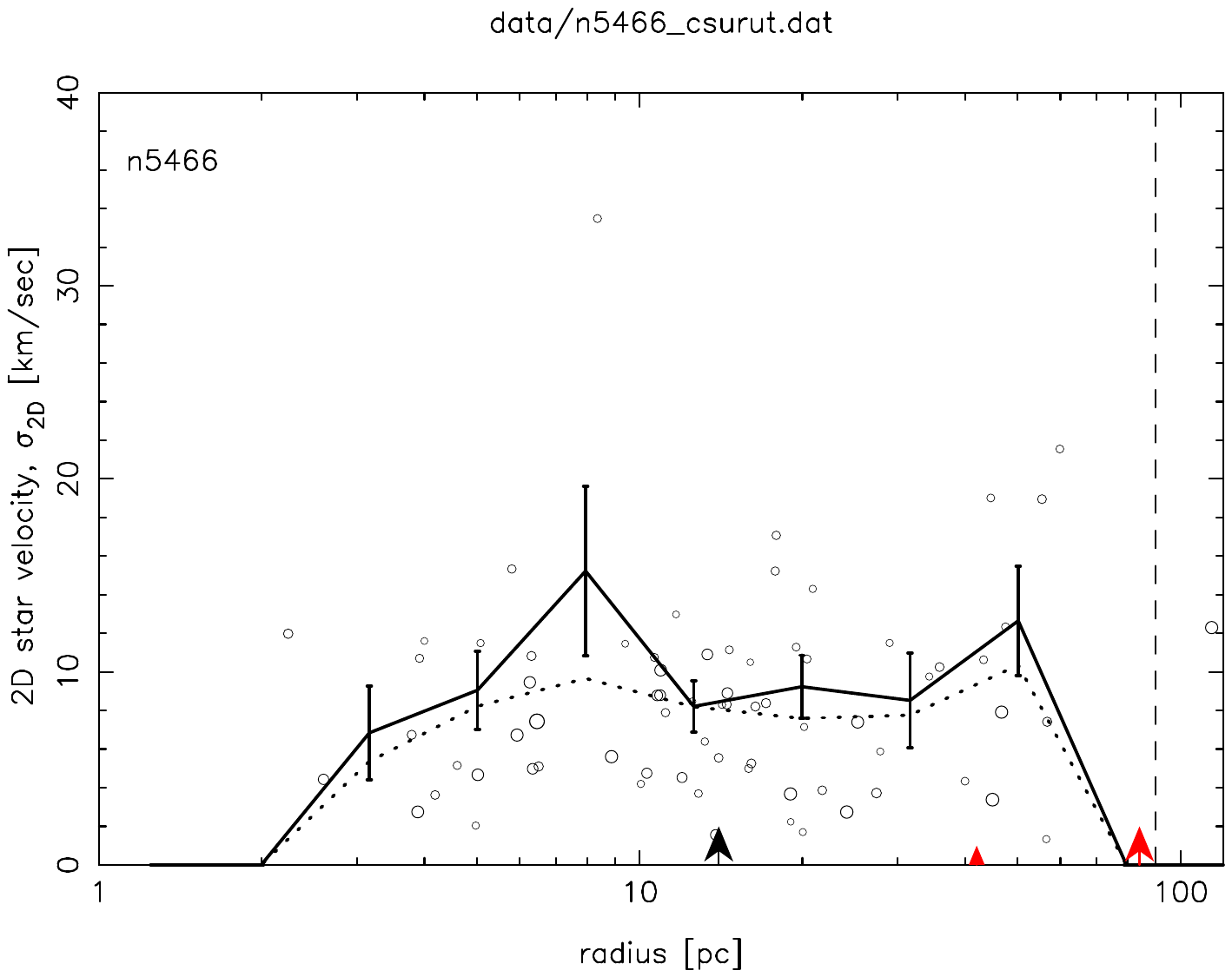}}
\put(5,0){\includegraphics[angle=0,scale=0.36,trim=30 10 100 80, clip=true]{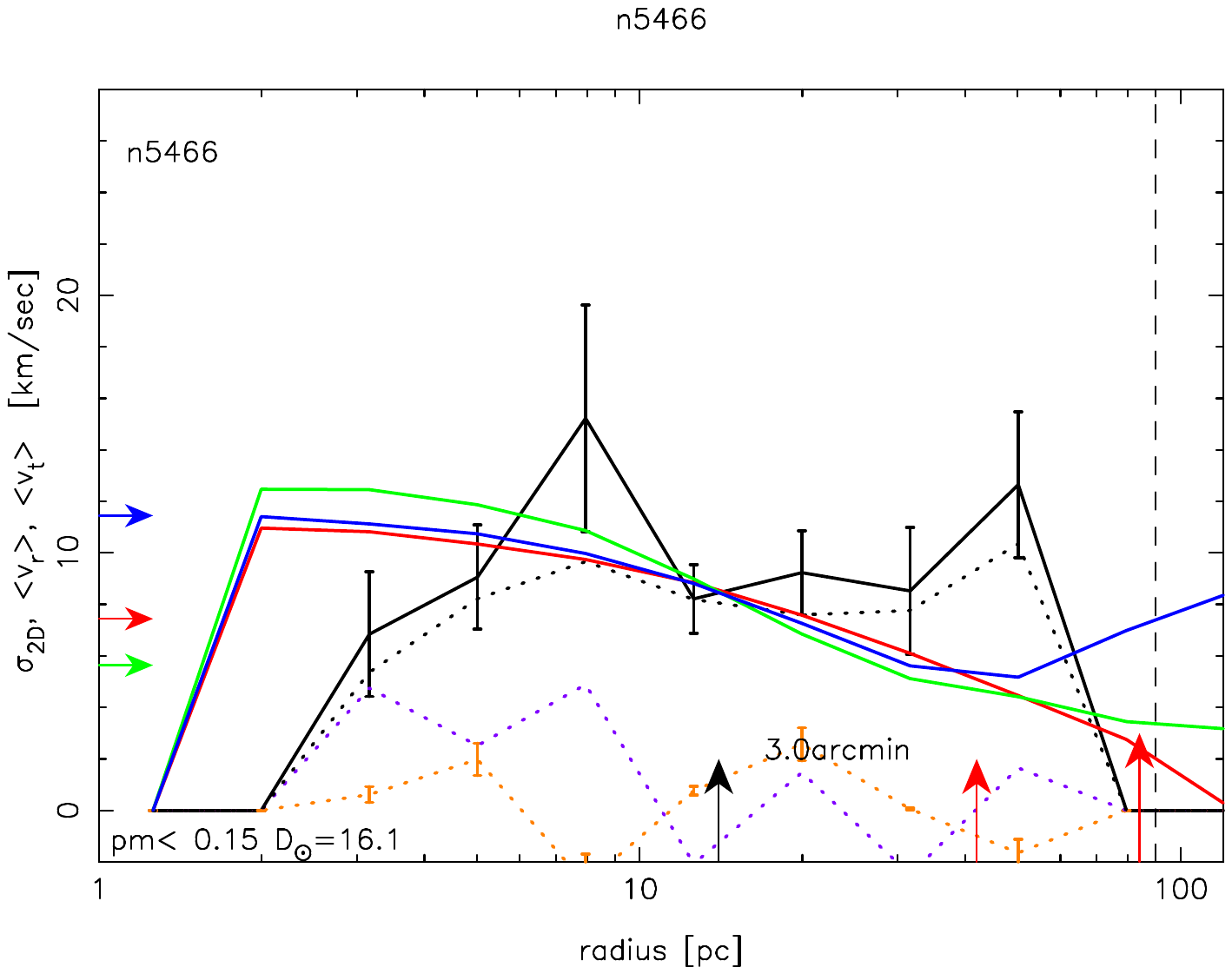}}
\put(120,120){\includegraphics[angle=0,scale=0.12,trim=30 10 100 80, clip=true]{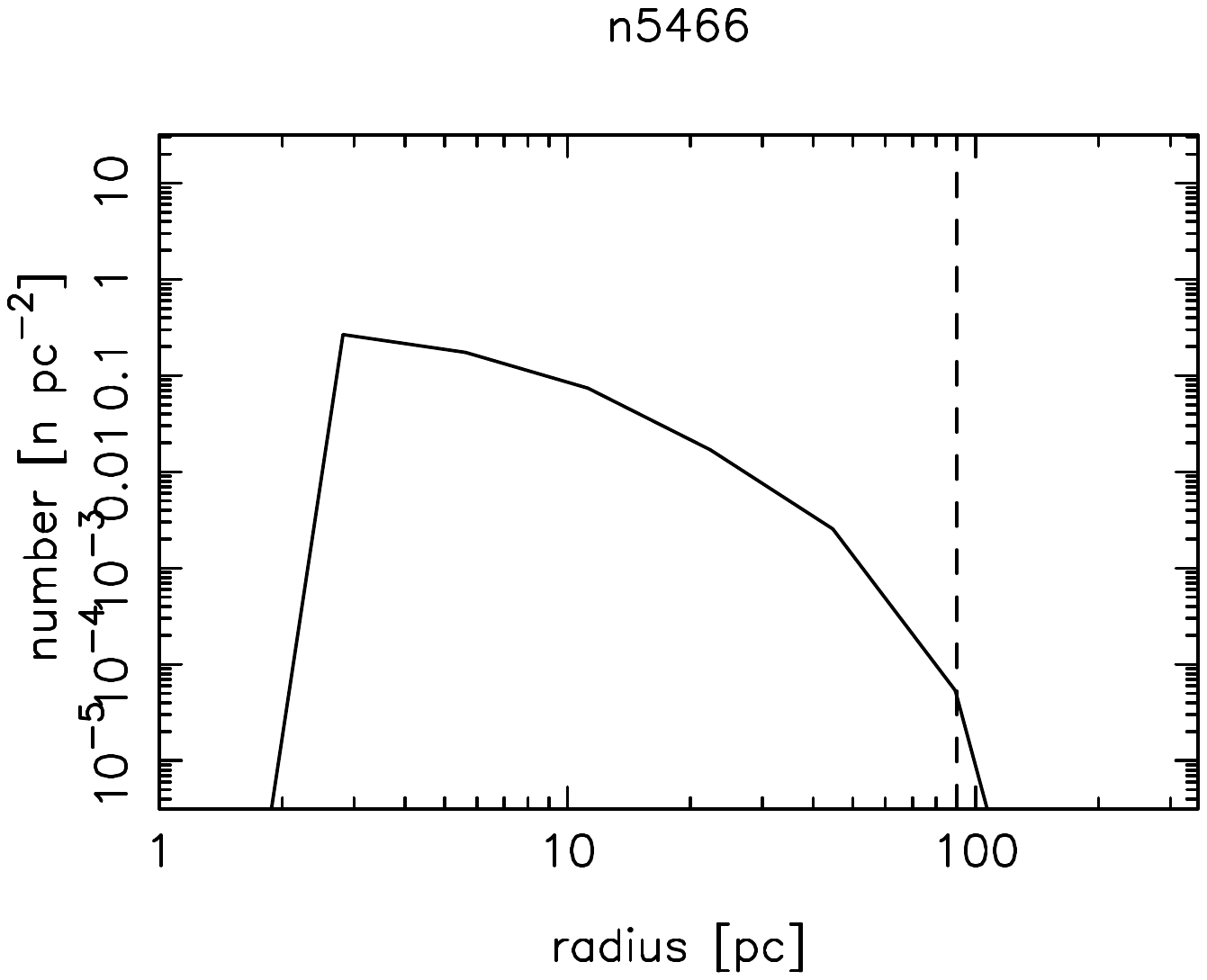}}

{\includegraphics[angle=0,scale=0.36,trim=30 10 100 80, clip=true]{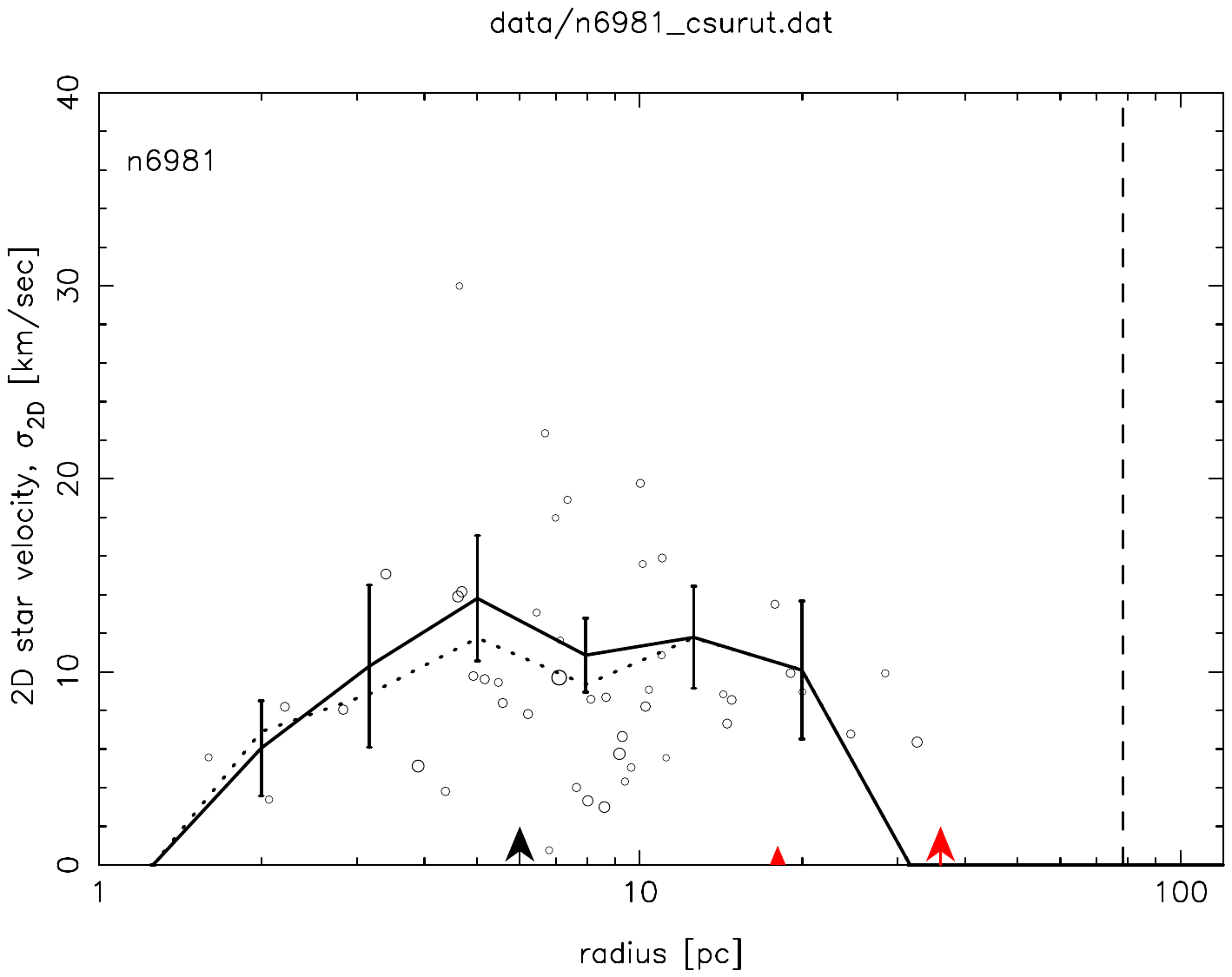}}
\put(5,0){\includegraphics[angle=0,scale=0.36,trim=30 10 100 80, clip=true]{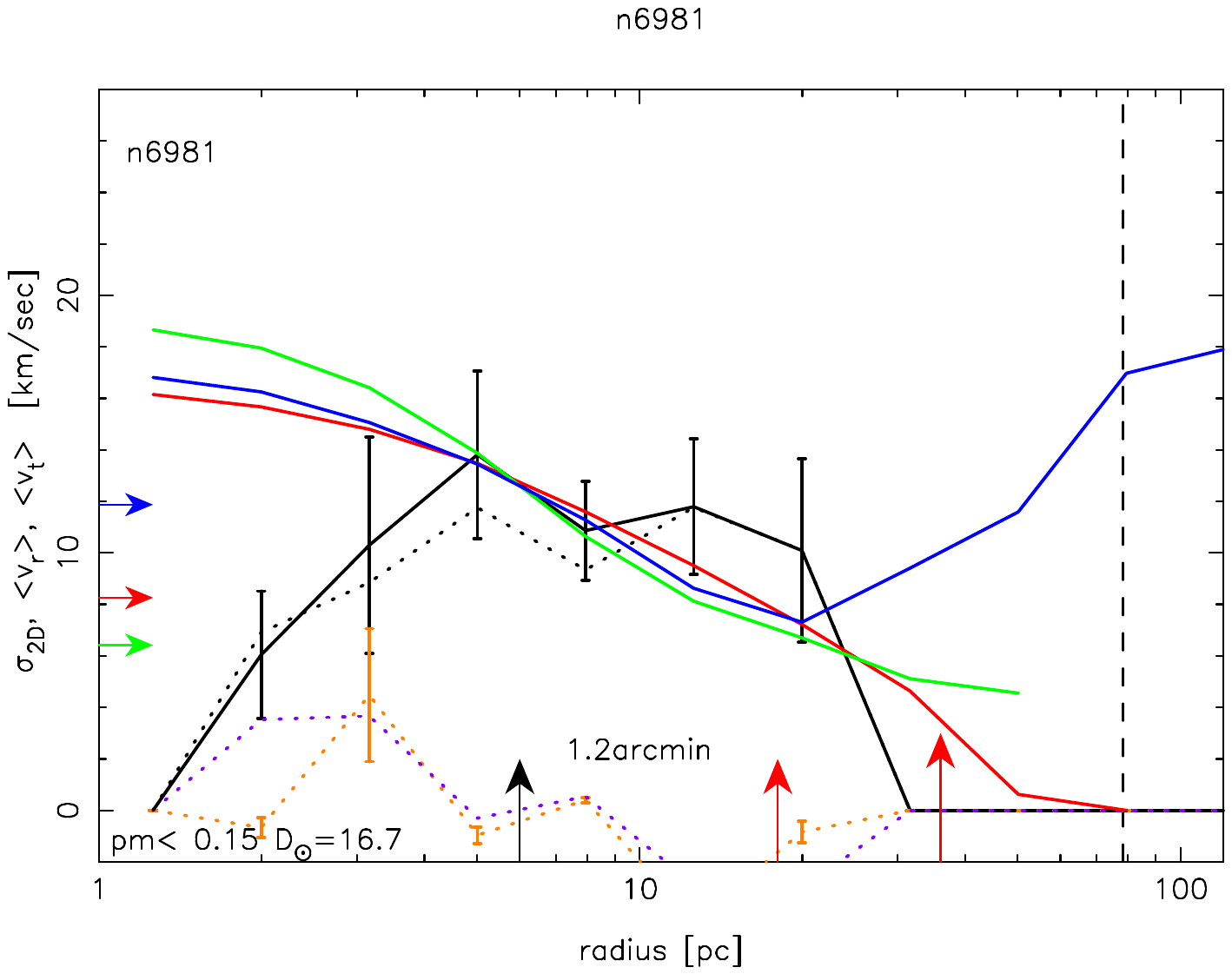}}
\put(120,120){\includegraphics[angle=0,scale=0.12,trim=30 10 100 80, clip=true]{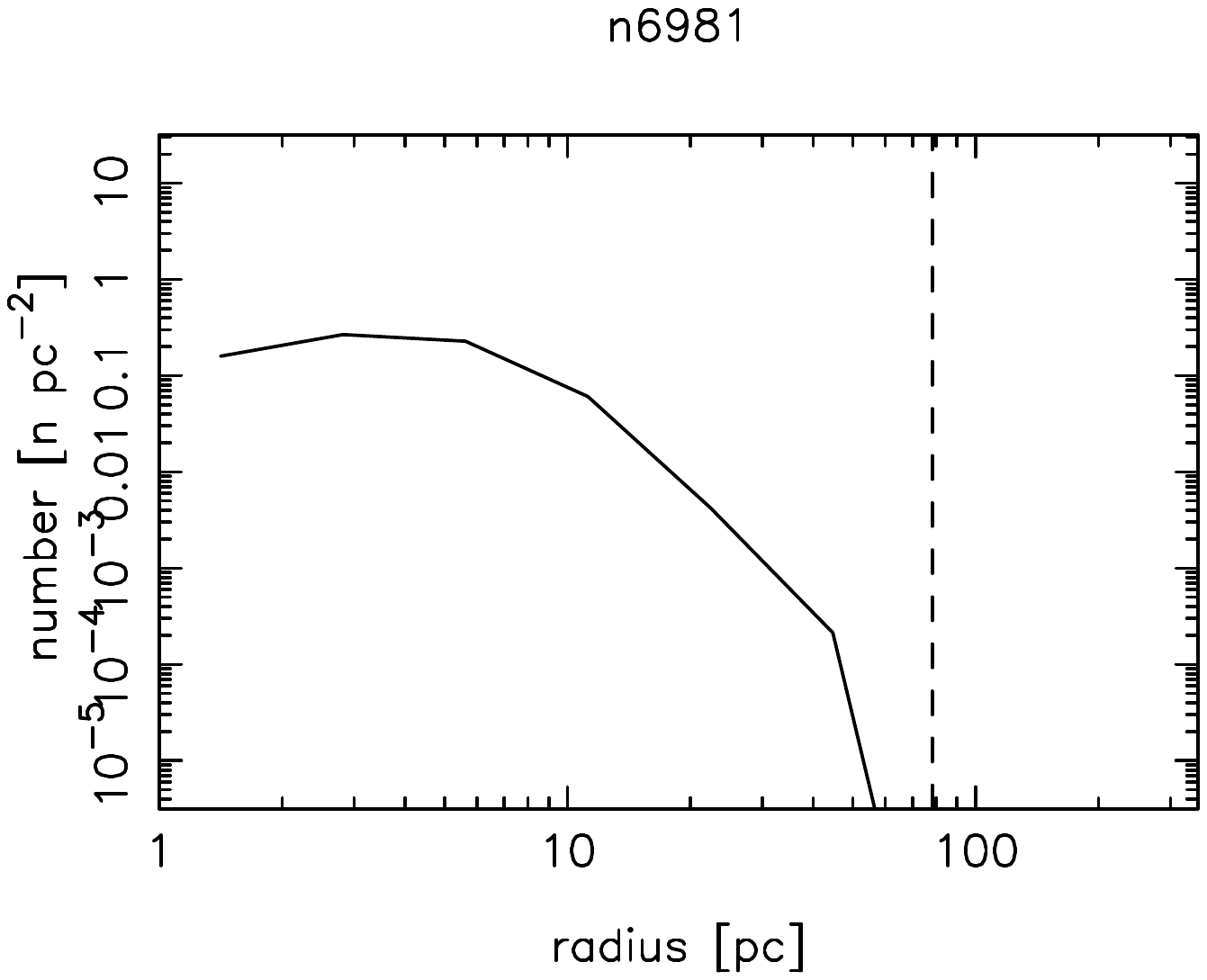}}
\caption{Velocities for the clusters NGC 1904, 5466, and 6981, top to bottom. Symbols as in Figure~\ref{fig_6752}.
}
\label{fig_6981}
\end{figure*}

\begin{figure*}
{\includegraphics[angle=0,scale=0.36,trim=30 10 100 80, clip=true]{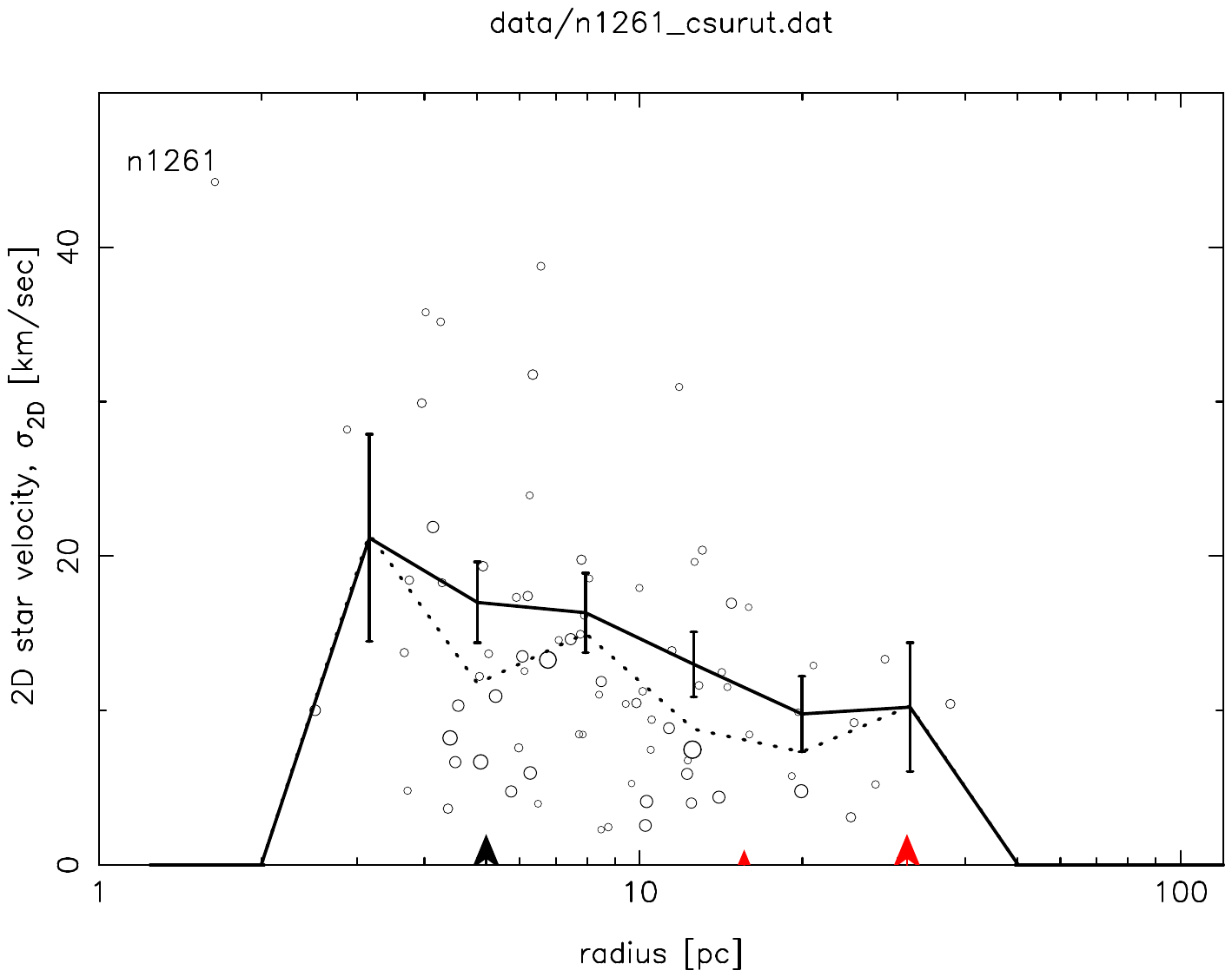}}
\put(5,0){\includegraphics[angle=0,scale=0.36,trim=30 10 100 80, clip=true]{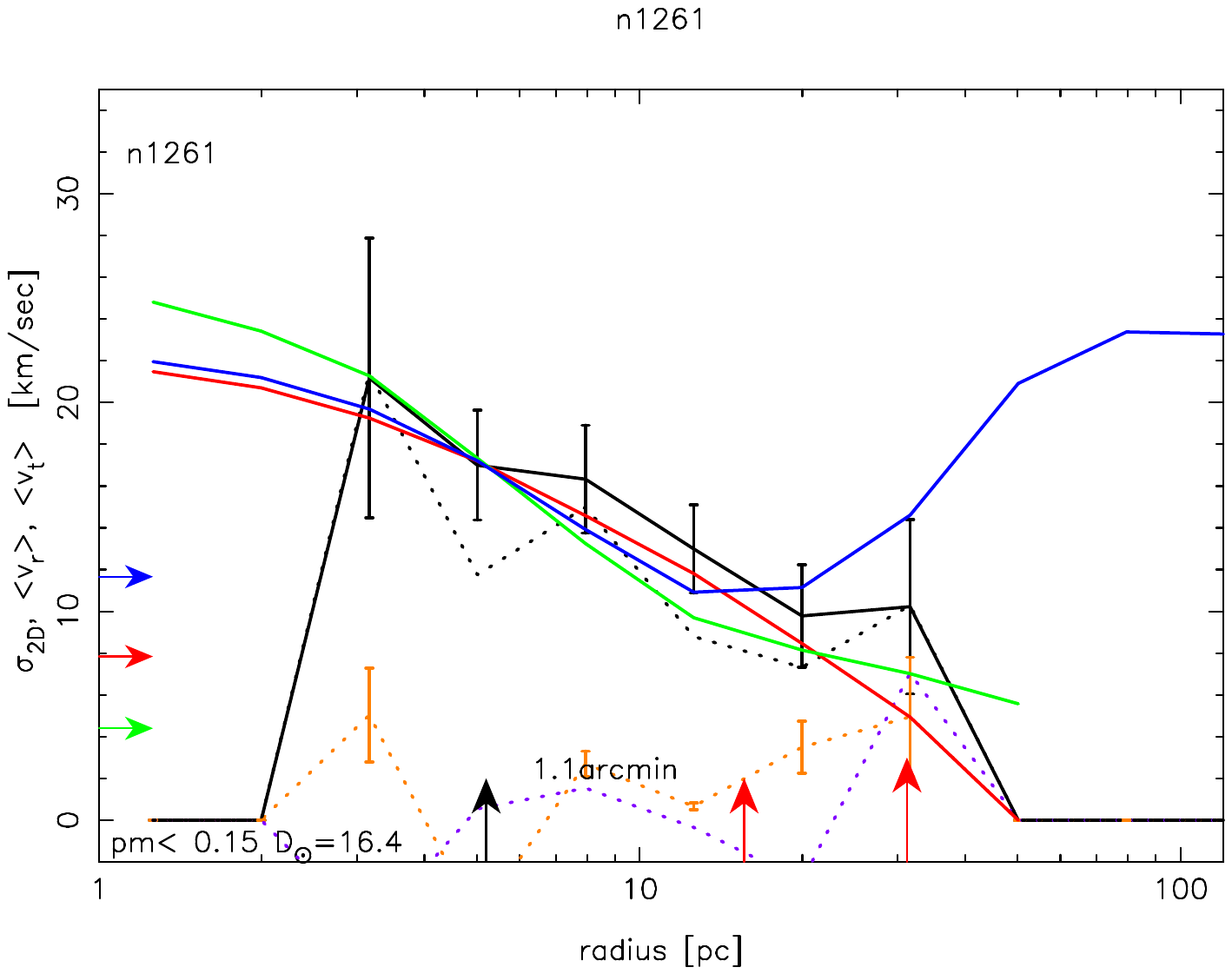}}
\put(120,120){\includegraphics[angle=0,scale=0.12,trim=30 10 100 80, clip=true]{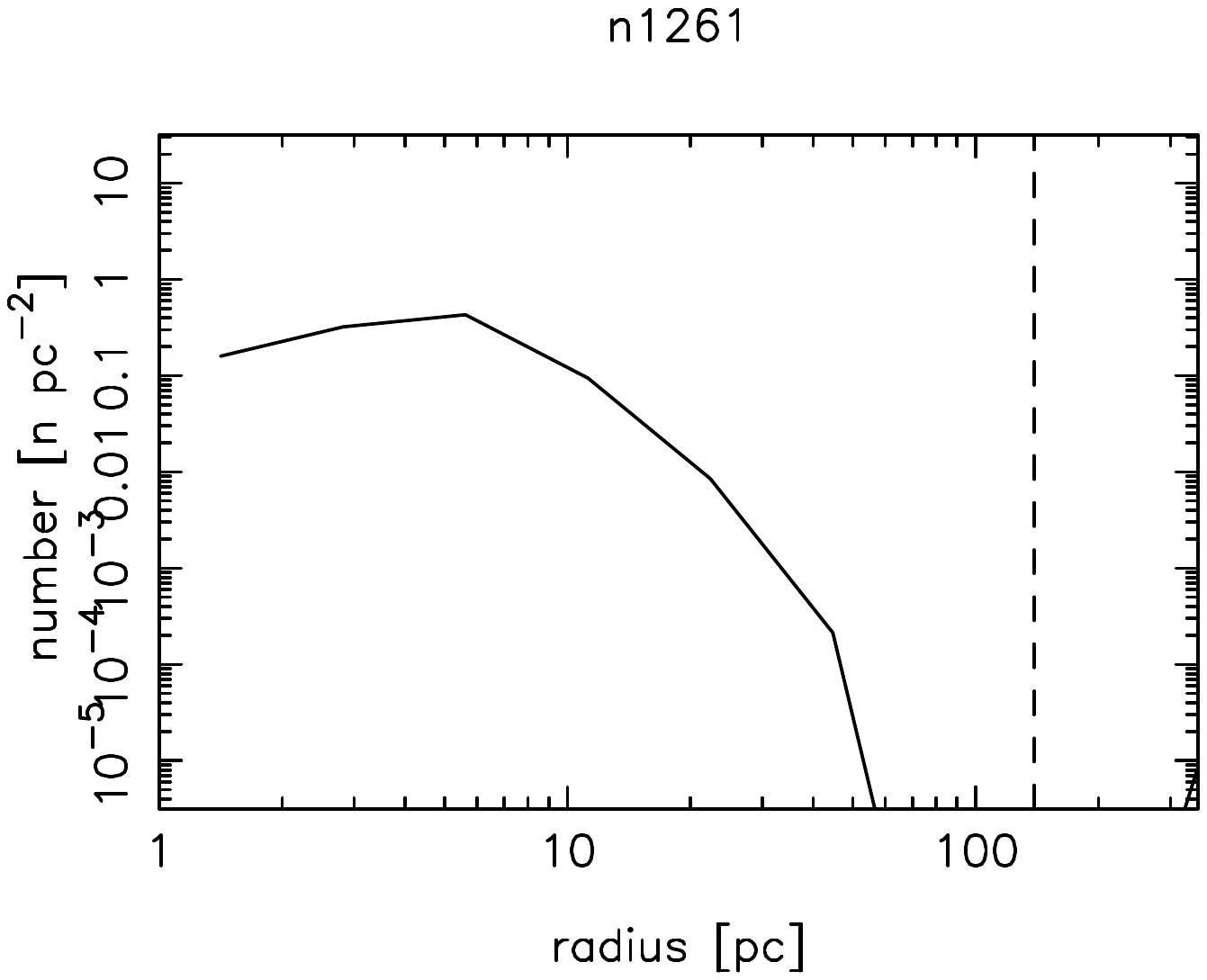}}

{\includegraphics[angle=0,scale=0.36,trim=30 10 100 80, clip=true]{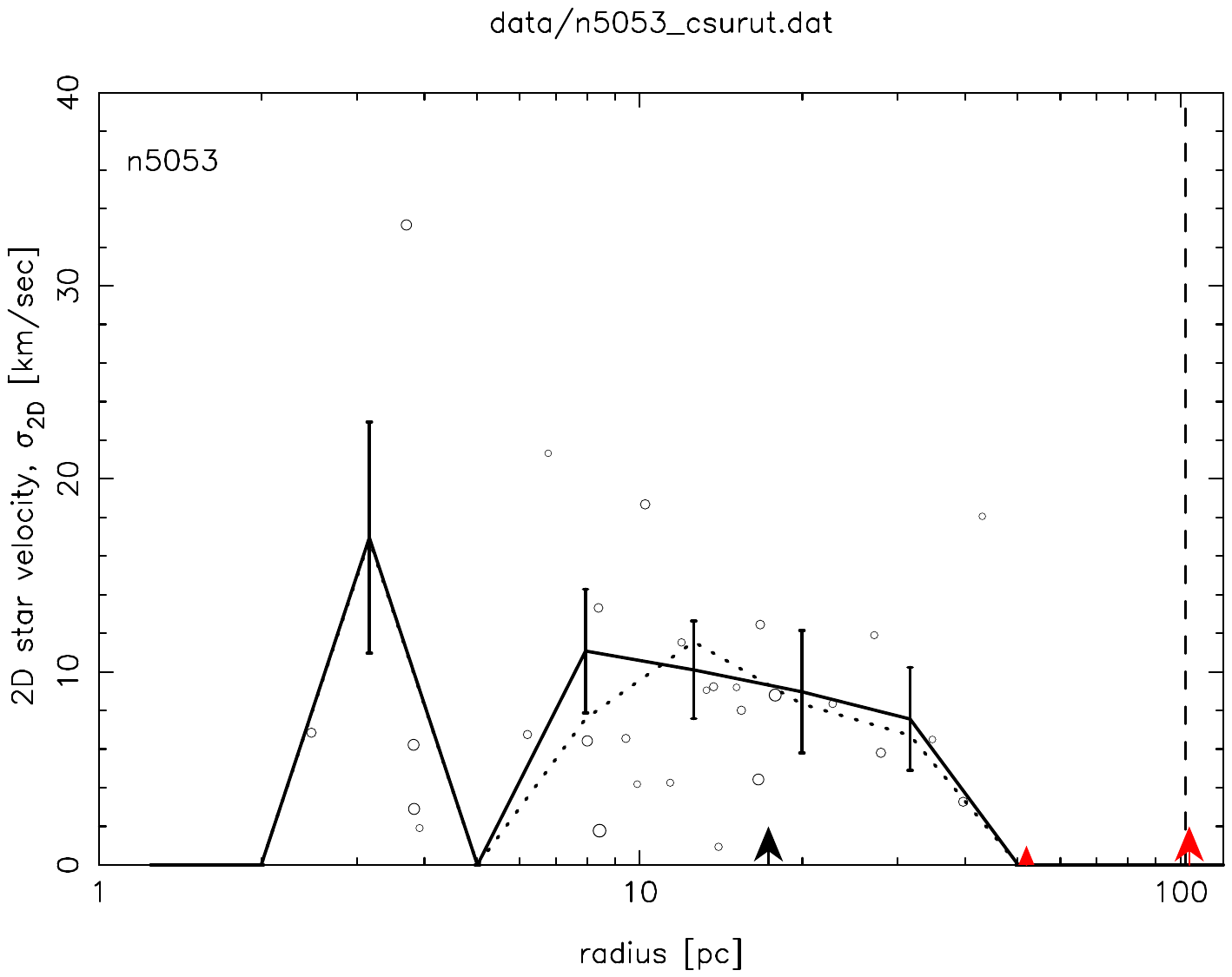}}
\put(5,0){\includegraphics[angle=0,scale=0.36,trim=30 10 100 80, clip=true]{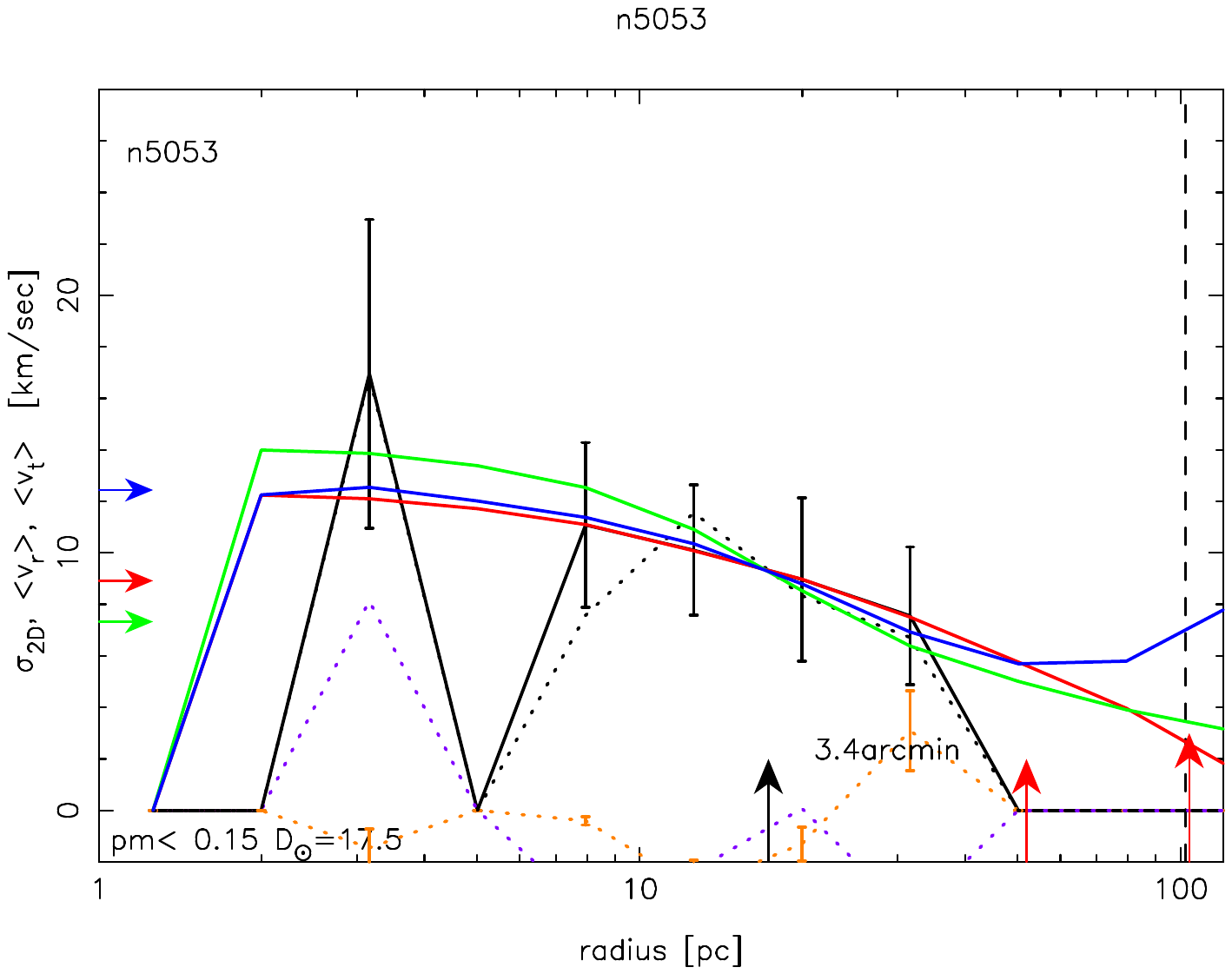}}
\put(120,120){\includegraphics[angle=0,scale=0.12,trim=30 10 100 80, clip=true]{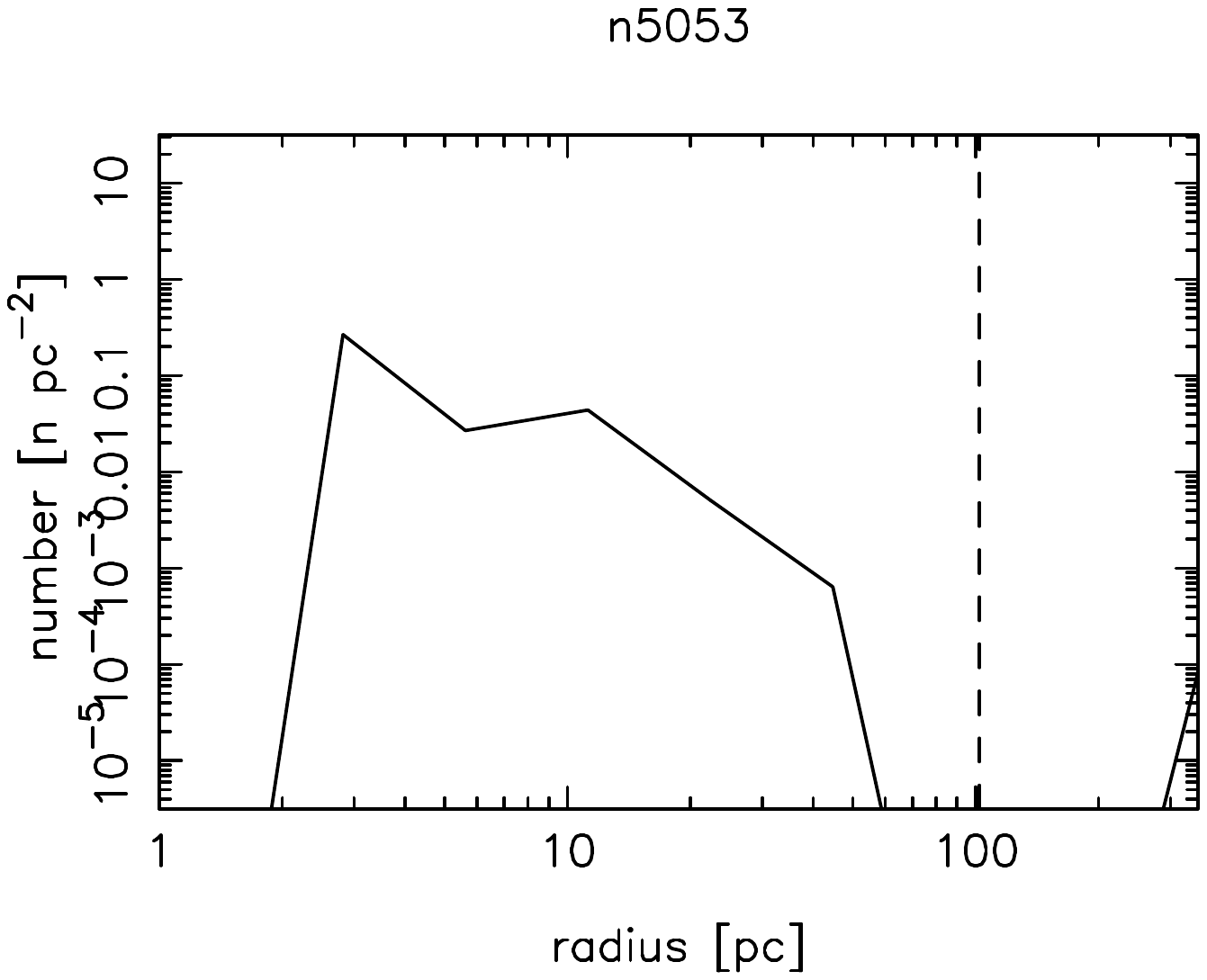}}

{\includegraphics[angle=0,scale=0.36,trim=30 10 100 80, clip=true]{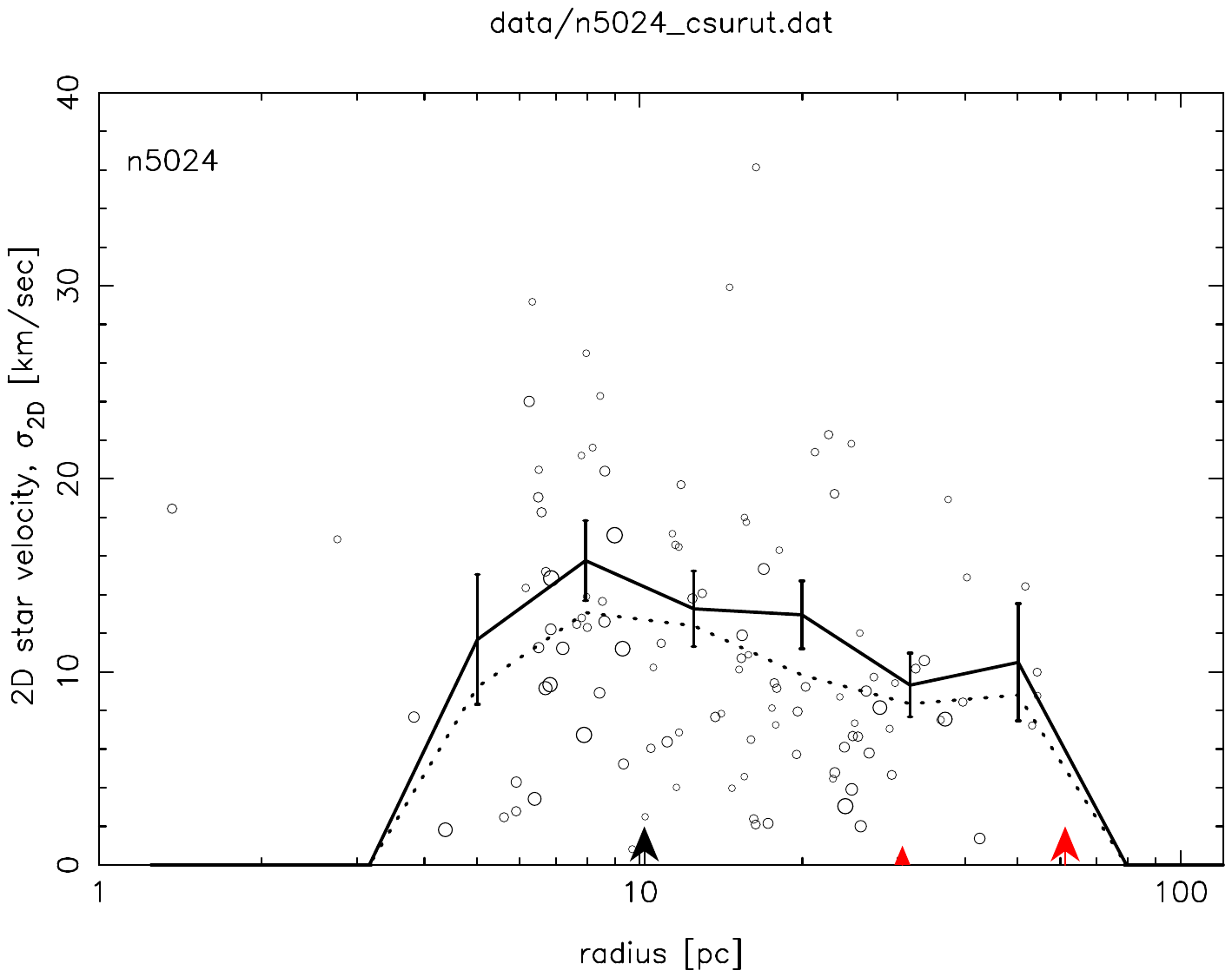}}
\put(5,0){\includegraphics[angle=0,scale=0.36,trim=30 10 100 80, clip=true]{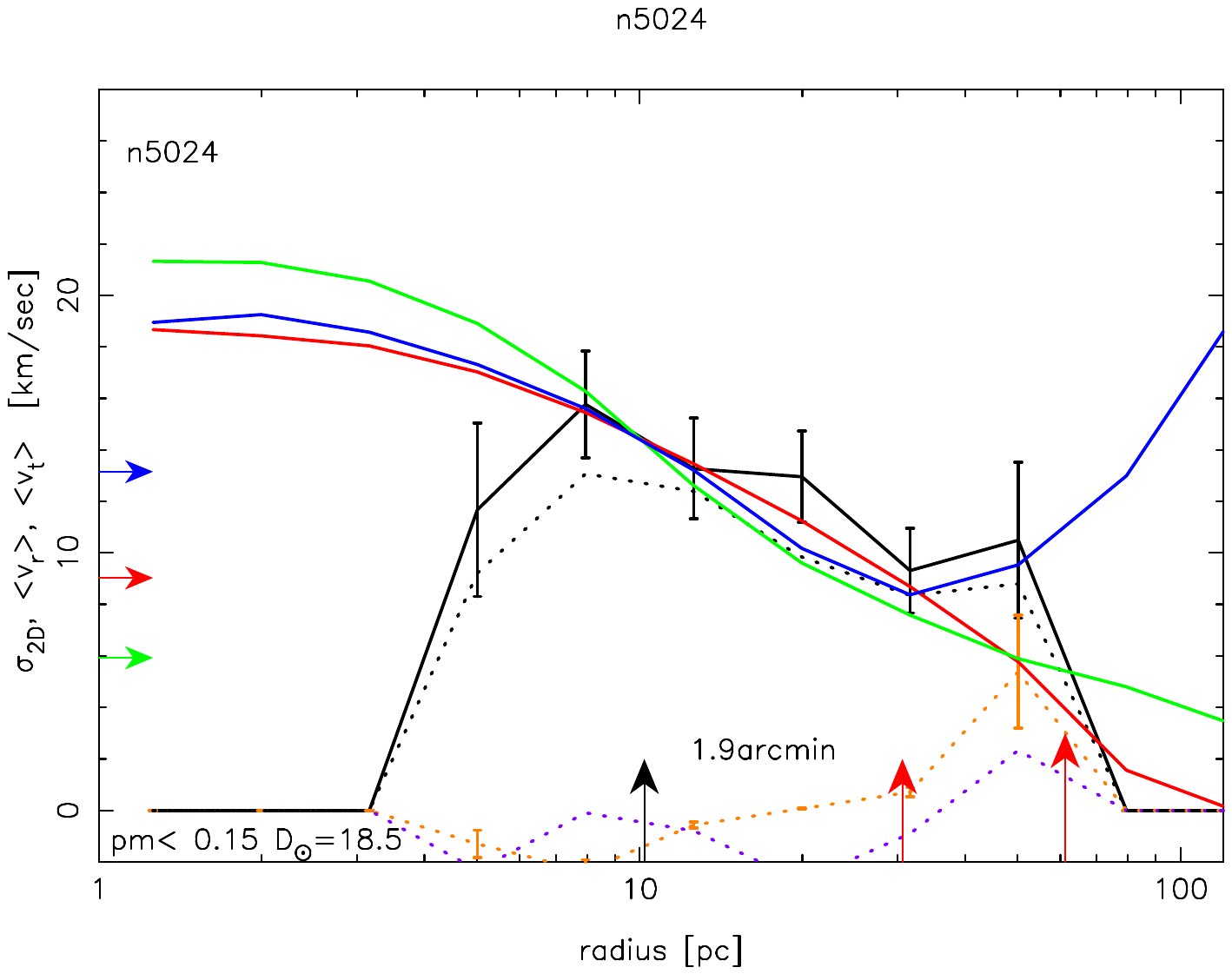}}
\put(120,120){\includegraphics[angle=0,scale=0.12,trim=30 10 100 80, clip=true]{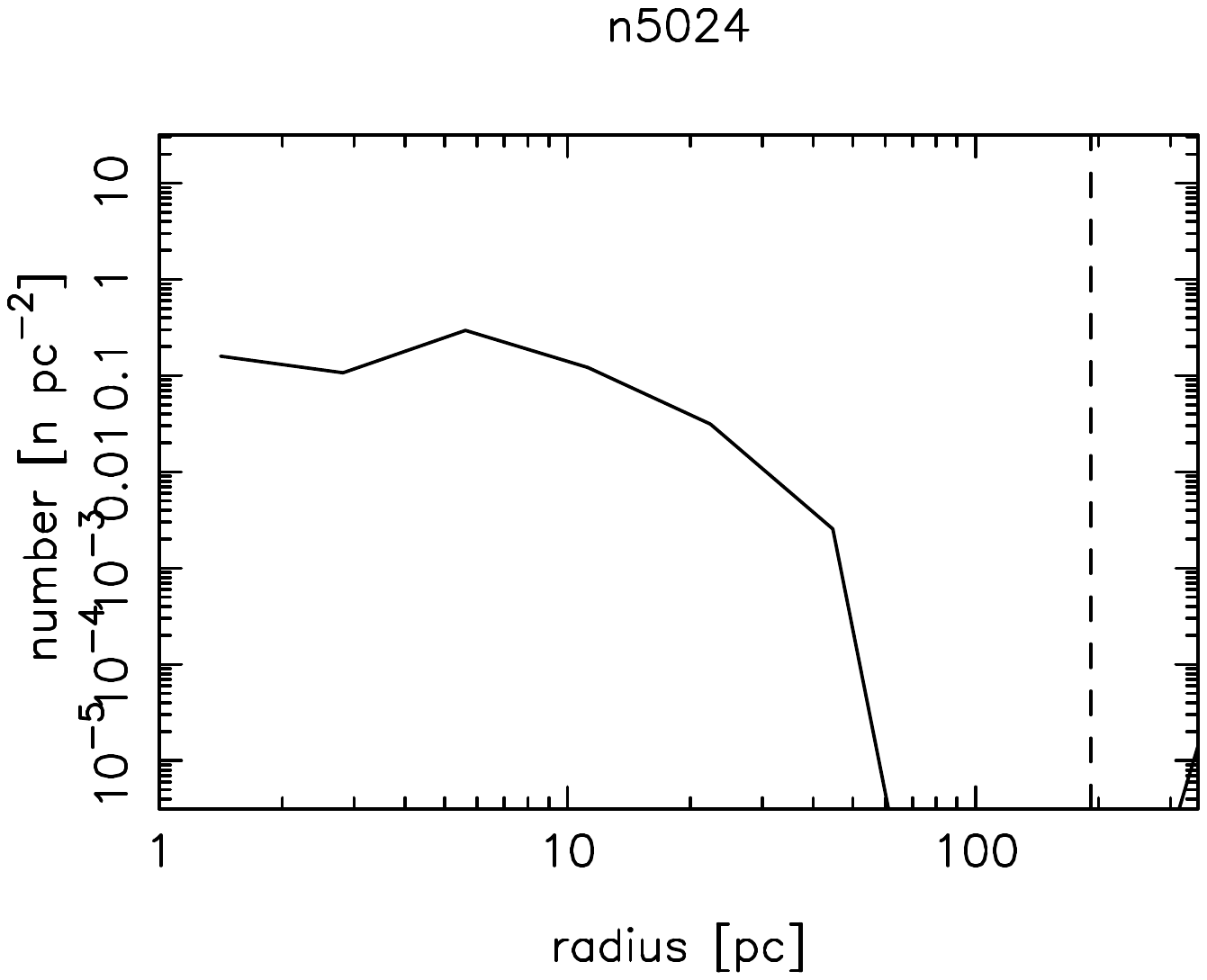}}
\caption{Velocities for the clusters NGC 1261, 5053 and 5024, top to bottom. Symbols as in Figure~\ref{fig_6752}.
}
\label{fig_5024}
\end{figure*}

\begin{figure*}
{\includegraphics[angle=0,scale=0.36,trim=30 10 100 80, clip=true]{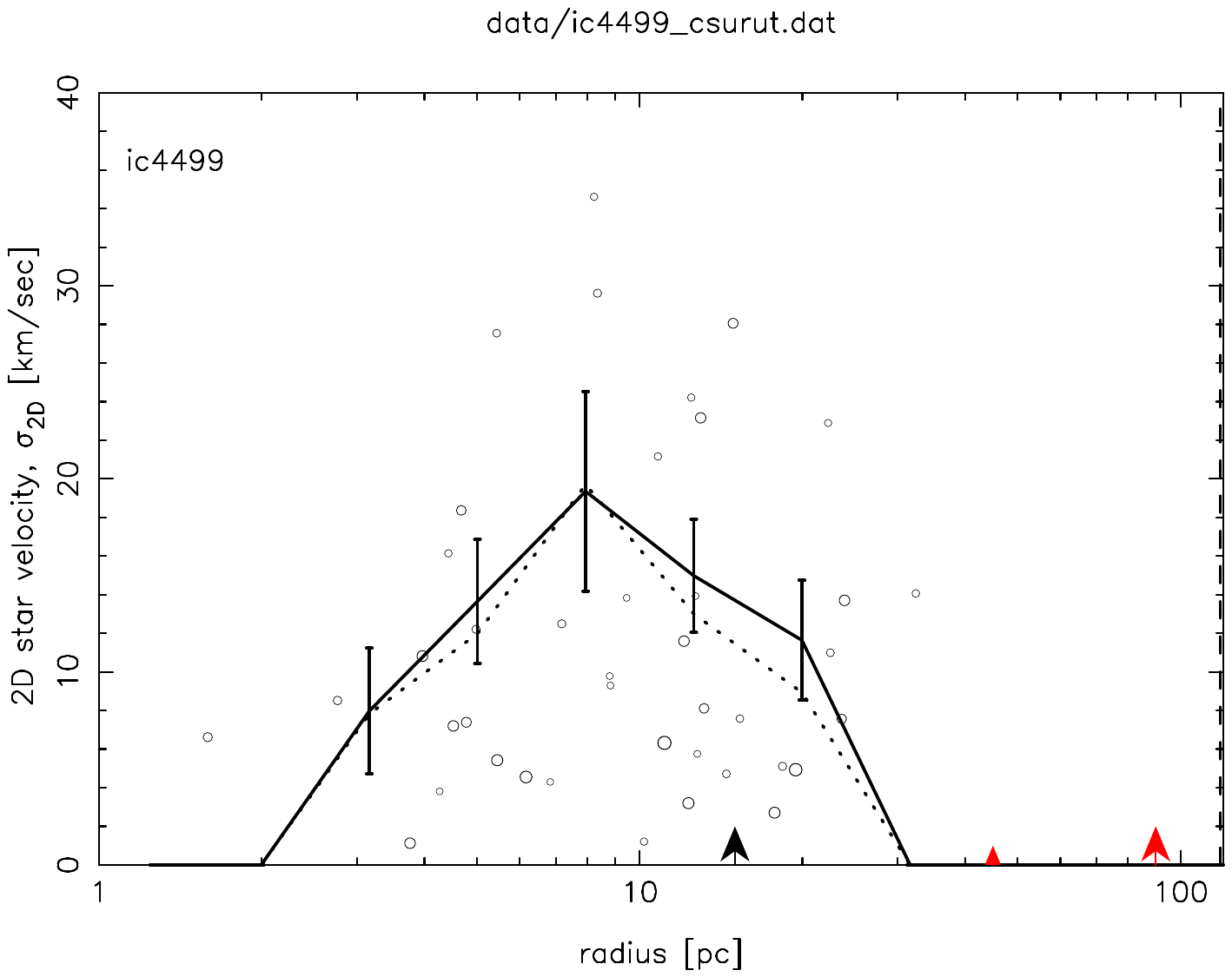}}
\put(5,0){\includegraphics[angle=0,scale=0.36,trim=30 10 100 80, clip=true]{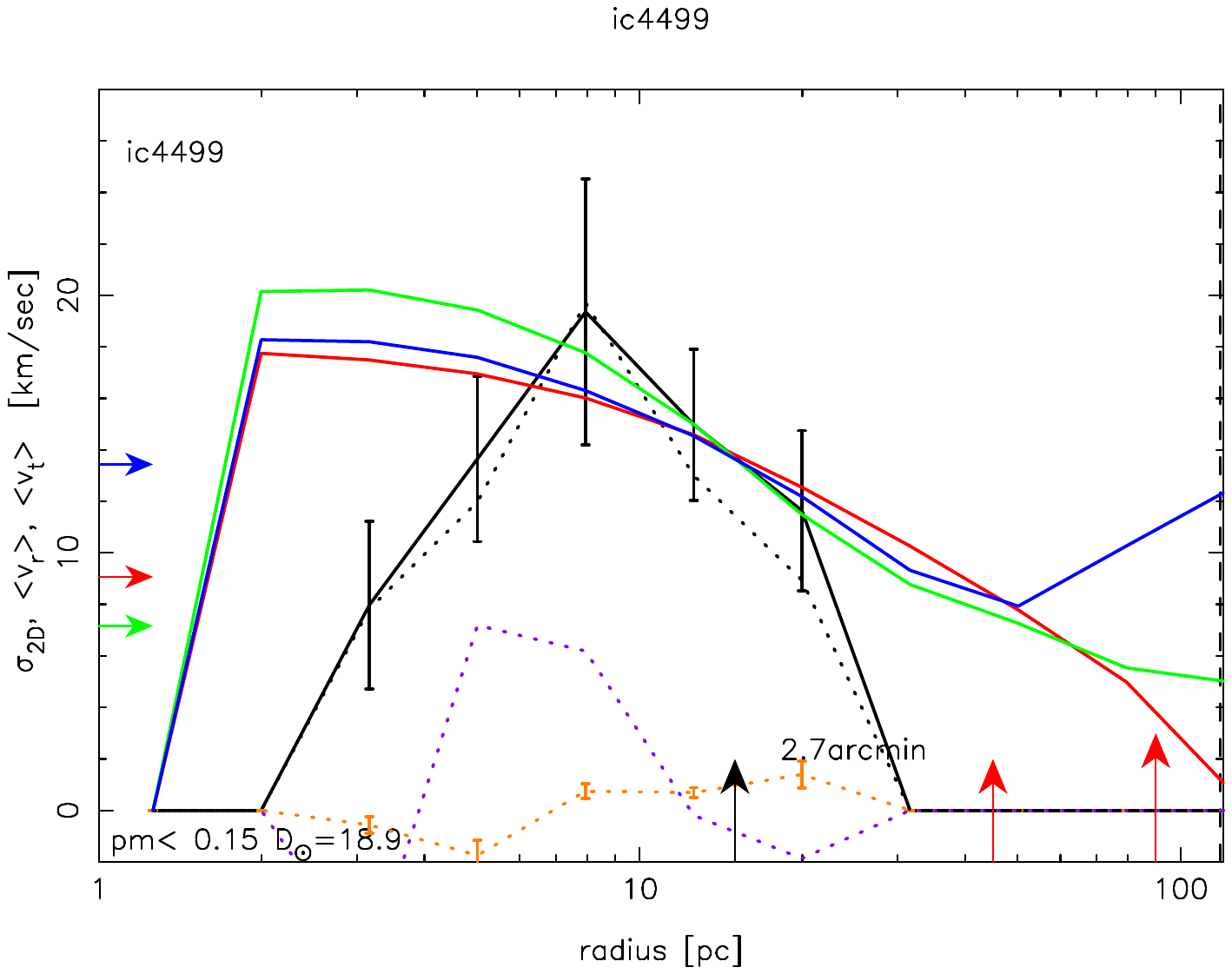}}
\put(120,120){\includegraphics[angle=0,scale=0.12,trim=30 10 100 80, clip=true]{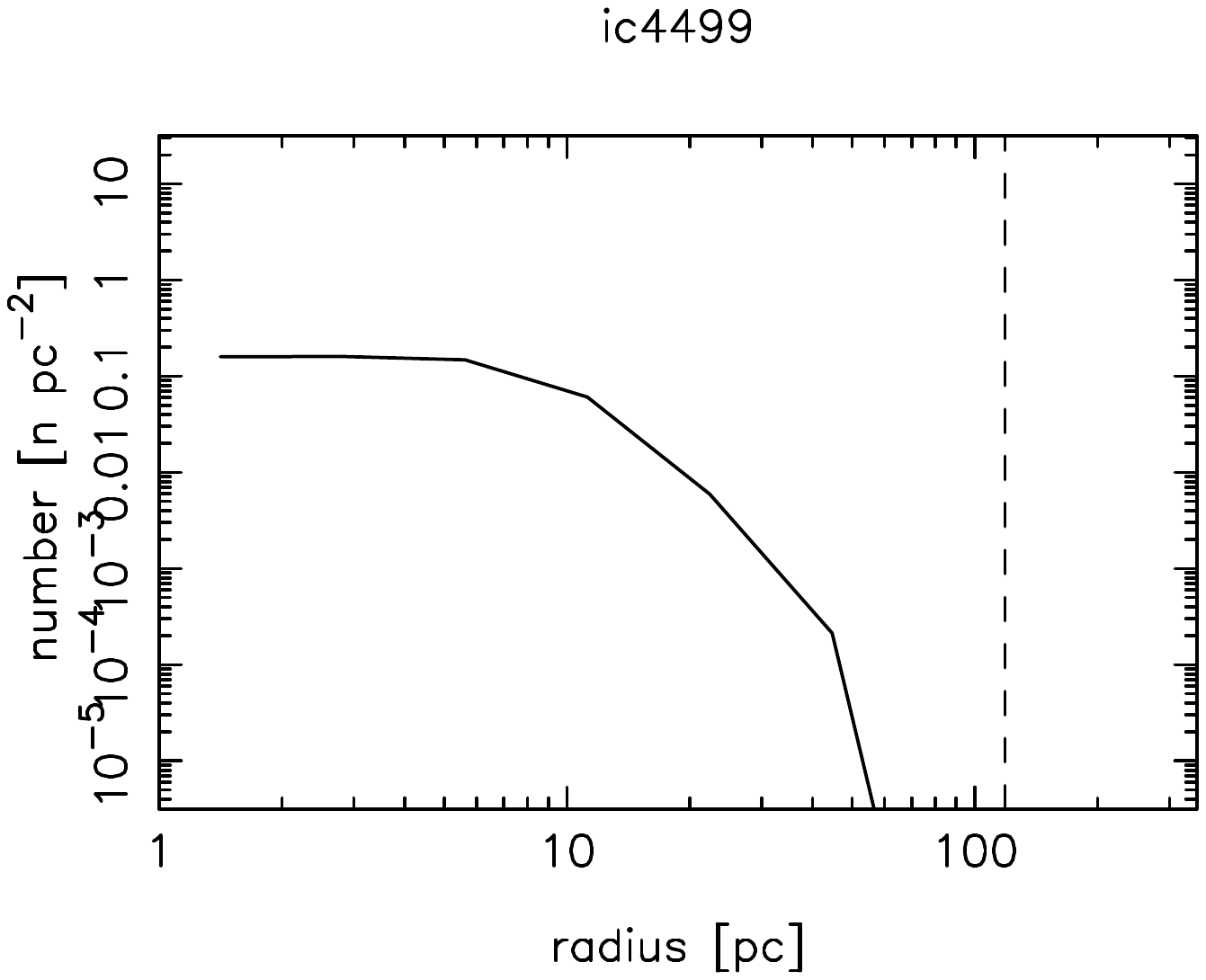}}

{\includegraphics[angle=0,scale=0.36,trim=30 10 100 80, clip=true]{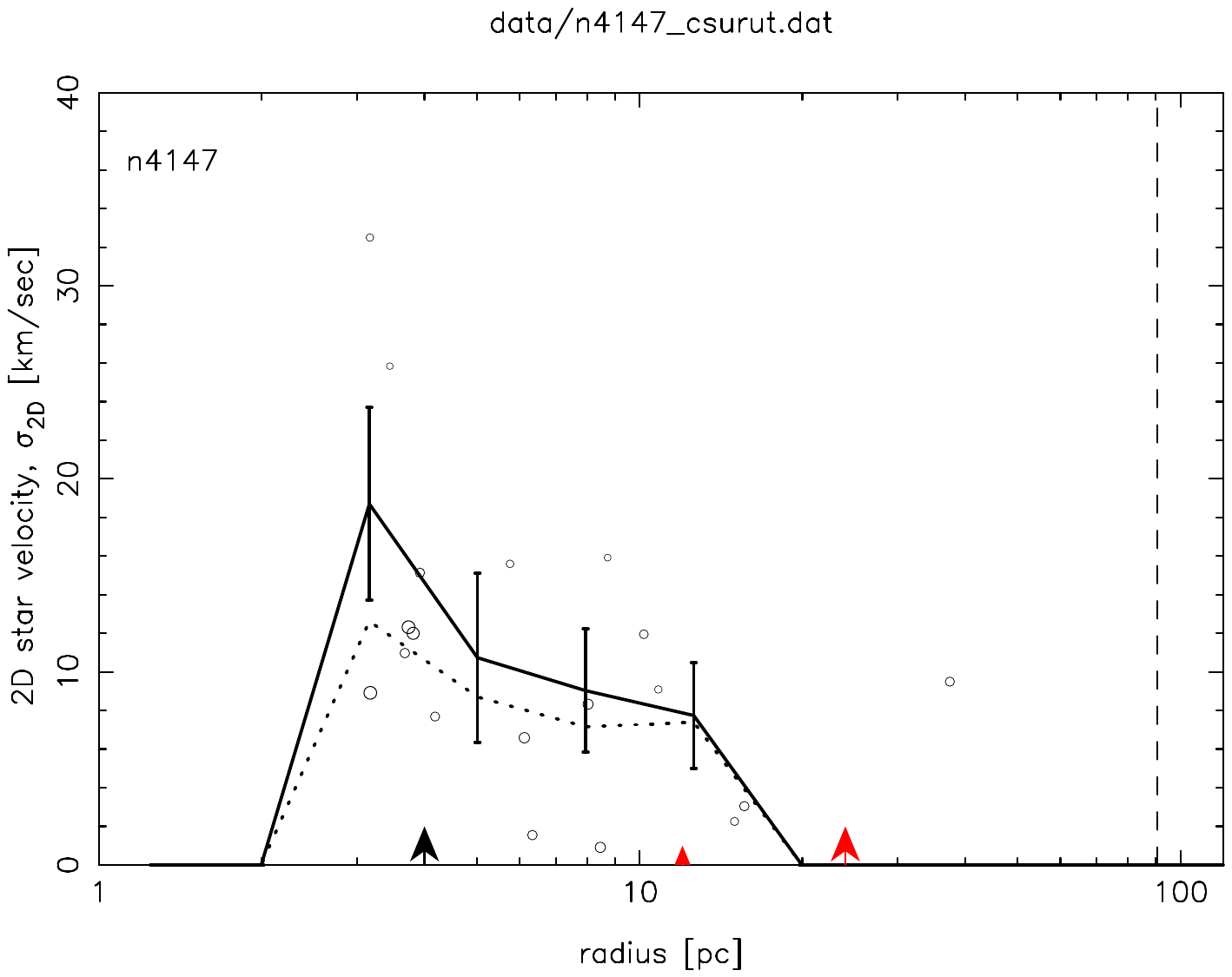}}
\put(5,0){\includegraphics[angle=0,scale=0.36,trim=30 10 100 80, clip=true]{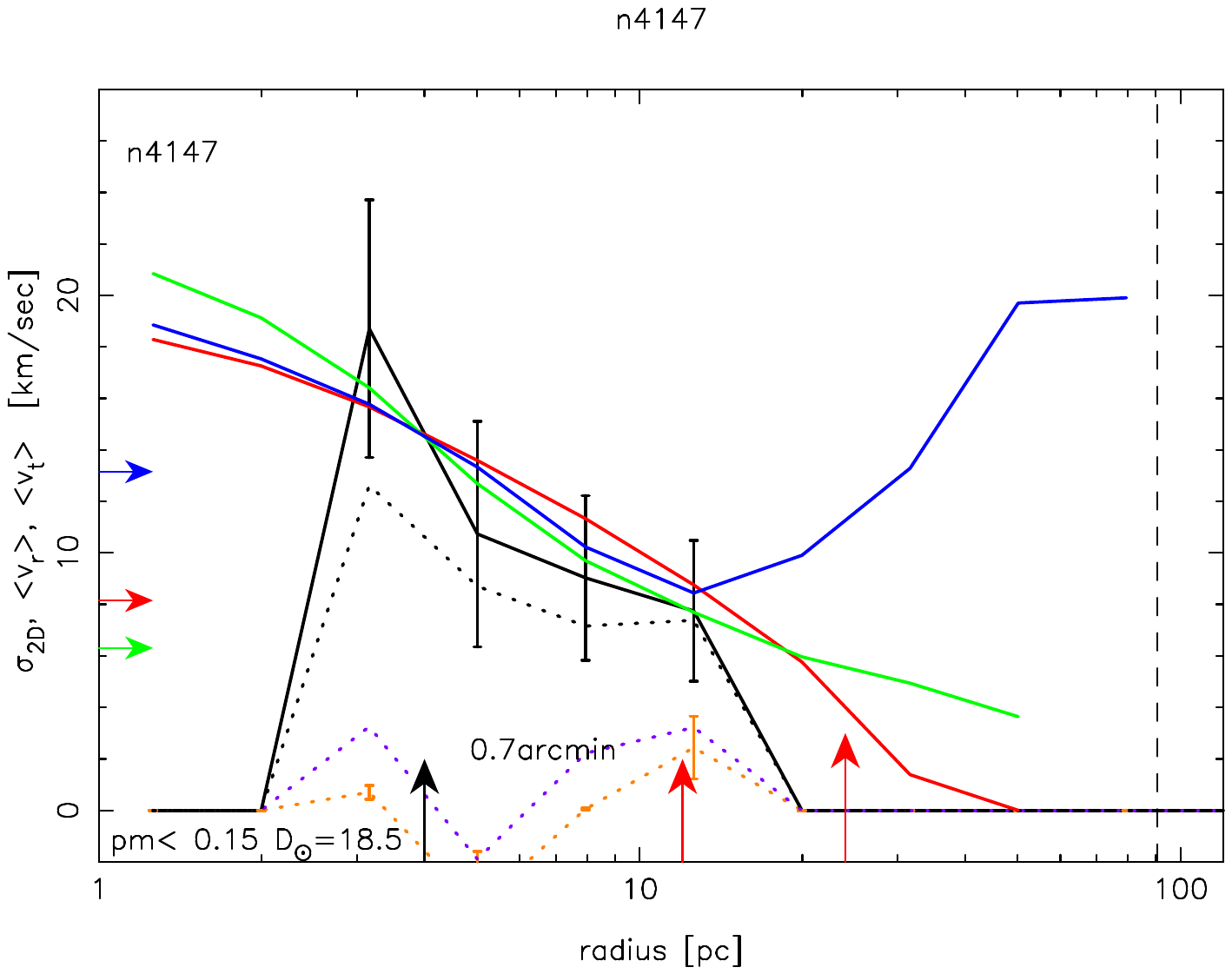}}
\put(120,120){\includegraphics[angle=0,scale=0.12,trim=30 10 100 80, clip=true]{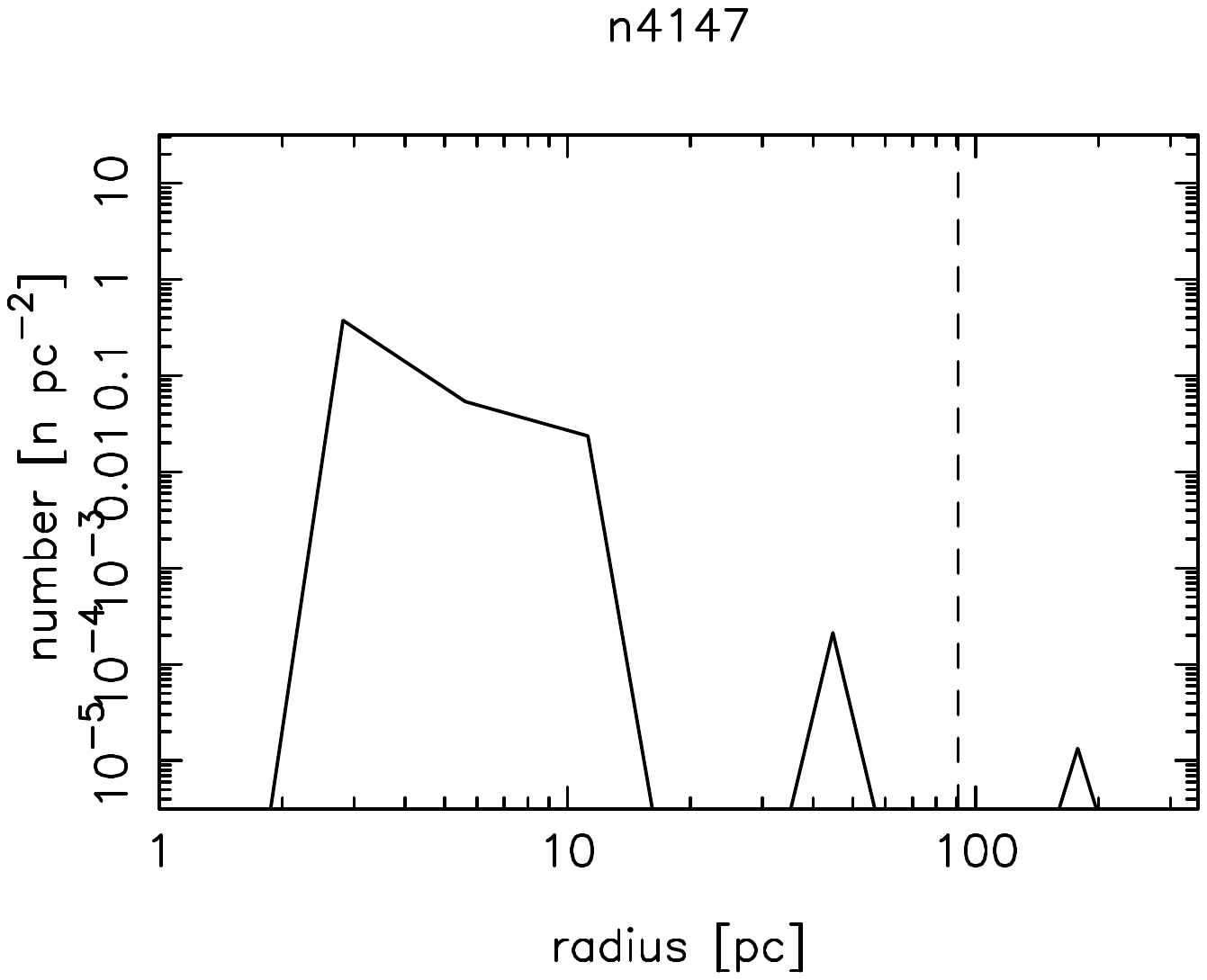}}
\caption{Velocities for the clusters IC 4499 and NGC 4147, top and bottom. Symbols as in Figure~\ref{fig_6752}.
}
\label{fig_4147}
\end{figure*}

\section{Velocity Dispersion Profile Analysis}

The {\it Gaia} proper motions of the selected stars are converted to cluster-centric velocities using the cluster distances and center of mass velocity components of \citet{2019MNRAS.482.5138B}. For simplicity, the two components of the velocity are added together to produce a total velocity in the plane of the cluster. While there is useful information in the velocity anisotropy at large distances, its measurement beyond about 3 half-mass radii is very uncertain with the current data. These clusters have been previously shown to have nearly isotropic planar velocities and relatively little rotation \citep{VB21}. All velocities  are measured with respect to the cluster center and the published velocity of the cluster. Our analysis assumes that all the color-magnitude selected stars are at the same distance.  The mean radial and tangential velocities are 1-2 \kms, generally consistent with zero and small enough to have no dynamical significance. 

Potential issues in estimating the velocity dispersion are the presence of unbound background stars and velocity errors that depend on brightness. Although the clusters do exhibit mass segregation \citep{Ebrahimi20} any variation of velocity dispersion with stellar mass should be small, since the cluster stars bright enough to have high precision proper motions are generally near the main sequence turn-off and above, so have comparable masses even though their luminosities vary substantially. The maximum velocity for inclusion in the cluster has no radial dependence and is chosen in the range of 20 to 50 \kms\ based on the velocity dispersion of the cluster, with the velocity indicated as the maximum of the velocity range of the scatter plots shown in the left-hand panels of the velocity plots. The velocity cuts were set to be at least 4 times the velocity dispersion at about 30 pc. The velocity cuts are at least 2.35 times the half-mass circular velocity, $\sqrt{GM_c/(2r_h)}$, which is above the escape velocity in the outskirts of the cluster. The color-magnitude selection and the velocity cut reduce the number of contaminating field stars in the clusters to very low levels.    

The two clusters with the largest remnant background counts, NGC~6254 and NGC~6752, have 6 background stars between 159 and 200 pc of the center. The background can be assumed to have a uniform surface density distribution of stars, which implies that in the radial range from 40 to 50 pc  there should be 3/8 of a background star. We actually find 6 stars meeting our selection criteria, a factor of 16  above the estimated background. For the bin extending from 32 to 40 pc we would expect 0.24 background stars. The 16 stars meeting our criteria therefore exceed the estimated background level by a factor of 66. Nevertheless, the higher velocity stars within the tidal radius should be prime targets for spectroscopic membership tests. Inset in the velocity dispersion plot is  the surface density profile of the selected stars within the velocity cut and proper motion accuracy cut. The surface density of cluster stars within about half the tidal radius is at least an order of magnitude above the background at large radius. The surface density plot also shows that crowding diminishes the numbers of selected stars at small radius, though the accuracy of the velocity dispersions at small radii is not important for this paper. 

The simplest approach to measuring the velocity dispersion is simply to calculate the variance of the velocities, with and without weighting of the stars. The error in the velocity dispersion $\sigma$ is estimated as $\sigma/\sqrt{2N}$. We tried a double Gaussian fit to the cluster stars and the background, but there are so few background stars predicted within half the tidal radius that the results were essentially identical to the two simple velocity dispersion estimates. The velocity dispersion measurements are displayed in Figures~\ref{fig_6752} to \ref{fig_4147}. The plots show the data with the unweighted velocity dispersion in the left panel and both measures of the velocity dispersion profile along with model curves in the right panel. The last column of Table~\ref{table_clusters} gives our assessment of whether the velocity dispersion is rising, falling, flat, or no stars at a projected distance in the range extending to around 6 times the half-mass radius. Of the 25 clusters there are two with rising velocity dispersion profiles, six with no stars at large radius, eleven with falling dispersion profiles, and six that are approximately flat. 

\begin{figure*}
{\includegraphics[angle=0,scale=0.36,trim=30 10 100 80, clip=true]{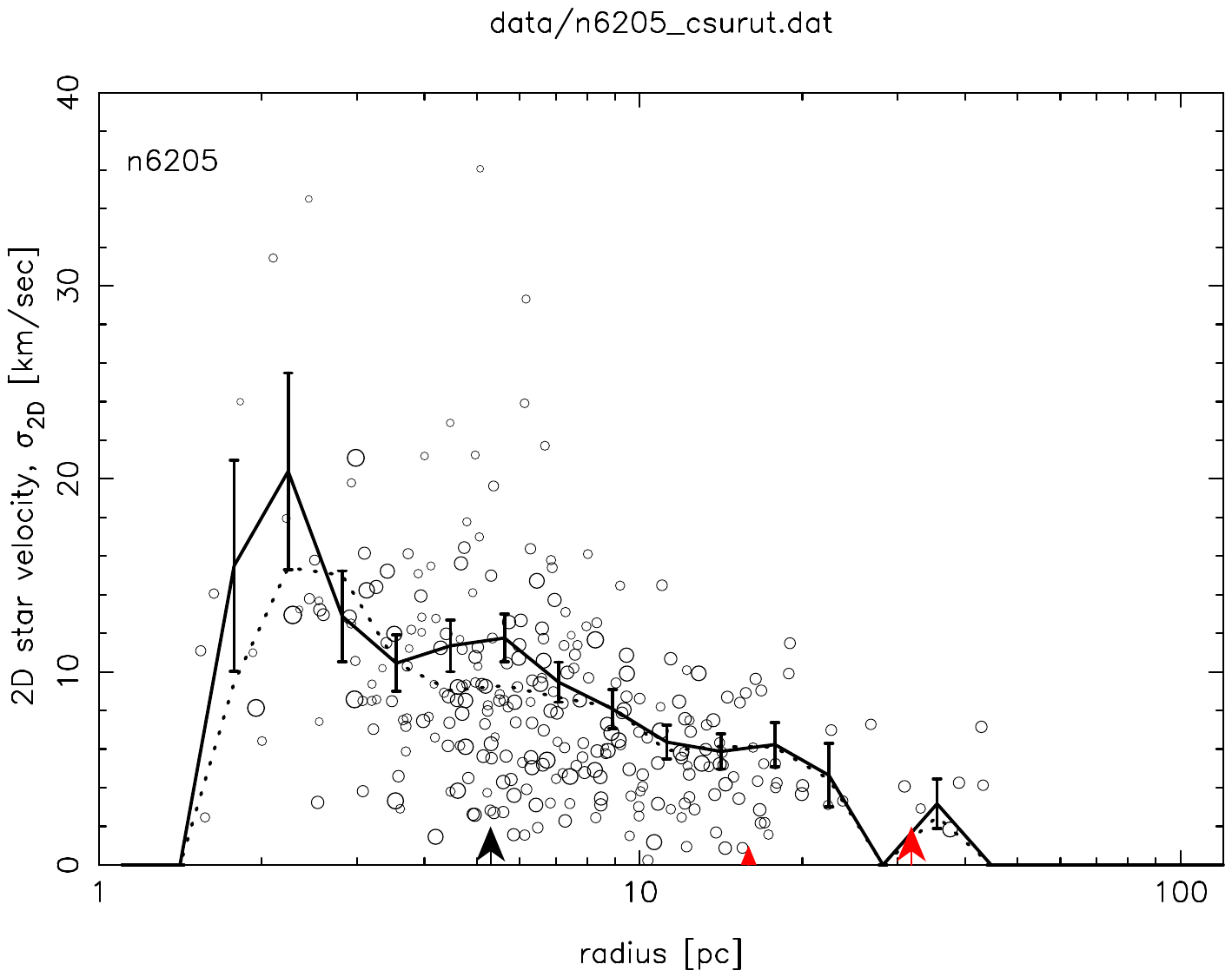}}
\put(5,0){\includegraphics[angle=0,scale=0.36,trim=30 10 100 80, clip=true]{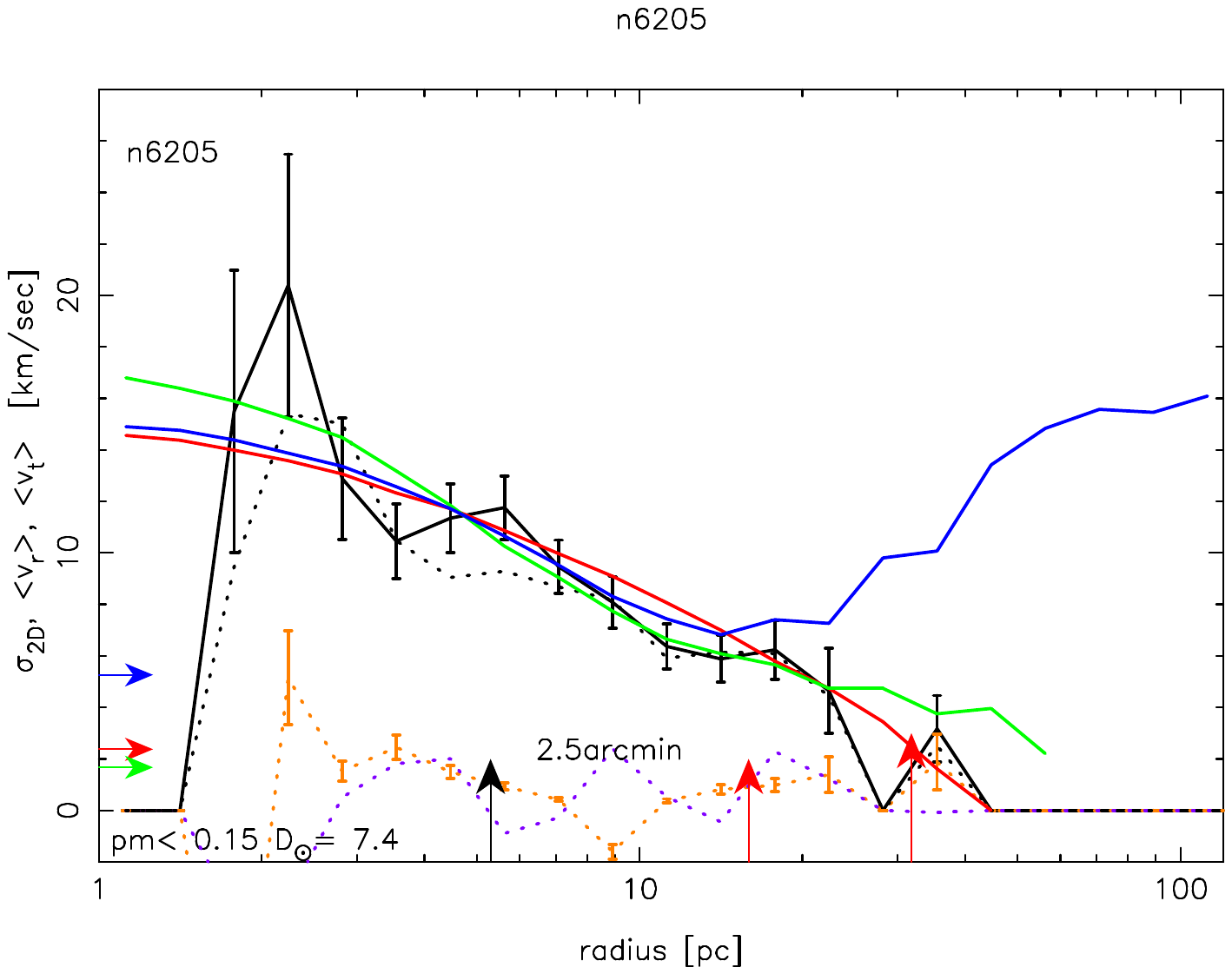}}

{\includegraphics[angle=0,scale=0.36,trim=30 10 100 80, clip=true]{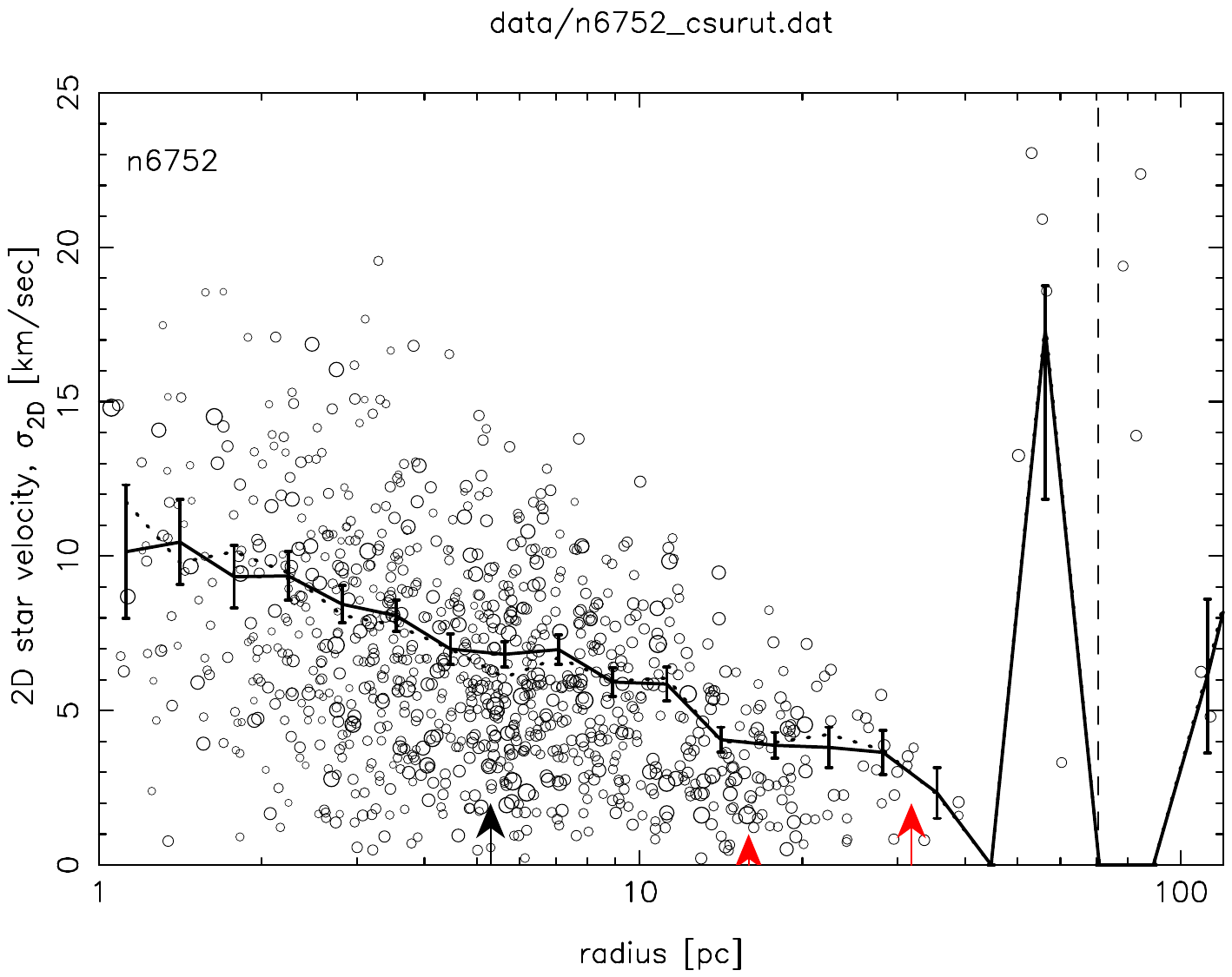}}
\put(5,0){\includegraphics[angle=0,scale=0.36,trim=30 10 100 80, clip=true]{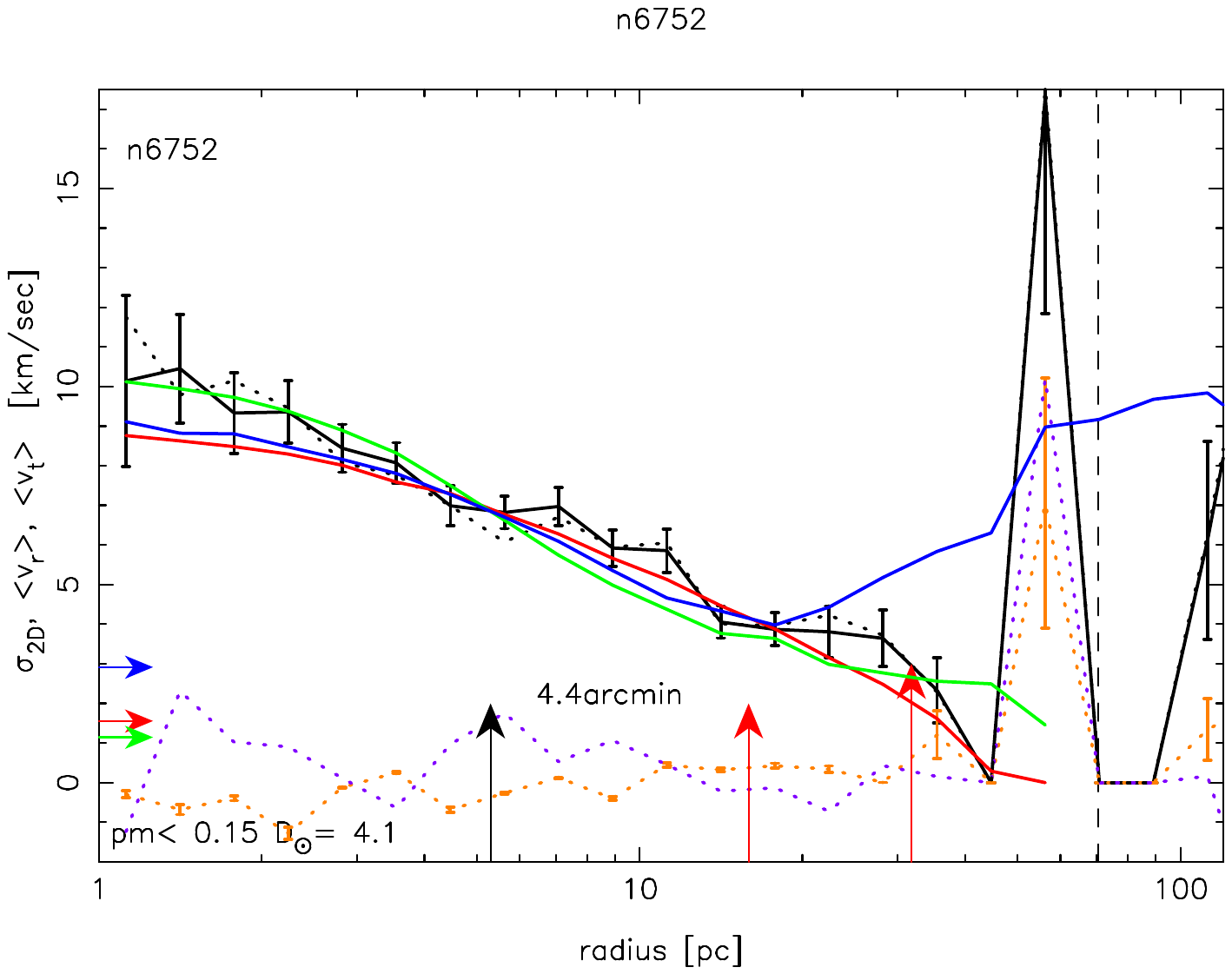}}
\caption{Velocities for stars brighter than $G=17$ mag, or absolute magnitudes 2.65 and 3.94  for the clusters NGC~6205 and 6752, respectively. Symbols as in Figure~\ref{fig_6752}. 
}
\label{fig_bright}
\end{figure*}

For comparison we plot projected velocity dispersion profiles for a W=7 King model \citep{1966AJ.....71...64K}, with velocities calculated from points generated with the McCluster model-maker \citep{2011MNRAS.417.2300K}. The King model gives densities and velocities that go to zero at a finite radius, hence is not expected to be a good representation of the outer density or velocity profile of a cluster that is undergoing slow mass loss. We also plot two models from a simulation of globular clusters in a cosmological model \citep{CK}. The green line shows the relation for a cluster with no dark matter other than the galactic background. The blue line is for a typical cluster at the center of its small dark matter halo. The dark matter mass density is nearly constant with radius over this range so there is essentially no detectable dark matter within the stellar cluster, but it begins to dominate at about 10 half mass radii. Beyond the tidal radius the dark matter is subject to tidal removal. The cluster simulations  were projected on the sky and measured in  the same as the observational data. The simulations find essentially a continuous range of possibilities for the large radius velocity dispersion profile. The model profiles are normalized to the velocity dispersion at the half-mass radius or nearby. There is no fitting of the models to the data at this stage.

The velocity dispersion profiles of NGC 6752 and NGC 6205 exhibit a rise in the velocity dispersion in the range 30-50 pc. In both cases the velocity anisotropy becomes radial at the onset of the rise, agreeing with \citet{2019MNRAS.482.5138B} for the radii in common. There is no clear indication in the current data that the rise in velocity dispersion in the 30-50 pc range is the result of background stars. For instance, for NGC~6205 the rise in velocity dispersion at 30-40 pc is dependent on three stars with velocities between 19 and 31 \kms. However the density of all background stars in this radial range is a factor of 80 lower, making these stars fairly high probability members.  Furthermore, the background has no stars in this range of velocities. A  concern is that the rising velocity dispersion at large radius is not present in stars brighter than $G=17$ mag as shown in Figure~\ref{fig_bright}, however this is entirely because there are very few bright stars at larger radii, likely a result of mass segregation \citep{Ebrahimi20}. That is, the bright stars do not sample the large radius velocity dispersion in these clusters.

\section{Discussion}

Testing for the presence of dark matter around a globular cluster requires pushing kinematic measurements  to as close to the tidal radius as possible where the dark matter density of a cosmological halo may begin to dominate over the stellar density \citep{Ibata13,CK}. Many globular clusters with high quality {\it Gaia} globular cluster proper motion have been modeled with great success inside 3-5 half mass radii, with no need to include a significant distributed dark matter component  \citep{2019MNRAS.482.5138B,VB21}.  However, for a typical cluster with a tidal radius of 100 pc, conclusively ruling out a local dark matter halo requires kinematic measurements at larger radii, where the surface density of cluster stars drops below the that of Galactic field stars. The approach here is to use a color magnitude diagram carefully tuned to each cluster to identify likely cluster stars, which are further culled of background stars using a velocity cut that eliminates stars in the outskirts that are not bound to the cluster. 

The outer velocity dispersion profile  depends on the orbit of the cluster in the galaxy.  Clusters on non-circular orbits are subject to regularly varying tides, which lead to a regular orbital phase dependence of  the velocity dispersion at large radius, whether the cluster is embedded in dark matter halo or not.  That is, as a cluster passes the orbital pericenter, stars are quickly accelerated, with the velocity boost increasing outward in proportion to their distance from the cluster center.  The stars move outwards settling into a new equilibrium after apocenter \citep{BT08}, with the unbound stars leaving in a tidal stream.

\begin{figure}
\begin{center}
{\includegraphics[angle=0,scale=0.35,trim=30 10 50 80,clip=true]{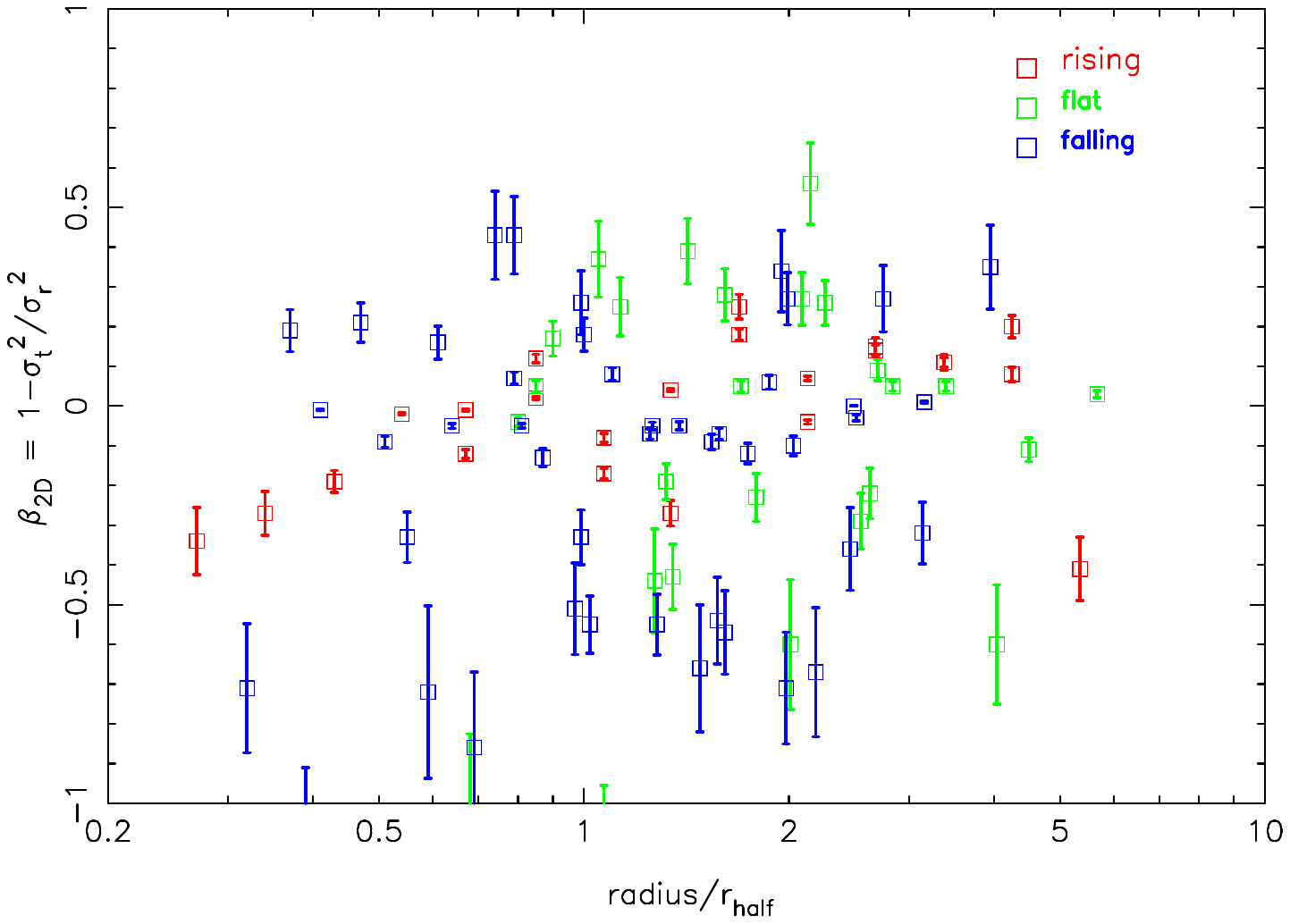}}
\caption{The velocity anisotropy parameter for the different classes of velocity dispersion profile plotted as a function of the radial distance normalized to the half mass radius circular velocity. Only radial bins with at least 20 velocities are plotted.
}
\end{center}
\label{fig_beta}
\end{figure}

The n-body simulations found that the velocity dispersion anisotropy is an indicator of the presence or absence of dark matter \citep{CK}. That is,  the velocity dispersion ellipsoid of cluster without dark matter tends to be quite radial in the outer region \citep{2017MNRAS.471.1181B}, but nearly isotropic if dark matter is present. Figure~14 
displays the 2D version of the velocity anisotropy parameter, $\beta_{2D}= 1-\sigma_t^2/\sigma_r^2$. The clusters identified as rising profile are conspicuously near zero velocity anisotropy, whereas the other clusters have a significantly larger spread in $\beta_{2D}$, with no clear indication of an offset to positive $\beta_{2D}$, (radial anisotropy).  All of the clusters in our sample have been previously studied at smaller radii using a different measure of velocity anisotropy, but also finding a spread around relatively weak anisotropy \citep{2019MNRAS.482.5138B,VB21}.

Clusters that have velocity dispersions in their outskirts that rise above the expectations of a self-gravitating cluster of stars are not new.
\citet{2018MNRAS.478.1520B} show results for 59 clusters with more than 100 stars with kinematics. Of those, 8 have velocity dispersion profiles at large radius that are significantly above the self-gravitating model, and NGC~6715/M54, has a rising velocity dispersion profile from 5 to 8 half-mass radii, a consequence of being the at the center of the Sagittarius dwarf \citep{Ibata97,Ibata13,SgrDwarf20}. 

We selected 25 nearby halo clusters, of which 6 have effectively no stars at large radii, hence with no evidence for or against a local dark matter halo. Of the remaining 19 clusters, 11 have a falling velocity dispersion with cluster radius. These clusters are consistent with having their origin in an environment with no dark matter above the galactic background at a radius of a few kpc in a high redshift dark matter halo. The 6 velocity dispersion profiles that are approximately flat at large radius are ambiguous at present. The \citet{CK} n-body simulations find that both clusters centered in their dark halos as well as significantly offset clusters lead to flat velocity dispersion profiles at large radius.


We find that two clusters, NGC 6205 and 6752, plausibly have a rising velocity dispersion at large radius. The rise is similar to what is seen in n-body simulations of star clusters with local dark matter halos and is tentative evidence of local dark matter halos for these two clusters. The rise in velocity dispersion profiles is subject to the measurement problem  of interloper background stars. For these two clusters the velocity dispersion profile falls for stars brighter than $G=17$ mag, but rises for stars between $G=17$ and 20, which are more radially extended as expected in these mass segregated clusters.

NGC 6205 and 6752 are at Galacto-centric distances, $R_{GC}$, of 8.3 and 5.2 kpc, respectively, with estimated peri-Galacticons of 1.55 and 3.23 kpc, and apo-Galacticons of 8.32 and 5.17 kpc, respectively \citep{2019MNRAS.482.5138B}. Estimating the tidal radius for a cluster of mass $M_c$ as $r_t=R_{GC}[GM_c/(2V_c^2)]^{1/3}$ \citep{2013ApJ...764..124W} where $V_c=220 \kms$ is the assumed inner galaxy circular velocity \citep{2012ApJ...759..131B}, gives tidal radii of 113 and 65 pc for NGC 6205 and 6752, respectively. These tidal radii are well beyond the radii at which the velocity dispersion profiles begin their rise.  Cluster membership criteria  will be strengthened with additional photometric and/or spectroscopic data to confirm that all the stars share the chemical abundance signatures of the cluster stars. Future Gaia data releases will also greatly reduce the kinematic errors. This increased accuracy will also enable more distant clusters to be examined for evidence of local dark halos. 

The cluster sample has a mean radial position relative to their apocenter of 0.78 and a mean orbital eccentricity of 0.64. The orbital distribution of the sample is almost certainly missing a few members closer to pericenter where they are below our Galactic latitude cut. Both NGC 6205 and 6794 have a negative radial velocity with respect to the Galactic center so they are just past their orbital apocenters, in which case we expect there to be minimal or no radial flow within the tidal radius, which is observed., although stars beyond the tidal radius NGC~6752 have an outward flow of as much as 5 \kms. Near pericenter a rise in velocities at large radius is much more likely than just past apocenter, where both of  these clusters are located. Nevertheless, the complications of orbital phase effects and the possibility that background stars contaminate the outer part of the velocity dispersion measurement means it is not yet possible to say whether the rising velocity dispersion profiles are the result of a local dark matter sub-halos or are the response of the outer stars in an eccentric galactic orbit.

A local dark matter sub-halo would deepen the gravitational potential of a cluster and could have some bearing on the multiple stellar populations of globular clusters \citep{2012A&ARv..20...50G}. Detailed abundance studies do show some differences from field stars in NGC 6752 \citep{1981ApJ...244..205N} but no clear differences for NGC 6205 \citep{2021A&A...646A...9C}. NGC 6752 stands out in a comparison of widths of the asymptotic giant branch and the red giant branch, but NGC 6205 does not \citep{2021ApJ...910....6L}. We conclude that there is no clear correlation of cluster metallicity trends with velocity dispersion profile within our small sample of clusters. In summary, the velocity dispersion measurements presented here provide consistent, but not compelling, evidence for the presence of remnant dark matter within the tidal radii of 2 of the 19 clusters, with a further 6 clusters with flat profiles being ambiguous. 

\section{Conclusions}

Determining which stars are globular cluster members is the central issue of measuring their velocity dispersion at large radii. The approach here is to use a color-magnitude relation tuned to the high precision {\it Gaia} photometry, which has the advantage that the primary membership criterion is completely independent of cluster kinematics. The method has been proven to work well in uncovering faint stellar streams \citep{2009ApJ...693.1118G}, but is even more powerful for globular clusters with known distances and large numbers of stars in the central region which can be used to accurately calibrate the color-magnitude diagram.  The color-magnitude cut is augmented with a velocity cut that is at least the 2.35 times the scale velocity of the cluster, the circular velocity at the half mass radius, $\sqrt{GM_c/(2r_h)}$.  The color-magnitude cut reduces the numbers of stars at least a factor of 2 and more typically a factor of 5 in the radial range of interest, as shown in Figure~2. 
Of the 25 clusters in our sample, 19 have sufficient numbers and accuracy of proper motion velocities at large radii to give a velocity dispersion profile to  5 half mass radii or more. Of the 19, 11 have velocity dispersion profiles that are falling between 3 and 6 half mass radii, with a further 6 with larger errors that are flat or falling. Therefore, one result of our analysis is that 11/19 (58\%) of the sample clusters show no evidence for significant local dark matter based on their falling velocity dispersion profiles at large radius. The 2 clusters with rising velocity dispersion profiles are consistent with local dark matter halos, but are not conclusive evidence. Those two, along with the 6 flat dispersion profile clusters are interesting targets for further observations and modeling.

\begin{acknowledgements}
This research was supported by NSERC of Canada. 
\end{acknowledgements}

\bibliography{GCedge}{}

\begin{thebibliography}{}
\expandafter\ifx\csname natexlab\endcsname\relax\def\natexlab#1{#1}\fi
\providecommand{\url}[1]{\href{#1}{#1}}
\providecommand{\dodoi}[1]{doi:~\href{http://doi.org/#1}{\nolinkurl{#1}}}
\providecommand{\doeprint}[1]{\href{http://ascl.net/#1}{\nolinkurl{http://ascl.net/#1}}}
\providecommand{\doarXiv}[1]{\href{https://arxiv.org/abs/#1}{\nolinkurl{https://arxiv.org/abs/#1}}}

\bibitem[{{Alfaro-Cuello} {et~al.}(2020){Alfaro-Cuello}, {Kacharov},
  {Neumayer}, {Bianchini}, {Mastrobuono-Battisti}, {L{\"u}tzgendorf}, {Seth},
  {B{\"o}ker}, {Kamann}, {Leaman}, {Watkins}, \& {van de Ven}}]{M54MUSE}
{Alfaro-Cuello}, M., {Kacharov}, N., {Neumayer}, N., {et~al.} 2020, \apj, 892,
  20, \dodoi{10.3847/1538-4357/ab77bb}

\bibitem[{{Baumgardt}(2017)}]{2017MNRAS.464.2174B}
{Baumgardt}, H. 2017, \mnras, 464, 2174, \dodoi{10.1093/mnras/stw2488}

\bibitem[{{Baumgardt} \& {Hilker}(2018)}]{2018MNRAS.478.1520B}
{Baumgardt}, H., \& {Hilker}, M. 2018, \mnras, 478, 1520,
  \dodoi{10.1093/mnras/sty1057}

\bibitem[{{Baumgardt} {et~al.}(2019){Baumgardt}, {Hilker}, {Sollima}, \&
  {Bellini}}]{2019MNRAS.482.5138B}
{Baumgardt}, H., {Hilker}, M., {Sollima}, A., \& {Bellini}, A. 2019, \mnras,
  482, 5138, \dodoi{10.1093/mnras/sty2997}

\bibitem[{{Bellazzini} {et~al.}(2008){Bellazzini}, {Ibata}, {Chapman},
  {Mackey}, {Monaco}, {Irwin}, {Martin}, {Lewis}, \&
  {Dalessandro}}]{Bellazzini08}
{Bellazzini}, M., {Ibata}, R.~A., {Chapman}, S.~C., {et~al.} 2008, \aj, 136,
  1147, \dodoi{10.1088/0004-6256/136/3/1147}

\bibitem[{{Bianchini} {et~al.}(2017){Bianchini}, {Sills}, \&
  {Miholics}}]{2017MNRAS.471.1181B}
{Bianchini}, P., {Sills}, A., \& {Miholics}, M. 2017, \mnras, 471, 1181,
  \dodoi{10.1093/mnras/stx1680}

\bibitem[{{Binney} \& {Tremaine}(2008)}]{BT08}
{Binney}, J., \& {Tremaine}, S. 2008, {Galactic Dynamics: Second Edition}
  (Princeton University Press)

\bibitem[{{Boldrini} {et~al.}(2020){Boldrini}, {Mohayaee}, \&
  {Silk}}]{2020MNRAS.492.3169B}
{Boldrini}, P., {Mohayaee}, R., \& {Silk}, J. 2020, \mnras, 492, 3169,
  \dodoi{10.1093/mnras/staa011}

\bibitem[{{Bovy} {et~al.}(2012){Bovy}, {Allende Prieto}, {Beers}, {Bizyaev},
  {da Costa}, {Cunha}, {Ebelke}, {Eisenstein}, {Frinchaboy}, {Garc{\'\i}a
  P{\'e}rez}, {Girardi}, {Hearty}, {Hogg}, {Holtzman}, {Maia}, {Majewski},
  {Malanushenko}, {Malanushenko}, {M{\'e}sz{\'a}ros}, {Nidever}, {O'Connell},
  {O'Donnell}, {Oravetz}, {Pan}, {Rocha-Pinto}, {Schiavon}, {Schneider},
  {Schultheis}, {Skrutskie}, {Smith}, {Weinberg}, {Wilson}, \&
  {Zasowski}}]{2012ApJ...759..131B}
{Bovy}, J., {Allende Prieto}, C., {Beers}, T.~C., {et~al.} 2012, \apj, 759,
  131, \dodoi{10.1088/0004-637X/759/2/131}

\bibitem[{{Carlberg} \& {Keating}(2021)}]{CK}
{Carlberg}, R.~G., \& {Keating}, L.~C. 2021, arXiv e-prints, arXiv:2105.13900.
\newblock \doarXiv{2105.13900}

\bibitem[{{Carretta} \& {Bragaglia}(2021)}]{2021A&A...646A...9C}
{Carretta}, E., \& {Bragaglia}, A. 2021, \aap, 646, A9,
  \dodoi{10.1051/0004-6361/202039392}

\bibitem[{{Daniel} {et~al.}(2017){Daniel}, {Heggie}, \&
  {Varri}}]{2017MNRAS.468.1453D}
{Daniel}, K.~J., {Heggie}, D.~C., \& {Varri}, A.~L. 2017, \mnras, 468, 1453,
  \dodoi{10.1093/mnras/stx571}

\bibitem[{{Ebrahimi} {et~al.}(2020){Ebrahimi}, {Sollima}, {Haghi}, {Baumgardt},
  \& {Hilker}}]{Ebrahimi20}
{Ebrahimi}, H., {Sollima}, A., {Haghi}, H., {Baumgardt}, H., \& {Hilker}, M.
  2020, \mnras, 494, 4226, \dodoi{10.1093/mnras/staa969}

\bibitem[{{Fukushige} \& {Heggie}(2000)}]{2000MNRAS.318..753F}
{Fukushige}, T., \& {Heggie}, D.~C. 2000, \mnras, 318, 753,
  \dodoi{10.1046/j.1365-8711.2000.03811.x}

\bibitem[{{Gaia Collaboration} {et~al.}(2016){Gaia Collaboration}, {Prusti},
  {de Bruijne}, {Brown}, {Vallenari}, {Babusiaux}, {Bailer-Jones}, {Bastian},
  {Biermann}, {Evans}, {Eyer}, {Jansen}, {Jordi}, {Klioner}, {Lammers},
  {Lindegren}, {Luri}, {Mignard}, {Milligan}, {Panem}, {Poinsignon},
  {Pourbaix}, {Randich}, {Sarri}, {Sartoretti}, {Siddiqui}, {Soubiran},
  {Valette}, {van Leeuwen}, {Walton}, {Aerts}, {Arenou}, {Cropper}, {Drimmel},
  {H{\o}g}, {Katz}, {Lattanzi}, {O'Mullane}, {Grebel}, {Holland}, {Huc},
  {Passot}, {Bramante}, {Cacciari}, {Casta{\~n}eda}, {Chaoul}, {Cheek}, {De
  Angeli}, {Fabricius}, {Guerra}, {Hern{\'a}ndez}, {Jean-Antoine-Piccolo},
  {Masana}, {Messineo}, {Mowlavi}, {Nienartowicz}, {Ord{\'o}{\~n}ez-Blanco},
  {Panuzzo}, {Portell}, {Richards}, {Riello}, {Seabroke}, {Tanga},
  {Th{\'e}venin}, {Torra}, {Els}, {Gracia-Abril}, {Comoretto},
  {Garcia-Reinaldos}, {Lock}, {Mercier}, {Altmann}, {Andrae}, {Astraatmadja},
  {Bellas-Velidis}, {Benson}, {Berthier}, {Blomme}, {Busso}, {Carry},
  {Cellino}, {Clementini}, {Cowell}, {Creevey}, {Cuypers}, {Davidson}, {De
  Ridder}, {de Torres}, {Delchambre}, {Dell'Oro}, {Ducourant}, {Fr{\'e}mat},
  {Garc{\'\i}a-Torres}, {Gosset}, {Halbwachs}, {Hambly}, {Harrison}, {Hauser},
  {Hestroffer}, {Hodgkin}, {Huckle}, {Hutton}, {Jasniewicz}, {Jordan},
  {Kontizas}, {Korn}, {Lanzafame}, {Manteiga}, {Moitinho}, {Muinonen},
  {Osinde}, {Pancino}, {Pauwels}, {Petit}, {Recio-Blanco}, {Robin}, {Sarro},
  {Siopis}, {Smith}, {Smith}, {Sozzetti}, {Thuillot}, {van Reeven}, {Viala},
  {Abbas}, {Abreu Aramburu}, {Accart}, {Aguado}, {Allan}, {Allasia},
  {Altavilla}, {{\'A}lvarez}, {Alves}, {Anderson}, {Andrei}, {Anglada Varela},
  {Antiche}, {Antoja}, {Ant{\'o}n}, {Arcay}, {Atzei}, {Ayache}, {Bach},
  {Baker}, {Balaguer-N{\'u}{\~n}ez}, {Barache}, {Barata}, {Barbier}, {Barblan},
  {Baroni}, {Barrado y Navascu{\'e}s}, {Barros}, {Barstow}, {Becciani},
  {Bellazzini}, {Bellei}, {Bello Garc{\'\i}a}, {Belokurov}, {Bendjoya},
  {Berihuete}, {Bianchi}, {Bienaym{\'e}}, {Billebaud}, {Blagorodnova},
  {Blanco-Cuaresma}, {Boch}, {Bombrun}, {Borrachero}, {Bouquillon}, {Bourda},
  {Bouy}, {Bragaglia}, {Breddels}, {Brouillet}, {Br{\"u}semeister},
  {Bucciarelli}, {Budnik}, {Burgess}, {Burgon}, {Burlacu}, {Busonero}, {Buzzi},
  {Caffau}, {Cambras}, {Campbell}, {Cancelliere}, {Cantat-Gaudin}, {Carlucci},
  {Carrasco}, {Castellani}, {Charlot}, {Charnas}, {Charvet}, {Chassat},
  {Chiavassa}, {Clotet}, {Cocozza}, {Collins}, {Collins}, {Costigan}, {Crifo},
  {Cross}, {Crosta}, {Crowley}, {Dafonte}, {Damerdji}, {Dapergolas}, {David},
  {David}, {De Cat}, {de Felice}, {de Laverny}, {De Luise}, {De March}, {de
  Martino}, {de Souza}, {Debosscher}, {del Pozo}, {Delbo}, {Delgado},
  {Delgado}, {di Marco}, {Di Matteo}, {Diakite}, {Distefano}, {Dolding}, {Dos
  Anjos}, {Drazinos}, {Dur{\'a}n}, {Dzigan}, {Ecale}, {Edvardsson}, {Enke},
  {Erdmann}, {Escolar}, {Espina}, {Evans}, {Eynard Bontemps}, {Fabre},
  {Fabrizio}, {Faigler}, {Falc{\~a}o}, {Farr{\`a}s Casas}, {Faye}, {Federici},
  {Fedorets}, {Fern{\'a}ndez-Hern{\'a}ndez}, {Fernique}, {Fienga}, {Figueras},
  {Filippi}, {Findeisen}, {Fonti}, {Fouesneau}, {Fraile}, {Fraser}, {Fuchs},
  {Furnell}, {Gai}, {Galleti}, {Galluccio}, {Garabato}, {Garc{\'\i}a-Sedano},
  {Gar{\'e}}, {Garofalo}, {Garralda}, {Gavras}, {Gerssen}, {Geyer}, {Gilmore},
  {Girona}, {Giuffrida}, {Gomes}, {Gonz{\'a}lez-Marcos},
  {Gonz{\'a}lez-N{\'u}{\~n}ez}, {Gonz{\'a}lez-Vidal}, {Granvik}, {Guerrier},
  {Guillout}, {Guiraud}, {G{\'u}rpide}, {Guti{\'e}rrez-S{\'a}nchez}, {Guy},
  {Haigron}, {Hatzidimitriou}, {Haywood}, {Heiter}, {Helmi}, {Hobbs},
  {Hofmann}, {Holl}, {Holland}, {Hunt}, {Hypki}, {Icardi}, {Irwin}, {Jevardat
  de Fombelle}, {Jofr{\'e}}, {Jonker}, {Jorissen}, {Julbe}, {Karampelas},
  {Kochoska}, {Kohley}, {Kolenberg}, {Kontizas}, {Koposov}, {Kordopatis},
  {Koubsky}, {Kowalczyk}, {Krone-Martins}, {Kudryashova}, {Kull}, {Bachchan},
  {Lacoste-Seris}, {Lanza}, {Lavigne}, {Le Poncin-Lafitte}, {Lebreton},
  {Lebzelter}, {Leccia}, {Leclerc}, {Lecoeur-Taibi}, {Lemaitre}, {Lenhardt},
  {Leroux}, {Liao}, {Licata}, {Lindstr{\o}m}, {Lister}, {Livanou}, {Lobel},
  {L{\"o}ffler}, {L{\'o}pez}, {Lopez-Lozano}, {Lorenz}, {Loureiro},
  {MacDonald}, {Magalh{\~a}es Fernandes}, {Managau}, {Mann}, {Mantelet},
  {Marchal}, {Marchant}, {Marconi}, {Marie}, {Marinoni}, {Marrese},
  {Marschalk{\'o}}, {Marshall}, {Mart{\'\i}n-Fleitas}, {Martino}, {Mary},
  {Matijevi{\v{c}}}, {Mazeh}, {McMillan}, {Messina}, {Mestre}, {Michalik},
  {Millar}, {Miranda}, {Molina}, {Molinaro}, {Molinaro}, {Moln{\'a}r},
  {Moniez}, {Montegriffo}, {Monteiro}, {Mor}, {Mora}, {Morbidelli}, {Morel},
  {Morgenthaler}, {Morley}, {Morris}, {Mulone}, {Muraveva}, {Musella},
  {Narbonne}, {Nelemans}, {Nicastro}, {Noval}, {Ord{\'e}novic},
  {Ordieres-Mer{\'e}}, {Osborne}, {Pagani}, {Pagano}, {Pailler}, {Palacin},
  {Palaversa}, {Parsons}, {Paulsen}, {Pecoraro}, {Pedrosa}, {Pentik{\"a}inen},
  {Pereira}, {Pichon}, {Piersimoni}, {Pineau}, {Plachy}, {Plum}, {Poujoulet},
  {Pr{\v{s}}a}, {Pulone}, {Ragaini}, {Rago}, {Rambaux}, {Ramos-Lerate},
  {Ranalli}, {Rauw}, {Read}, {Regibo}, {Renk}, {Reyl{\'e}}, {Ribeiro},
  {Rimoldini}, {Ripepi}, {Riva}, {Rixon}, {Roelens}, {Romero-G{\'o}mez},
  {Rowell}, {Royer}, {Rudolph}, {Ruiz-Dern}, {Sadowski}, {Sagrist{\`a}
  Sell{\'e}s}, {Sahlmann}, {Salgado}, {Salguero}, {Sarasso}, {Savietto},
  {Schnorhk}, {Schultheis}, {Sciacca}, {Segol}, {Segovia}, {Segransan},
  {Serpell}, {Shih}, {Smareglia}, {Smart}, {Smith}, {Solano}, {Solitro},
  {Sordo}, {Soria Nieto}, {Souchay}, {Spagna}, {Spoto}, {Stampa}, {Steele},
  {Steidelm{\"u}ller}, {Stephenson}, {Stoev}, {Suess}, {S{\"u}veges}, {Surdej},
  {Szabados}, {Szegedi-Elek}, {Tapiador}, {Taris}, {Tauran}, {Taylor},
  {Teixeira}, {Terrett}, {Tingley}, {Trager}, {Turon}, {Ulla}, {Utrilla},
  {Valentini}, {van Elteren}, {Van Hemelryck}, {van Leeuwen}, {Varadi},
  {Vecchiato}, {Veljanoski}, {Via}, {Vicente}, {Vogt}, {Voss}, {Votruba},
  {Voutsinas}, {Walmsley}, {Weiler}, {Weingrill}, {Werner}, {Wevers},
  {Whitehead}, {Wyrzykowski}, {Yoldas}, {{\v{Z}}erjal}, {Zucker}, {Zurbach},
  {Zwitter}, {Alecu}, {Allen}, {Allende Prieto}, {Amorim},
  {Anglada-Escud{\'e}}, {Arsenijevic}, {Azaz}, {Balm}, {Beck}, {Bernstein},
  {Bigot}, {Bijaoui}, {Blasco}, {Bonfigli}, {Bono}, {Boudreault}, {Bressan},
  {Brown}, {Brunet}, {Bunclark}, {Buonanno}, {Butkevich}, {Carret}, {Carrion},
  {Chemin}, {Ch{\'e}reau}, {Corcione}, {Darmigny}, {de Boer}, {de Teodoro}, {de
  Zeeuw}, {Delle Luche}, {Domingues}, {Dubath}, {Fodor}, {Fr{\'e}zouls},
  {Fries}, {Fustes}, {Fyfe}, {Gallardo}, {Gallegos}, {Gardiol}, {Gebran},
  {Gomboc}, {G{\'o}mez}, {Grux}, {Gueguen}, {Heyrovsky}, {Hoar}, {Iannicola},
  {Isasi Parache}, {Janotto}, {Joliet}, {Jonckheere}, {Keil}, {Kim},
  {Klagyivik}, {Klar}, {Knude}, {Kochukhov}, {Kolka}, {Kos}, {Kutka}, {Lainey},
  {LeBouquin}, {Liu}, {Loreggia}, {Makarov}, {Marseille}, {Martayan},
  {Martinez-Rubi}, {Massart}, {Meynadier}, {Mignot}, {Munari}, {Nguyen},
  {Nordlander}, {Ocvirk}, {O'Flaherty}, {Olias Sanz}, {Ortiz}, {Osorio},
  {Oszkiewicz}, {Ouzounis}, {Palmer}, {Park}, {Pasquato}, {Peltzer}, {Peralta},
  {P{\'e}turaud}, {Pieniluoma}, {Pigozzi}, {Poels}, {Prat}, {Prod'homme},
  {Raison}, {Rebordao}, {Risquez}, {Rocca-Volmerange}, {Rosen}, {Ruiz-Fuertes},
  {Russo}, {Sembay}, {Serraller Vizcaino}, {Short}, {Siebert}, {Silva},
  {Sinachopoulos}, {Slezak}, {Soffel}, {Sosnowska}, {Strai{\v{z}}ys}, {ter
  Linden}, {Terrell}, {Theil}, {Tiede}, {Troisi}, {Tsalmantza}, {Tur},
  {Vaccari}, {Vachier}, {Valles}, {Van Hamme}, {Veltz}, {Virtanen}, {Wallut},
  {Wichmann}, {Wilkinson}, {Ziaeepour}, \& {Zschocke}}]{2016A&A...595A...1G}
{Gaia Collaboration}, {Prusti}, T., {de Bruijne}, J.~H.~J., {et~al.} 2016,
  \aap, 595, A1, \dodoi{10.1051/0004-6361/201629272}

\bibitem[{{Gaia Collaboration} {et~al.}(2021){Gaia Collaboration}, {Brown},
  {Vallenari}, {Prusti}, {de Bruijne}, {Babusiaux}, {Biermann}, {Creevey},
  {Evans}, {Eyer}, {Hutton}, {Jansen}, {Jordi}, {Klioner}, {Lammers},
  {Lindegren}, {Luri}, {Mignard}, {Panem}, {Pourbaix}, {Randich}, {Sartoretti},
  {Soubiran}, {Walton}, {Arenou}, {Bailer-Jones}, {Bastian}, {Cropper},
  {Drimmel}, {Katz}, {Lattanzi}, {van Leeuwen}, {Bakker}, {Cacciari},
  {Casta{\~n}eda}, {De Angeli}, {Ducourant}, {Fabricius}, {Fouesneau},
  {Fr{\'e}mat}, {Guerra}, {Guerrier}, {Guiraud}, {Jean-Antoine Piccolo},
  {Masana}, {Messineo}, {Mowlavi}, {Nicolas}, {Nienartowicz}, {Pailler},
  {Panuzzo}, {Riclet}, {Roux}, {Seabroke}, {Sordo}, {Tanga}, {Th{\'e}venin},
  {Gracia-Abril}, {Portell}, {Teyssier}, {Altmann}, {Andrae}, {Bellas-Velidis},
  {Benson}, {Berthier}, {Blomme}, {Brugaletta}, {Burgess}, {Busso}, {Carry},
  {Cellino}, {Cheek}, {Clementini}, {Damerdji}, {Davidson}, {Delchambre},
  {Dell'Oro}, {Fern{\'a}ndez-Hern{\'a}ndez}, {Galluccio}, {Garc{\'\i}a-Lario},
  {Garcia-Reinaldos}, {Gonz{\'a}lez-N{\'u}{\~n}ez}, {Gosset}, {Haigron},
  {Halbwachs}, {Hambly}, {Harrison}, {Hatzidimitriou}, {Heiter},
  {Hern{\'a}ndez}, {Hestroffer}, {Hodgkin}, {Holl}, {Jan{\ss}en}, {Jevardat de
  Fombelle}, {Jordan}, {Krone-Martins}, {Lanzafame}, {L{\"o}ffler}, {Lorca},
  {Manteiga}, {Marchal}, {Marrese}, {Moitinho}, {Mora}, {Muinonen}, {Osborne},
  {Pancino}, {Pauwels}, {Petit}, {Recio-Blanco}, {Richards}, {Riello},
  {Rimoldini}, {Robin}, {Roegiers}, {Rybizki}, {Sarro}, {Siopis}, {Smith},
  {Sozzetti}, {Ulla}, {Utrilla}, {van Leeuwen}, {van Reeven}, {Abbas}, {Abreu
  Aramburu}, {Accart}, {Aerts}, {Aguado}, {Ajaj}, {Altavilla}, {{\'A}lvarez},
  {{\'A}lvarez Cid-Fuentes}, {Alves}, {Anderson}, {Anglada Varela}, {Antoja},
  {Audard}, {Baines}, {Baker}, {Balaguer-N{\'u}{\~n}ez}, {Balbinot}, {Balog},
  {Barache}, {Barbato}, {Barros}, {Barstow}, {Bartolom{\'e}}, {Bassilana},
  {Bauchet}, {Baudesson-Stella}, {Becciani}, {Bellazzini}, {Bernet}, {Bertone},
  {Bianchi}, {Blanco-Cuaresma}, {Boch}, {Bombrun}, {Bossini}, {Bouquillon},
  {Bragaglia}, {Bramante}, {Breedt}, {Bressan}, {Brouillet}, {Bucciarelli},
  {Burlacu}, {Busonero}, {Butkevich}, {Buzzi}, {Caffau}, {Cancelliere},
  {C{\'a}novas}, {Cantat-Gaudin}, {Carballo}, {Carlucci}, {Carnerero},
  {Carrasco}, {Casamiquela}, {Castellani}, {Castro-Ginard}, {Castro Sampol},
  {Chaoul}, {Charlot}, {Chemin}, {Chiavassa}, {Cioni}, {Comoretto}, {Cooper},
  {Cornez}, {Cowell}, {Crifo}, {Crosta}, {Crowley}, {Dafonte}, {Dapergolas},
  {David}, {David}, {de Laverny}, {De Luise}, {De March}, {De Ridder}, {de
  Souza}, {de Teodoro}, {de Torres}, {del Peloso}, {del Pozo}, {Delbo},
  {Delgado}, {Delgado}, {Delisle}, {Di Matteo}, {Diakite}, {Diener},
  {Distefano}, {Dolding}, {Eappachen}, {Edvardsson}, {Enke}, {Esquej}, {Fabre},
  {Fabrizio}, {Faigler}, {Fedorets}, {Fernique}, {Fienga}, {Figueras},
  {Fouron}, {Fragkoudi}, {Fraile}, {Franke}, {Gai}, {Garabato},
  {Garcia-Gutierrez}, {Garc{\'\i}a-Torres}, {Garofalo}, {Gavras}, {Gerlach},
  {Geyer}, {Giacobbe}, {Gilmore}, {Girona}, {Giuffrida}, {Gomel}, {Gomez},
  {Gonzalez-Santamaria}, {Gonz{\'a}lez-Vidal}, {Granvik},
  {Guti{\'e}rrez-S{\'a}nchez}, {Guy}, {Hauser}, {Haywood}, {Helmi}, {Hidalgo},
  {Hilger}, {H{\l}adczuk}, {Hobbs}, {Holland}, {Huckle}, {Jasniewicz},
  {Jonker}, {Juaristi Campillo}, {Julbe}, {Karbevska}, {Kervella}, {Khanna},
  {Kochoska}, {Kontizas}, {Kordopatis}, {Korn}, {Kostrzewa-Rutkowska},
  {Kruszy{\'n}ska}, {Lambert}, {Lanza}, {Lasne}, {Le Campion}, {Le Fustec},
  {Lebreton}, {Lebzelter}, {Leccia}, {Leclerc}, {Lecoeur-Taibi}, {Liao},
  {Licata}, {Lindstr{\o}m}, {Lister}, {Livanou}, {Lobel}, {Madrero Pardo},
  {Managau}, {Mann}, {Marchant}, {Marconi}, {Marcos Santos}, {Marinoni},
  {Marocco}, {Marshall}, {Martin Polo}, {Mart{\'\i}n-Fleitas}, {Masip},
  {Massari}, {Mastrobuono-Battisti}, {Mazeh}, {McMillan}, {Messina},
  {Michalik}, {Millar}, {Mints}, {Molina}, {Molinaro}, {Moln{\'a}r},
  {Montegriffo}, {Mor}, {Morbidelli}, {Morel}, {Morris}, {Mulone}, {Munoz},
  {Muraveva}, {Murphy}, {Musella}, {Noval}, {Ord{\'e}novic}, {Orr{\`u}},
  {Osinde}, {Pagani}, {Pagano}, {Palaversa}, {Palicio}, {Panahi}, {Pawlak},
  {Pe{\~n}alosa Esteller}, {Penttil{\"a}}, {Piersimoni}, {Pineau}, {Plachy},
  {Plum}, {Poggio}, {Poretti}, {Poujoulet}, {Pr{\v{s}}a}, {Pulone}, {Racero},
  {Ragaini}, {Rainer}, {Raiteri}, {Rambaux}, {Ramos}, {Ramos-Lerate}, {Re
  Fiorentin}, {Regibo}, {Reyl{\'e}}, {Ripepi}, {Riva}, {Rixon}, {Robichon},
  {Robin}, {Roelens}, {Rohrbasser}, {Romero-G{\'o}mez}, {Rowell}, {Royer},
  {Rybicki}, {Sadowski}, {Sagrist{\`a} Sell{\'e}s}, {Sahlmann}, {Salgado},
  {Salguero}, {Samaras}, {Sanchez Gimenez}, {Sanna}, {Santove{\~n}a},
  {Sarasso}, {Schultheis}, {Sciacca}, {Segol}, {Segovia}, {S{\'e}gransan},
  {Semeux}, {Shahaf}, {Siddiqui}, {Siebert}, {Siltala}, {Slezak}, {Smart},
  {Solano}, {Solitro}, {Souami}, {Souchay}, {Spagna}, {Spoto}, {Steele},
  {Steidelm{\"u}ller}, {Stephenson}, {S{\"u}veges}, {Szabados}, {Szegedi-Elek},
  {Taris}, {Tauran}, {Taylor}, {Teixeira}, {Thuillot}, {Tonello}, {Torra},
  {Torra}, {Turon}, {Unger}, {Vaillant}, {van Dillen}, {Vanel}, {Vecchiato},
  {Viala}, {Vicente}, {Voutsinas}, {Weiler}, {Wevers}, {Wyrzykowski}, {Yoldas},
  {Yvard}, {Zhao}, {Zorec}, {Zucker}, {Zurbach}, \&
  {Zwitter}}]{2021A&A...649A...1G}
{Gaia Collaboration}, {Brown}, A.~G.~A., {Vallenari}, A., {et~al.} 2021, \aap,
  649, A1, \dodoi{10.1051/0004-6361/202039657}

\bibitem[{{Girardi} {et~al.}(2004){Girardi}, {Grebel}, {Odenkirchen}, \&
  {Chiosi}}]{2004A&A...422..205G}
{Girardi}, L., {Grebel}, E.~K., {Odenkirchen}, M., \& {Chiosi}, C. 2004, \aap,
  422, 205, \dodoi{10.1051/0004-6361:20040250}

\bibitem[{{Gratton} {et~al.}(2012){Gratton}, {Carretta}, \&
  {Bragaglia}}]{2012A&ARv..20...50G}
{Gratton}, R.~G., {Carretta}, E., \& {Bragaglia}, A. 2012, \aapr, 20, 50,
  \dodoi{10.1007/s00159-012-0050-3}

\bibitem[{{Grillmair}(2009)}]{2009ApJ...693.1118G}
{Grillmair}, C.~J. 2009, \apj, 693, 1118, \dodoi{10.1088/0004-637X/693/2/1118}

\bibitem[{{Grillmair} \& {Dionatos}(2006)}]{2006ApJ...643L..17G}
{Grillmair}, C.~J., \& {Dionatos}, O. 2006, \apjl, 643, L17,
  \dodoi{10.1086/505111}

\bibitem[{{Ibata} {et~al.}(2013){Ibata}, {Nipoti}, {Sollima}, {Bellazzini},
  {Chapman}, \& {Dalessandro}}]{Ibata13}
{Ibata}, R., {Nipoti}, C., {Sollima}, A., {et~al.} 2013, \mnras, 428, 3648,
  \dodoi{10.1093/mnras/sts302}

\bibitem[{{Ibata} {et~al.}(2009){Ibata}, {Bellazzini}, {Chapman},
  {Dalessandro}, {Ferraro}, {Irwin}, {Lanzoni}, {Lewis}, {Mackey}, {Miocchi},
  \& {Varghese}}]{Ibata09}
{Ibata}, R., {Bellazzini}, M., {Chapman}, S.~C., {et~al.} 2009, \apjl, 699,
  L169, \dodoi{10.1088/0004-637X/699/2/L169}

\bibitem[{{Ibata} {et~al.}(1997){Ibata}, {Wyse}, {Gilmore}, {Irwin}, \&
  {Suntzeff}}]{Ibata97}
{Ibata}, R.~A., {Wyse}, R. F.~G., {Gilmore}, G., {Irwin}, M.~J., \& {Suntzeff},
  N.~B. 1997, \aj, 113, 634, \dodoi{10.1086/118283}

\bibitem[{{King}(1966)}]{1966AJ.....71...64K}
{King}, I.~R. 1966, \aj, 71, 64, \dodoi{10.1086/109857}

\bibitem[{{K{\"u}pper} {et~al.}(2011){K{\"u}pper}, {Maschberger}, {Kroupa}, \&
  {Baumgardt}}]{2011MNRAS.417.2300K}
{K{\"u}pper}, A. H.~W., {Maschberger}, T., {Kroupa}, P., \& {Baumgardt}, H.
  2011, \mnras, 417, 2300, \dodoi{10.1111/j.1365-2966.2011.19412.x}

\bibitem[{{Kuzma} {et~al.}(2018){Kuzma}, {Da Costa}, \&
  {Mackey}}]{2018MNRAS.473.2881K}
{Kuzma}, P.~B., {Da Costa}, G.~S., \& {Mackey}, A.~D. 2018, \mnras, 473, 2881,
  \dodoi{10.1093/mnras/stx2353}

\bibitem[{{Lagioia} {et~al.}(2021){Lagioia}, {Milone}, {Marino}, {Tailo},
  {Renzini}, {Carlos}, {Cordoni}, {Dondoglio}, {Jang}, {Karakas}, \&
  {Dotter}}]{2021ApJ...910....6L}
{Lagioia}, E.~P., {Milone}, A.~P., {Marino}, A.~F., {et~al.} 2021, \apj, 910,
  6, \dodoi{10.3847/1538-4357/abdfcf}

\bibitem[{{Neumayer} {et~al.}(2020){Neumayer}, {Seth}, \&
  {B{\"o}ker}}]{NSCReview2020}
{Neumayer}, N., {Seth}, A., \& {B{\"o}ker}, T. 2020, \aapr, 28, 4,
  \dodoi{10.1007/s00159-020-00125-0}

\bibitem[{{Norris} {et~al.}(1981){Norris}, {Cottrell}, {Freeman}, \& {Da
  Costa}}]{1981ApJ...244..205N}
{Norris}, J., {Cottrell}, P.~L., {Freeman}, K.~C., \& {Da Costa}, G.~S. 1981,
  \apj, 244, 205, \dodoi{10.1086/158698}

\bibitem[{{Odenkirchen} {et~al.}(2003){Odenkirchen}, {Grebel}, {Dehnen}, {Rix},
  {Yanny}, {Newberg}, {Rockosi}, {Mart{\'\i}nez-Delgado}, {Brinkmann}, \&
  {Pier}}]{2003AJ....126.2385O}
{Odenkirchen}, M., {Grebel}, E.~K., {Dehnen}, W., {et~al.} 2003, \aj, 126,
  2385, \dodoi{10.1086/378601}

\bibitem[{{Pancino} {et~al.}(2017){Pancino}, {Bellazzini}, {Giuffrida}, \&
  {Marinoni}}]{2017MNRAS.467..412P}
{Pancino}, E., {Bellazzini}, M., {Giuffrida}, G., \& {Marinoni}, S. 2017,
  \mnras, 467, 412, \dodoi{10.1093/mnras/stx079}

\bibitem[{{Pe{\~n}arrubia} {et~al.}(2017){Pe{\~n}arrubia}, {Varri}, {Breen},
  {Ferguson}, \& {S{\'a}nchez-Janssen}}]{2017MNRAS.471L..31P}
{Pe{\~n}arrubia}, J., {Varri}, A.~L., {Breen}, P.~G., {Ferguson}, A. M.~N., \&
  {S{\'a}nchez-Janssen}, R. 2017, \mnras, 471, L31,
  \dodoi{10.1093/mnrasl/slx094}

\bibitem[{{Peebles}(1984)}]{1984ApJ...277..470P}
{Peebles}, P.~J.~E. 1984, \apj, 277, 470, \dodoi{10.1086/161714}

\bibitem[{{Portegies Zwart} {et~al.}(2010){Portegies Zwart}, {McMillan}, \&
  {Gieles}}]{2010ARA&A..48..431P}
{Portegies Zwart}, S.~F., {McMillan}, S. L.~W., \& {Gieles}, M. 2010, \araa,
  48, 431, \dodoi{10.1146/annurev-astro-081309-130834}

\bibitem[{{Riello} {et~al.}(2021){Riello}, {De Angeli}, {Evans}, {Montegriffo},
  {Carrasco}, {Busso}, {Palaversa}, {Burgess}, {Diener}, {Davidson}, {Rowell},
  {Fabricius}, {Jordi}, {Bellazzini}, {Pancino}, {Harrison}, {Cacciari}, {van
  Leeuwen}, {Hambly}, {Hodgkin}, {Osborne}, {Altavilla}, {Barstow}, {Brown},
  {Castellani}, {Cowell}, {De Luise}, {Gilmore}, {Giuffrida}, {Hidalgo},
  {Holland}, {Marinoni}, {Pagani}, {Piersimoni}, {Pulone}, {Ragaini}, {Rainer},
  {Richards}, {Sanna}, {Walton}, {Weiler}, \& {Yoldas}}]{2021A&A...649A...3R}
{Riello}, M., {De Angeli}, F., {Evans}, D.~W., {et~al.} 2021, \aap, 649, A3,
  \dodoi{10.1051/0004-6361/202039587}

\bibitem[{{Rockosi} {et~al.}(2002){Rockosi}, {Odenkirchen}, {Grebel}, {Dehnen},
  {Cudworth}, {Gunn}, {York}, {Brinkmann}, {Hennessy}, \&
  {Ivezi{\'c}}}]{2002AJ....124..349R}
{Rockosi}, C.~M., {Odenkirchen}, M., {Grebel}, E.~K., {et~al.} 2002, \aj, 124,
  349, \dodoi{10.1086/340957}

\bibitem[{{Schlafly} \& {Finkbeiner}(2011)}]{2011ApJ...737..103S}
{Schlafly}, E.~F., \& {Finkbeiner}, D.~P. 2011, \apj, 737, 103,
  \dodoi{10.1088/0004-637X/737/2/103}

\bibitem[{{Schlegel} {et~al.}(1998){Schlegel}, {Finkbeiner}, \&
  {Davis}}]{1998ApJ...500..525S}
{Schlegel}, D.~J., {Finkbeiner}, D.~P., \& {Davis}, M. 1998, \apj, 500, 525,
  \dodoi{10.1086/305772}

\bibitem[{{Searle} \& {Zinn}(1978)}]{1978ApJ...225..357S}
{Searle}, L., \& {Zinn}, R. 1978, \apj, 225, 357, \dodoi{10.1086/156499}

\bibitem[{{Sollima}(2020)}]{2020MNRAS.495.2222S}
{Sollima}, A. 2020, \mnras, 495, 2222, \dodoi{10.1093/mnras/staa1209}

\bibitem[{{Spitzer}(1987)}]{1987degc.book.....S}
{Spitzer}, L. 1987, {Dynamical evolution of globular clusters} (Princeton
  University Press)

\bibitem[{{Sun} {et~al.}(2018){Sun}, {Leroy}, {Schruba}, {Rosolowsky},
  {Hughes}, {Kruijssen}, {Meidt}, {Schinnerer}, {Blanc}, {Bigiel}, {Bolatto},
  {Chevance}, {Groves}, {Herrera}, {Hygate}, {Pety}, {Querejeta}, {Usero}, \&
  {Utomo}}]{2018ApJ...860..172S}
{Sun}, J., {Leroy}, A.~K., {Schruba}, A., {et~al.} 2018, \apj, 860, 172,
  \dodoi{10.3847/1538-4357/aac326}

\bibitem[{{Tiongco} {et~al.}(2016){Tiongco}, {Vesperini}, \&
  {Varri}}]{2016MNRAS.461..402T}
{Tiongco}, M.~A., {Vesperini}, E., \& {Varri}, A.~L. 2016, \mnras, 461, 402,
  \dodoi{10.1093/mnras/stw1341}

\bibitem[{{VandenBerg} {et~al.}(2013){VandenBerg}, {Brogaard}, {Leaman}, \&
  {Casagrande}}]{2013ApJ...775..134V}
{VandenBerg}, D.~A., {Brogaard}, K., {Leaman}, R., \& {Casagrande}, L. 2013,
  \apj, 775, 134, \dodoi{10.1088/0004-637X/775/2/134}

\bibitem[{{Vasiliev} \& {Baumgardt}(2021)}]{VB21}
{Vasiliev}, E., \& {Baumgardt}, H. 2021, \mnras, 505, 5978,
  \dodoi{10.1093/mnras/stab1475}

\bibitem[{{Vasiliev} \& {Belokurov}(2020)}]{SgrDwarf20}
{Vasiliev}, E., \& {Belokurov}, V. 2020, \mnras, 497, 4162,
  \dodoi{10.1093/mnras/staa2114}

\bibitem[{{Wan} {et~al.}(2021){Wan}, {Oliver}, {Baumgardt}, {Lewis}, {Gieles},
  {H{\'e}nault-Brunet}, {de Boer}, {Balbinot}, {Da Costa}, \&
  {Mackey}}]{2021MNRAS.502.4513W}
{Wan}, Z., {Oliver}, W.~H., {Baumgardt}, H., {et~al.} 2021, \mnras, 502, 4513,
  \dodoi{10.1093/mnras/stab306}

\bibitem[{{Webb} {et~al.}(2013){Webb}, {Harris}, {Sills}, \&
  {Hurley}}]{2013ApJ...764..124W}
{Webb}, J.~J., {Harris}, W.~E., {Sills}, A., \& {Hurley}, J.~R. 2013, \apj,
  764, 124, \dodoi{10.1088/0004-637X/764/2/124}

\end{thebibliography}
\bibliographystyle{aasjournal}
\end{document}